\documentclass[aps,preprint,floatfix,nofootinbib,showpacs]{revtex4-1}
\pdfoutput=1
\usepackage{graphicx,color}
\usepackage{hyperref}
\begin{document}

{\small
\begin{flushright}
CNU-HEP-13-01
\end{flushright} }

\title{Higgs Precision (Higgcision) Era begins}

\renewcommand{\thefootnote}{\arabic{footnote}}

\author{
Kingman Cheung$^{1,2}$, Jae Sik Lee$^3$, and
Po-Yan Tseng$^1$}
\affiliation{
$^1$ Department of Physics, National Tsing Hua University,
Hsinchu 300, Taiwan \\
$^2$ Division of Quantum Phases and Devices, School of Physics, 
Konkuk University, Seoul 143-701, Republic of Korea \\
$^3$ Department of Physics, Chonnam National University, \\
300 Yongbong-dong, Buk-gu, Gwangju, 500-757, Republic of Korea
}
\date{\today}

\begin{abstract}
  After the discovery of the Higgs boson at the LHC, it is natural to
  start the research program on the precision study of the Higgs-boson
  couplings to various standard model (SM) particles.  We provide a
  generic framework for the deviations of the couplings from their SM
  values by introducing a number of parameters.  We show that a large
  number of models beyond the SM can be covered, including
  two-Higgs-doublet models, supersymmetric models, little-Higgs
  models, extended Higgs sectors with singlets, and fourth generation
  models.  We perform global fits to the most updated data from CMS,
  ATLAS, and Tevatron under various initial conditions of the
  parameter set.
  In particular, we have made explicit comparisons between the 
  fitting results {\it before} and {\it after} the
    Moriond 2013 meetings.
  Highlights of the results include: (i) the
  nonstandard decay branching ratio of the Higgs boson is less than
  22\%; (ii) the most efficient way to achieve the best
  fit for the data before the Moriond update is to introduce
  additional particle contributions to the triangular-loop functions
  of $H\gamma\gamma$ and $Hgg$ vertices; (iii) the $1\sigma$ allowed
  range of the relative coupling of $HVV$ is
  $1.01\,^{+0.13}_{-0.14}$, which means that the
  electroweak-symmetry breaking contribution from the observed Higgs
  boson leaves only a small room for other Higgs bosons; (iv) the
  current data do not rule out pseudoscalar couplings nor pseudoscalar
  contributions to the $H\gamma\gamma$ and $Hgg$ vertices; and (v)
  the SM Higgs boson provides the best fit to all the
  current Higgs data.
\end{abstract}

\maketitle

\section{Introduction}

It is of very high expectation that the observed particle at the Large
Hadron Collider (LHC) \cite{atlas,cms} is the long-sought Higgs boson
of the standard model (SM), which was proposed in 1960s \cite{higgs}.
At the end of 2011, both the ATLAS and CMS \cite{cms-atlas}
experiments at the LHC have seen some excess of events of a possible
Higgs boson candidate in the decay modes of $H\to \gamma\gamma$, $H\to
WW^* \to \ell^+ \nu \ell^- \bar \nu$, and $H \to Z Z^* \to 4\ell$
channels.  Finally, the discovery was announced in July 2012 by ATLAS
\cite{atlas} and CMS \cite{cms}. The channels $WW$ and $ZZ$ are
consistent with the predictions of the SM Higgs boson, while the
$\gamma\gamma$ rate is somewhat higher than expectation.  Some
evidence is seen in the $b\bar b$ mode at the Tevatron
\cite{tevatron}, but the mass range is quite wide.  
On the other hand,  
the $\tau^+\tau^-$ mode appears to be suppressed before the very recent 
update, although the data contain large uncertainties.

A previous update was presented at the {\it Hadron Collider Physics
Symposium 2012} \cite{hpc,tev} and in a series of experimental notes 
\cite{tev_h_bb,atlas_h_aa,atlas_com,atlas_h_tau,atlas_h_zz,cms_com,cms_h_zz,eff}
at the end of 2012. 
At that time, the $\tau^+\tau^-$ data began to appear, but still too
early to say something concrete.
The diphoton production rate was somewhat higher than the 
SM prediction by a factor of $1.4-1.8$.
Nevertheless, the deviations are only $1-2~\sigma$.
A large number of models have 
been put forward to account for the observed particle at 125 GeV,
including the SM, supersymmetric models such as the minimal supersymmetric
standard model \cite{ex-susy}, the 
next-to-minimal supersymmetric standard model \cite{ex-nmssm}, 
the $U(1)$-extended
minimal supersymmetric standard model \cite{ex-umssm}, 
fermiophobic Higgs boson \cite{ex-fp}, 
two-Higgs-doublet models (2HDM) of various types \cite{ex-2hdm}, 
Randall-Sundrum radion \cite{ex-rs}, inert-Higgs doublet model \cite{ex-in},
etc
(a summary of various models can be found in Ref.~\cite{ours}.)
They all can explain the enhanced diphoton rate with some choices of 
parameter space.
Yet, more data are needed in order to firmly establish the excess in the
diphoton channel.
The most recent update was during the Moriond 2013 meetings \cite{moriond}.
The updated data can be found in a number of conference note
from the ATLAS \cite{atlas_h_aa_2013,atlas_com_2013} and CMS
\cite{cms_aa_2013,cms_zz_2013,cms_ww_2013,cms_tau_2013}.

Currently, a number of decay and production channels are available. On 
the production side, there are gluon fusion (ggF), vector-boson fusion (VBF),
and associated production with a $V=W/Z$ boson (VH) 
and top quarks (ttH); 
while the decay 
channels include $\gamma\gamma$, $ZZ^* \to 4\ell$, $WW^* \to \ell^+ \nu \ell^-
\bar \nu$, $b\bar b$, and $\tau^+ \tau^-$. 
One can extract useful information on the size of the Higgs boson couplings
from the available data. However, in order to do that
the dependence of
various production and decay modes on the Higgs couplings 
has to be taken into account correctly.
For example, the ggF depends on the Higgs couplings to a pair
of top ($H\bar tt$) and bottom ($H\bar bb$) quarks,
as well as 
possible existence of exotic colored particles running in the loop;
while the VBF and VH depend only on 
the Higgs coupling to a pair of vector bosons ($HVV$).
Also, the decays into $WW^*$ and
$ZZ^*$ simply depend on $HVV$, and the decays into $b\bar b$ and $\tau^+
\tau^-$ depend on $H\bar bb$ and $H \bar\tau\tau$ respectively; but 
the decay into $\gamma\gamma$ involves all of the above couplings and
perhaps new electrically-charged particles in the triangular loop.
A global analysis of all the Higgs couplings using all the available data
would be extremely useful to identify the observed Higgs boson. 
Once we disentangle each of the Higgs couplings from the global data set,
we can use the result to compare with models.
This approach is in contrast to those top-down approaches,
which usually start with a model, calculate the signal strengths, and then 
find the allowed parameter space to fit to the data.

Indeed, the Higgs precision era just begins. There have been a number 
of works in the past few months going in this direction,
in a more or less model-independent framework 
\cite{r1,r2,r3,r4,r5,r6,r7,r8,r9,r10,r11,r12,r13,r14,r15,r16,r17,r18,anom1,anom2,anom3},
in 2HDM framework \cite{2hdm0,2hdm1,2hdm2,2hdm3,2hdm4,2hdm5,2hdm6,2hdm7}, and
in supersymmetry \cite{susy1,susy2,susy3,susy4,susy5}.
Also, there are studies toward the determination of the spin-parity
nature of the Higgs boson (see Ref.~\cite{hy} for more references in
literature) that cannot be obtained from the signal strengths.

About a couple of weeks after we posted the first version of our paper
to arXiv, both ATLAS and CMS Collaborations have updated their Higgs
data during the Moriond 2013 meetings \cite{moriond}. 
They have released a series of conference notes 
\cite{atlas_com_2013,cms_aa_2013,cms_zz_2013,cms_ww_2013,cms_tau_2013} 
on the new data.  In particular, the most striking
is the change of the diphoton data by the CMS \cite{cms_aa_2013}. 
Because of this change the overall fits also change dramatically.  
In the following, we will show the results {\it before} and {\it after} 
the Moriond 2013 meetings.

The characteristic features of our analysis are summarized as follows.
\begin{enumerate}
\item 
We allow the Yukawa couplings to the charged-lepton ($H\bar\tau\tau$) and
the down-quark ($H\bar bb$) sectors to vary independently. 
This can be realized in some versions of the 2HDM. 
This has also been adopted in a few previous works.

\item 
We allow an independent deviation in the total decay width of the Higgs boson,
in addition to the parameters of the Yukawa and $HVV$ couplings. This can be
realized in nonstandard decays of the Higgs boson, e.g., $h \to
\widetilde{\chi}^0_1 \widetilde{\chi}^0_1$ in SUSY models, $H \to a a$ where
$a$ stands for some lighter Higgs bosons in the model.  
This has been included in some previous works.
An interesting result
is obtained because of this improvement. The nonstandard decay width is
constrained to be less than $1.2$ MeV at 95\% CL, 
which accounts for a branching ratio of about 22\%.

\item  
In the loop vertices of $H\gamma\gamma$ and $Hgg$, we allow 
new parameters $\Delta S^\gamma$ and $\Delta S^g$ respectively, which 
can most conveniently account for the effects of new particles 
contributing to the loops. 

\item 
Preliminarily, the pseudoscalar interpretation of the observed Higgs boson
is disfavored. 
However, it may not apply to the case of CP-mixed state
carrying both scalar-type and pseudoscalar-type couplings simultaneously.
We perform a whole new analysis
including both scalar-type and pseudoscalar-type Yukawa couplings, and
both radiatively-induced scalar and pseudoscalar 
$H\gamma\gamma$ and $Hgg$ vertices. 
We show that pseudoscalar couplings are actually not ruled
out based on signal strengths only,
by giving equivalently good fits compared to the CP-conserving case.

\end{enumerate}

The organization of the paper 
is as follows. In the next section, we describe the interactions
of the Higgs boson, including deviations in the Yukawa couplings and
deviations in the loop functions of $H\gamma\gamma$, $Hgg$, and $HZ\gamma$
vertices, as well as the notation used in the analysis. In Sec. III, we list
the Higgs boson data both {\it before} and {\it after} Moriond 2013
meetings that we use in this analysis. We present the results
of various fits in Sec. IV, and the readers can see
how the fits change when the Higgs data are changed.
In Sec. V, we present the results using both
scalar-type and pseudoscalar-type couplings.
We conclude in Sec. VI.

\section{Formalism}

\subsection{Higgs Couplings}
We follow the conventions and notations of 
{\tt CPsuperH}~\cite{Lee:2003nta,Lee:2007gn,Lee:2012wa}
for the Higgs couplings to the SM particles assuming the Higgs boson is 
a generally CP-mixed state without carrying any definite CP--parity.

\begin{itemize}
%
\item\underline{Higgs couplings to fermions}:
\begin{equation}
\label{eq1}
{\cal L}_{H\bar{f}f}\ =\ - \sum_{f=u,d,l}\,\frac{g m_f}{2 M_W}\,
\sum_{i=1}^3\, H\, \bar{f}\,\Big( g^S_{H\bar{f}f}\, +\,
ig^P_{H\bar{f}f}\gamma_5 \Big)\, f\ .
\end{equation}
For the SM couplings, $g^S_{H\bar{f}f}=1$ and $g^P_{H\bar{f}f}=0$.
\item\underline{Higgs couplings to the massive vector bosons}:
\begin{equation}
\label{eq2}
{\cal L}_{HVV}  =  g\,M_W \, \left(
g_{_{HWW}} W^+_\mu W^{- \mu}\ + \
g_{_{HZZ}} \frac{1}{2c_W^2}\,Z_\mu Z^\mu\right) \, H\,.
\end{equation}
For the SM couplings, 
we have $g_{_{HWW}}=g_{_{HZZ}}\equiv g_{_{HVV}}=1$, respecting 
the custodial symmetry.
\item\underline{Higgs couplings to two photons}:
The amplitude for the decay process
$H \rightarrow \gamma\gamma$ can be written as
\begin{equation} \label{hipp}
{\cal M}_{\gamma\gamma H}=-\frac{\alpha M_{H}^2}{4\pi\,v}
\bigg\{S^\gamma(M_{H})\,
\left(\epsilon^*_{1\perp}\cdot\epsilon^*_{2\perp}\right)
 -P^\gamma(M_{H})\frac{2}{M_{H}^2}
\langle\epsilon^*_1\epsilon^*_2 k_1k_2\rangle
\bigg\}\,,
\end{equation}
where $k_{1,2}$ are the momenta of the two photons and
$\epsilon_{1,2}$ the wave vectors of the corresponding photons,
$\epsilon^\mu_{1\perp} = \epsilon^\mu_1 - 2k^\mu_1 (k_2 \cdot
\epsilon_1) / M^2_{H}$, $\epsilon^\mu_{2\perp} = \epsilon^\mu_2 -
2k^\mu_2 (k_1 \cdot \epsilon_2) / M^2_{H}$ and $\langle \epsilon_1
\epsilon_2 k_1 k_2 \rangle \equiv \epsilon_{\mu\nu\rho\sigma}\,
\epsilon_1^\mu \epsilon_2^\nu k_1^\rho k_2^\sigma$. 
The decay rate of $H\to \gamma\gamma$ is 
proportional to $|S^\gamma|^2 + |P^\gamma|^2$.
Including some additional loop contributions from new particles,
the scalar and
pseudoscalar form factors, retaining only the dominant loop
contributions from the third--generation fermions and $W^\pm$,
are given by
\footnote{
For the loop functions of $F_{sf,pf,1}(\tau)$, 
we refer to, for example, Ref.~\cite{Lee:2003nta}.}
\begin{eqnarray}
S^\gamma(M_{H})&=&2\sum_{f=b,t,\tau} N_C\,
Q_f^2\, g^{S}_{H\bar{f}f}\,F_{sf}(\tau_{f}) 
- g_{_{HWW}}F_1(\tau_{W}) 
+ \Delta S^\gamma \,, \nonumber \\
P^\gamma(M_{H})&=&2\sum_{f=b,t,\tau}
N_C\,Q_f^2\,g^{P}_{H\bar{f}f}\,F_{pf}(\tau_{f})
+ \Delta P^\gamma \,, 
\end{eqnarray}
where $\tau_{x}=M_{H}^2/4m_x^2$, $N_C=3$ for quarks and $N_C=1$ for
taus, respectively.
The additional contributions $\Delta S^\gamma$ and $\Delta P^\gamma$
are assumed to be real in our work,
as there are unlikely any new
{\it charged} particles lighter than $M_H/2$.

Taking $M_H=125.5$ GeV, we find that
\begin{eqnarray}
S^\gamma &\simeq& 
-8.35\,g_{HWW}+1.76\, g^S_{H\bar{t}t}
+(-0.015+0.017\,i)\,g^S_{H\bar{b}b} \nonumber \\ &&
+(-0.024+0.021\,i)\,g^S_{H\bar{\tau}\tau}
+(-0.007+0.005\,i)\,g^S_{H\bar{c}c}+\Delta S^\gamma
\nonumber \\[2mm]
P^\gamma &\simeq& 2.78\, g^P_{H\bar{t}t} 
+(-0.018+0.018\,i)\,g^P_{H\bar{b}b} \nonumber \\ &&
+(-0.025+0.022\,i)\,g^P_{H\bar{\tau}\tau}
+(-0.007+0.005\,i)\,g^P_{H\bar{c}c}+\Delta P^\gamma  \label{5}
\end{eqnarray}
giving
$S^\gamma_{\rm SM}=-6.64+0.043\,i$ and $P^\gamma_{\rm SM}=0$.  
%
\item\underline{Higgs couplings to two gluons}: 
Similar to $H\to\gamma\gamma$, 
the amplitude for the decay process
$H \rightarrow gg$ can be written as
\begin{equation} \label{higg}
{\cal M}_{gg H}=-\frac{\alpha_s\,M_{H}^2\,\delta^{ab}}{4\pi\,v}
\bigg\{S^g(M_{H})
\left(\epsilon^*_{1\perp}\cdot\epsilon^*_{2\perp}\right)
 -P^g(M_{H})\frac{2}{M_{H}^2}
\langle\epsilon^*_1\epsilon^*_2 k_1k_2\rangle
\bigg\}\,,
\end{equation}
where $a$ and $b$ ($a,b=1$ to 8) are indices of the eight $SU(3)$
generators in the adjoint representation.
The decay rate of $H\to gg $ is 
proportional to $|S^g|^2 + |P^g|^2$.
Again, including some additional loop contributions from new particles,
the scalar and pseudoscalar form factors are given by
\begin{eqnarray}
S^g(M_{H})&=&\sum_{f=b,t}
g^{S}_{H\bar{f}f}\,F_{sf}(\tau_{f}) +  
\Delta S^g\,,
\nonumber \\
P^g(M_{H})&=&\sum_{f=b,t}
g^{P}_{H\bar{f}f}\,F_{pf}(\tau_{f}) +
\Delta P^g
\,.
\end{eqnarray}
The additional contributions $\Delta S^g$ and $\Delta P^g$
are assumed to be real again. 

Taking $M_H=125.5$ GeV, we find that
\begin{eqnarray}
S^g &\simeq& 
0.688\, g^S_{H\bar{t}t}
+(-0.037+0.050\,i)\,g^S_{H\bar{b}b} + \Delta S^g\,, \nonumber \\ 
P^g &\simeq& 
1.047\, g^P_{H\bar{t}t} 
+(-0.042+0.050\,i)\,g^P_{H\bar{b}b} + \Delta P^g\,, \label{8}
\end{eqnarray}
giving $S^g_{\rm SM}=0.651+0.050\,i$ and $P^g_{\rm SM}=0$.
\item\underline{Higgs couplings to $Z$ and $\gamma$}: 
The amplitude for the decay process $H \to
Z(k_1,\epsilon_1)\
\gamma(k_2,\epsilon_2)$ can be written as
\begin{equation}
{\cal M}_{Z\gamma H} = -\,\frac{\alpha}{2\pi v}\left\{
S^{Z\gamma}(M_{H})\,
\left[ k_1\cdot k_2\,\epsilon_1^*\cdot\epsilon_2^*
-k_1\cdot\epsilon_2^*\,k_2\cdot\epsilon_1^* \right] \ - \
P^{Z\gamma}(M_{H})\,
\langle \epsilon_1^*\epsilon_2^* k_1 k_2\rangle
\right\}
\end{equation}
where $k_{1,2}$ are the momenta of the $Z$ boson and the photon (we note that
$2k_1\cdot k_2 = M_{H}^2-M_Z^2$),
$\epsilon_{1,2}$ are their polarization vectors.
The scalar and pseudoscalar form factors are given by
\begin{eqnarray}
S^{Z\gamma}(M_{H})\, &=& 
2 \sum_{f=t,b,\tau} Q_f N_C^f m_f^2\
\frac{I_3^f-2s_W^2 Q_f}{s_Wc_W}\ g^S_{H\bar{f}f}\ F_f^{(0)}
+M_Z^2 \cot\theta_W g_{_{HWW}} F_W 
+ \Delta S^{Z\gamma}\,,
\nonumber \\
P^{Z\gamma}(M_{H})\, &=& 
2 \sum_{f=t,b,\tau} Q_f N_C^f m_f^2\
\frac{I_3^f-2s_W^2 Q_f}{s_Wc_W}\ g^P_{H\bar{f}f}\ F_f^{(5)} 
+ \Delta P^{Z\gamma}\,.
\end{eqnarray}
The additional contributions $\Delta S^{Z\gamma}$ and $\Delta P^{Z\gamma}$
are assumed to be real. 
The loop functions are
\footnote{For the functions of $C_{0,2}(m^2)$,
we refer to~\cite{Djouadi:1996yq}.}
\begin{eqnarray}
F_f^{(0)}&=& C_0(m_f^2) + 4 C_2(m_f^2)\,, \nonumber \\
F_f^{(5)}&=& C_0(m_f^2) \,, \nonumber \\
F_W &=& 2\left[\frac{M_{H}^2}{M_W^2}(1-2c_W^2)+2(1-6c_W^2)\right]
C_2(M_W^2)+4(1-4c_W^2)C_0(M_W^2)\,.
\end{eqnarray}

Taking $M_H=125.5$ GeV, we find
\begin{eqnarray}
S^{Z\gamma} &\simeq& 
-11.966\,g_{HWW}+0.615\, g^S_{H\bar{t}t}
+(-0.008+0.004\,i)\,g^S_{H\bar{b}b} \nonumber \\ &&
+(-0.0004+0.0002\,i)\,g^S_{H\bar{\tau}\tau}
+\Delta S^{Z\gamma}\,,
\nonumber \\[2mm]
P^{Z\gamma} &\simeq& 0.933\, g^P_{H\bar{t}t} 
+(-0.009+0.004\,i)\,g^P_{H\bar{b}b}  \nonumber \\ &&
+(-0.0004+0.0002\,i)\,g^S_{H\bar{\tau}\tau}
+\Delta P^{Z\gamma}\,,
\end{eqnarray}
giving $S^{Z\gamma}_{\rm SM}=-11.358+0.004\,i$ and $P^{Z\gamma}_{\rm SM}=0$.

In passing,
we recall that the $Z$-boson couplings to the quarks and leptons are given by
\begin{eqnarray}
{\cal L}_{Z\bar{f}f} = -\,g_Z\,
\bar{f}\,\gamma^\mu\,\left(v_{Z\bar{f}f} - a_{Z\bar{f}f}
\gamma_5\right)\,f\,Z_\mu \, ,
\end{eqnarray}
where  $g_Z=e/(s_Wc_W)$,
$v_{Z\bar{f}f}=I_3^f/2-Q_f s_W^2$ and $a_{Z\bar{f}f}=I_3^f/2$
with $I_3^{u}=1/2$ and $I_3^{d,l}=-1/2$.
%
\end{itemize}
Finally, we define
the ratios of the effective 
Higgs couplings to $gg$, $\gamma\gamma$, and
$Z\gamma$ relative to the SM ones as follows:
\begin{equation}
C_g\equiv\sqrt{\frac{\left|S^g\right|^2+\left|P^g\right|^2}
{\left|S^g_{\rm SM}\right|^2}}\,; \ \
C_\gamma\equiv\sqrt{\frac{\left|S^\gamma\right|^2+\left|P^\gamma\right|^2}
{\left|S^\gamma_{\rm SM}\right|^2}}\,; \ \
C_{Z\gamma}\equiv\sqrt{\frac{\left|S^{Z\gamma}\right|^2+\left|P^{Z\gamma}\right|^2}
{\left|S^{Z\gamma}_{\rm SM}\right|^2}}\,. 
\end{equation}
Note that the ratios of decay rates relative to the SM are 
given by $|C_g|^2$, $|C_\gamma|^2$, and $|C_{Z\gamma}|^2$, respectively.

\subsection{Signal strengths}
The theoretical signal strength may be written  as the product
\begin{equation}
\widehat\mu({\cal P},{\cal D}) \simeq
\widehat\mu({\cal P})\ \widehat\mu({\cal D}) 
\end{equation}
where ${\cal P}={\rm ggF}, {\rm VBF}, {\rm VH}, {\rm ttH}$ denote the production mechanisms
and ${\cal D}=\gamma\gamma , ZZ, WW, b\bar{b}, \tau\bar\tau$
the decay channels.
More explicitly, we are taking
\begin{eqnarray}
\widehat\mu({\rm ggF}) &=&
\frac{\left|S^g(M_H)\right|^2+\left|P^g(M_H)\right|^2}
{\left|S^g_{\rm SM}(M_H)\right|^2}\,, \nonumber \\[2mm]
\widehat\mu({\rm VBF}) &=& g_{_{HWW,HZZ}}^2\,, \nonumber \\[2mm]
\widehat\mu({\rm VH}) &=& g_{_{HWW,HZZ}}^2\,, \nonumber \\[2mm]
\widehat\mu({\rm ttH}) &=& \left(g^S_{H\bar{t}t}\right)^2 
+\left(g^P_{H\bar{t}t}\right)^2\,; 
\end{eqnarray}
and
\begin{equation}
\widehat\mu({\cal D}) = \frac{B(H\to {\cal D})}{B(H_{\rm SM}\to {\cal D})}
\end{equation}
with
\begin{equation}
\label{eq:dgam}
B(H\to {\cal D})=\frac{\Gamma(H\to{\cal D})}
{\Gamma_{\rm tot}(H)+\Delta\Gamma_{\rm tot}}
\end{equation}
Note that we introduce an arbitrary non-SM contribution $\Delta\Gamma_{\rm tot}$
to the total decay width. Incidentally,
$\Gamma_{\rm tot}(H)$ becomes the SM total decay width
when 
$g^S_{H\bar{f}f}=1$,
$g^P_{H\bar{f}f}=0$,
$g_{_{HWW,HZZ}}=1$,
$\Delta S^{\gamma,g,Z\gamma}=
\Delta P^{\gamma,g,Z\gamma}=0$.

The experimentally observed signal strength 
should be compared to the theoretical one summed
over all production mechanisms:
\begin{equation}
\label{eq:cpq}
\mu({\cal Q},{\cal D}) =
\sum_{{\cal P}={\rm ggF, VBF, VH, ttH}}\ C_{{\cal Q} {\cal P}}\
\widehat\mu({\cal P},{\cal D}) 
\end{equation}
where ${\cal Q}$ denote the experimentally defined
channel involved with the decay ${\cal D}$ and
the decomposition coefficients $C_{{\cal Q} {\cal P}}$ may depend
on the relative Higgs production cross sections for a given 
Higgs-boson mass, experimental cuts, etc.

The $\chi^2$ associated with an uncorrelated observable is
\begin{equation}
\chi^2({\cal Q},{\cal D}) = 
\frac{\left[\mu({\cal Q},{\cal D})-\mu^{\rm EXP}({\cal Q},{\cal D})\right]^2}
{\left[\sigma^{\rm EXP}({\cal Q},{\cal D})\right]^2}\,,
\end{equation}
where $\sigma^{\rm EXP}({\cal Q},{\cal D})$ denotes the experimental error.
For two correlated observables, we use
\begin{eqnarray}
\chi^2({\cal Q}_1,{\cal D};{\cal Q}_2,{\cal D}) &=& \Bigg\{
\frac{\left[\mu({\cal Q}_1,{\cal D})-\mu^{\rm EXP}({\cal Q}_1,{\cal D})\right]^2}
{\left[\sigma^{\rm EXP}({\cal Q}_1,{\cal D})\right]^2} +
\frac{\left[\mu({\cal Q}_2,{\cal D})-\mu^{\rm EXP}({\cal Q}_2,{\cal D})\right]^2}
{\left[\sigma^{\rm EXP}({\cal Q}_2,{\cal D})\right]^2} \nonumber \\[2mm]
&&-2\rho\
\frac{\left[\mu({\cal Q}_1,{\cal D})-\mu^{\rm EXP}({\cal Q}_1,{\cal D})\right]
\left[\mu({\cal Q}_2,{\cal D})-\mu^{\rm EXP}({\cal Q}_2,{\cal D})\right]}
{\left[\sigma^{\rm EXP}({\cal Q}_1,{\cal D})\right]
\left[\sigma^{\rm EXP}({\cal Q}_2,{\cal D})\right]}  \Bigg\}\Bigg/(1-\rho^2) 
\nonumber \\
\end{eqnarray}
where $\rho$ is the correlation coefficient.

\subsection {Parameters using in the fits}

Without loss of generality we use the following notation for the 
parameters in the fits:
\begin{eqnarray}
&&
C_u^S=g^S_{H\bar uu}\,, \ \
C_d^S=g^S_{H\bar dd}\,, \ \
C_\ell^S=g^S_{H\bar ll}\,; \ \
C_v=g_{_{HVV}}\,; \nonumber \\
&&
C_u^P=g^P_{H\bar uu}\,, \ \
C_d^P=g^P_{H\bar dd}\,, \ \
C_\ell^P=g^P_{H\bar ll}\,. 
\end{eqnarray}
Here we assume generation independence and also custodial 
symmetry between the $W$ and $Z$ bosons. Note that the tree-level pseudoscalar
couplings to $W$ and $Z$ bosons are zero.
The first and second
generation fermion couplings to the Higgs boson is rather small, but if 
in the near future the $H \to \mu^+ \mu^-$ can be measured, one may set 
independent parameters $C_\mu$ and $C_\tau$ (for the present work we 
consider them to be the same.)

In the fits,
we further use 
\begin{equation}
\Delta S^g\,, \ \ \Delta S^\gamma\,;  \ \ \ 
\Delta P^g\,, \ \ \Delta P^\gamma\,
\end{equation}
which are real
quantities assuming that any new particles running in the triangular loop
are heavier than one half of the Higgs boson mass. Note that the 
quantities $S^g$ and $S^\gamma$ are in general complex in both the SM
and beyond the SM. 
For the most direct comparison with experimental data we use $C_\gamma$ 
and $C_g$ in the plots.
\footnote
{The quantities $\Delta C_\gamma$ and $\Delta C_g$ defined in Ref.~\cite{r17}
are in general complex. 
It would be more difficult to use them as fitting parameters.}

Another important quantity is the additional contributions
to the width of the Higgs boson, $\Delta\Gamma_{\rm tot}$
 as in Eq.~(\ref{eq:dgam}).
The parameter $\Delta \Gamma_{\rm tot}$ takes into account the 
nonstandard decays
of the Higgs boson, e.g., $h_1 \to a_1 a_1 $ in NMSSM, $h_1 \to
\tilde{\chi}^0_1 \tilde{\chi}^0_1$ in SUSY, $h_2 \to h_1 h_1$ in other
extended Higgs models. The current data still allow a small amount of
Higgs invisible decay branching ratio.

We first use the following scalar-type couplings:
\[
 C_u^S,\; C_d^S,\; C_\ell^S,\; C_v,\; \Delta S^g,\; \Delta S^\gamma,\;
   \Delta \Gamma_{\rm tot}
\]
to fit to the Higgs data in Sec. IV, where we focus on the SM-like Higgs 
boson. In Sec. V, where we also consider the possibility that
the observed Higgs boson can allow some level of pseudoscalar-type couplings
\[
C_u^P,\; C_d^P,\; C_\ell^P,\; \Delta P^{\gamma},\; \Delta P^g \;,
\]
in addition to the scalar ones.

\subsection{Two-Higgs Doublet Models}
Two-Higgs-doublet models employ two Higgs doublets
in the process of electroweak-symmetry breaking (EWSB).
A discrete $Z_2$ symmetry is usually imposed 
in order to avoid dangerous tree-level flavor-changing neutral currents.
The most studied are the type I and type II models. They can easily be
covered by the framework presented in this paper.
We illustrate using the model II.

The Higgs sector consists of two Higgs doublets
$H_u = \left( H_u^+ \, H^0_u \right)^T$ and 
$H_d = \left( H_d^+ \, H^0_d \right)^T$
where the subscripts $u,d$ denote the right-handed quark singlet fields
that the Higgs doublets couple to. 
After EWSB,
there are two CP-even, one CP-odd, and a pair of charged Higgs bosons.
The parameters of the model in the CP-conserving case can be chosen as
\[
  m_h,\; m_H,\; 
m_A, \; m_{H^+},\; 
\tan\beta \equiv \frac{v_u}{v_d},\;
 \alpha
\]
where $\alpha$ is the mixing angle between the two CP-even Higgs bosons.
The couplings of the lighter CP-even Higgs boson $h$ (assumed to be 
the observed boson) to the tau, bottom, top quarks, and $W/Z$ boson relative
to their corresponding SM values are given by
\[
  \begin{tabular}{ccccc}
   &  $\tau^- \tau^+$ &  $b \bar b$ & $t \bar t$ & $W^+W^-/ZZ$ \\
$h$: \quad & $\; {- \sin\alpha/\cos\beta}\;\;  $ & $\; \; 
{- \sin\alpha/\cos\beta}
\;\;  $ & $\; \; {\cos\alpha/\sin\beta} \;\;$ & $\sin(\beta-\alpha)$
  \end{tabular}
\]
Therefore, we can equate these quantities with the definitions of 
$C^S_u,C^S_d,C^S_\ell,C_v$, given by
\begin{equation}
C^S_u = \frac{\cos\alpha}{\sin\beta},\;\;
C^S_d = C^S_\ell = - \frac{\sin\alpha}{\cos\beta},\;\;
C_v = \sin(\beta-\alpha)  \;.
\end{equation}
{}From the above relations, one may derive the following consistency relations
which should hold in the type II model:
\begin{eqnarray}
&& 
\cos^2\beta=\frac{C_v-C^S_u}{C^S_{d,\ell}-C^S_u}\,, \ \ \ 
\sin^2\beta=\frac{C^S_{d,\ell}-C_v}{C^S_{d,\ell}-C^S_u}\,;
\nonumber \\[2mm]
&& 
\sin^2\alpha=\frac{C_v-1/C^S_u}{1/C^S_{d,\ell}-1/C^S_u}\,, \ \ \
\cos^2\alpha=\frac{1/C^S_{d,\ell}-C_v}{1/C^S_{d,\ell}-1/C^S_u}\,.
\end{eqnarray}

On the other hand,
the only additional particle that can run in the triangular loop
of $h\gamma\gamma$ is the charged Higgs boson. Also, there are no
new particles other than the SM particles that the Higgs boson $h$
can decay into. Therefore, the other quantities
\begin{equation}
\Delta S^\gamma \not\!\!{=}~0,\;\;
\Delta S^g = 0,\;\;
\Delta \Gamma_{\rm tot} = 0  \;.
\end{equation}
Thus, the two-Higgs doublet models, in general, can be covered by
our framework.

In more complicated Higgs sectors, e.g., with additional singlets,
there may be lighter Higgs bosons that the observed Higgs boson can decay
into. In such a case, the additional decay modes will contribute to
$\Delta \Gamma_{\rm tot}$. 

\subsection{Models with singlet Higgs bosons}
Simple extensions of the SM Higgs sector with one or more Higgs singlet 
fields are attractive, because they can often provide a dark matter
candidate once some kinds of discrete symmetries are imposed on the
extra fields. Some variants can be found in Refs.~\cite{sin1,sin2,sin3,sin4}.
In the simplest version \cite{sin4}, the Higgs sector consists of the usual
SM Higgs doublet $\Phi$ and a real singlet Higgs field $\chi$. 
They couple to each other via a renormalizable interaction $\rho \chi^2 
\Phi^\dagger \Phi$. A discrete $Z_2$ symmetry is imposed on $\chi \to -\chi$
such that $\chi$ cannot develop the vacuum expectation value (VEV) and 
becomes a dark matter candidate. After $\Phi$ develops the VEV,
$\chi$ couples to the $H$ via the interactions $\chi^2 H$ and $\chi^2 H^2$.
Therefore, the Higgs boson, in addition to the couplings to the SM fermions,
also couples to a pair of $\chi$s. The only modification in our framework
is the total decay width, accommodated by $\Delta \Gamma_{\rm tot}$.

\subsection{Supersymmetric models }
There are many varieties in supersymmetric models which contain at least
two Higgs doublets. In the minimal supersymmetric extension of the SM (MSSM),
there are three neutral Higgs states and, in principle,
any of them can be the candidate 
for the observed particle at 125 GeV. If the $i$-th ($i=1,2,3$) Higgs state
is assumed to be the observed particle, we have
\begin{eqnarray}
\label{eq:mssm}
&&
C_u^S=\frac{O_{\phi_2 i}}{\sin\beta}\,, \ \
C_d^S=C_\ell^S=\frac{O_{\phi_1 i}}{\cos\beta}\,, \ \
C_v=O_{\phi_1 i}\cos\beta+O_{\phi_2 i}\sin\beta\,;
\nonumber \\[2mm]
&&
C_u^P=-\frac{\cos\beta}{\sin\beta}\,O_{a i}\,, \ \
C_d^P=C_\ell^P=-\frac{\sin\beta}{\cos\beta}\,O_{a i}\,, \ \
\end{eqnarray}
where $O_{\phi_1 i\,,\phi_2 i}$ and $O_{a i}$ denote the CP-even and CP-odd
components of the $i$-th Higgs state, respectively
\footnote{For the precise definition of the orthogonal $3\times 3$ 
Higgs-boson-mixing
matrix $O$, we refer to Ref.~\cite{Lee:2003nta}.}, and $\cos\beta$
and $\sin\beta$ are defined in the same way as in the type-II 2HDM.
Including the threshold corrections to the third-generation
Yukawa couplings, the above tree-level relations undergo some changes.
Nevertheless, this case can be covered because we are treating
$C_d^{S,P}$ and $C_\ell^{S,P}$ independently in our framework. 

Beyond the MSSM there could be more than three neutral Higgs states. 
In this case, it is straightforward to find similar relations 
as in Eq.~(\ref{eq:mssm}). For example, in the next-to-minimal
supersymmetric standard model (NMSSM) the neutral Higgs bosons 
have a $5\times 5$ mixing matrix~\cite{Cheung:2010ba}.

Any bosonic and fermionic contributions of SUSY particles to
the $H\gamma\gamma$ and $Hgg$ vertices, including the charged Higgs-boson
contribution, can be nicely accommodated by
using the parameters $\Delta S^{g,\gamma}$ and $\Delta P^{g,\gamma}$. Also,
the parameter $\Delta\Gamma_{\rm tot}$ can take into account any
Higgs decays into the non-SM particles.

\subsection{Little Higgs models}
Little Higgs models belong to a class of models in which the quadratic 
divergences to the Higgs boson are cancelled by a set of particles having 
the same spin statistics as the SM particles. For each SM particle 
there corresponds a little-Higgs (LH) partner with the coupling to the Higgs 
boson specifically designed in such a way that the quadratic divergence
is cancelled. For example, the $W$ boson has the LH partner $W_H$. 
A phenomenological interesting example is the littlest Higgs model
\cite{lh0}, the phenomenology of which 
was described in details in Ref.~\cite{lh-han}.

In general, the Yukawa couplings of the SM-like Higgs boson could be
different from the SM, depending on the gauge structure of the LH model;
so are the couplings to the $W/Z$ bosons. Other heavy LH partners, if
they are electrically charged, can run in the triangular loop of 
$H\gamma\gamma$ vertex, and if they carry color they will contribute to
$Hgg$ vertex. If there are other light Higgs bosons that the SM-like
Higgs boson can decay into, they will increase the decay width 
$\Gamma_{\rm tot}$.
Thus, the LH models can be accommodated in the 
present framework by $C_{u,d,\ell}^{S,P}$, $C_v$, $\Delta S^\gamma$, $\Delta S^g$,
and $\Delta \Gamma_{\rm tot}$.

Recent analyses of little Higgs models with respect to current Higgs data
can be found in Refs.~\cite{lh1,lh2}.

\subsection{Fourth Generation model}
The sequential fourth generation model is a simple extension of the SM
by adding an analogous repetition of a generation of fermions. The new
charged leptons and quarks can run in the loop of $H\gamma\gamma$ while the
colored quarks run in the loop of $Hgg$. Thus, the fourth generation 
contributes to $\Delta S^\gamma$ and $\Delta S^g$ only while all the
Yukawa couplings are the same as the SM and $\Delta \Gamma_{\rm tot} =0$.

\section{Higgs Data}
Current Higgs data focus on a few decay channels of the Higgs boson:
(i) $h\to \gamma\gamma$, (ii) $h\to Z Z^* \to \ell^+ \ell^- \ell^+ \ell^-$, 
(iii) $h\to W W^* \to \ell^+ \bar \nu \ell^- \nu$, 
(iv) $h \to b\bar b$, and (v) $h \to \tau^+ \tau^-$. Within each decay mode
both CMS and ATLAS have reported a number of channels, such as  inclusive,
vector-boson-fusion tagged, and/or VH tagged. All the available
ATLAS, CMS, and Tevatron data in these five decay channels are shown
in Tables \ref{t1}--\ref{t5}, respectively.  
Both sets of data {\it before} and {\it after} the Moriond 2013 meetings
are listed in the tables.
We have used 22 data points
in our analysis.  
The chi-square of all these 22 data points relative
to the SM is about 17.5 and 18.94 for the data set {\it before} and
{\it after} Moriond, respectively, and so the chi-square per degree of freedom 
(dof) is about $17.5/22 = 0.80$ and $18.94/22=0.86$,
which means that the SM is a reasonably good fit to the Higgs data.
The goodness of the fit, measured by the $p$-value, is about
$p=0.74$ and $p=0.65$ for the data {\it before} and {\it after} Moriond,
respectively. 
\footnote
{Assuming the goodness-of-fit statistics follows a $\chi^2$ probability density 
function, the $p$-value for the hypothesis is given by \cite{pdg}
\[
 p = \int_{\chi^2}^{\infty} f(z;n) dz
\]
where $n$ is the degrees of freedom and 
\[
 f(z;n) = \frac{z^{n/2-1} e^{-z/2} }{ 2^{n/2} \Gamma(n/2) }\,.
\] 
}

There are four production modes: gluon fusion (ggF), vector-boson fusion (VBF),
associated production with a $W/Z$ boson (VH), and associated production
with a $t\bar t$ pair (ttH).  The
production cross sections for each production modes at the LHC could be
found in Ref.\cite{lhc_production}. For $\sqrt{s}=7$ TeV and Higgs-boson
mass $M_H= 126$ GeV, the cross sections are:
\begin{equation}
\sigma({\rm ggF})=15.080,\; \sigma({\rm VBF})=1.211,\; 
\sigma({\rm VH})=0.8653,\; 
\sigma({\rm ttH})=0.0843\; {\rm pb},
\end{equation}
where $\sigma({\rm VH})=\sigma(WH)+\sigma(ZH)$. 
For $\sqrt{s}=8$ TeV and Higgs-boson mass $M_H=126$ GeV the cross sections are:
\begin{equation}
\sigma({\rm ggF})=19.220,\; \sigma({\rm VBF})=1.568,\; \sigma({\rm VH})=1.0625,\; 
\sigma({\rm ttH})=0.1271\; {\rm pb}.
\end{equation}

Since the ATLAS and CMS from the above tables combined both
$\sqrt{s}=7$ TeV and $\sqrt{s}=8$ TeV, we take the luminosities
for the 7 and 8 TeV data as weights to recalculate the cross section of
each production mode.  Take the VH mode of ``untagged'' channel of ``CMS
($5.1$ fb$^{-1}$ at 7 TeV + $19.6$ fb$^{-1}$ at 8 TeV)'' 
(after Moriond 2013)
from Table~\ref{t1} as
an example, the weighted cross section of the VH mode is
\begin{equation}
\sigma({\rm VH})_{weighted}=\frac{5.1~{\rm fb}^{-1} \times 0.8653~{\rm pb} + 
19.6~{\rm fb}^{-1} \times 1.0625~{\rm pb}}
{5.1~{\rm fb}^{-1}+19.6~{\rm fb}^{-1}}=1.022~{\rm pb}\;.
\end{equation}
For other production modes we get
\begin{equation}
\sigma({\rm ggF})_{weighted}=18.365,\; 
\sigma({\rm VBF})_{weighted}=1.494,\;
\sigma({\rm ttH})_{weighted}=0.118\;{\rm pb}\;.
\end{equation}
By using the weighted cross sections, we obtain
the decomposition coefficients $C_{{\cal Q} {\cal P}}$ (\ref{eq:cpq}) for:
\begin{itemize}
\item[($i$)] ``untagged'' channel of CMS in Table~\ref{t1},
\item[($ii$)] ``Inclusive'' channels in Table~\ref{t2} and Table~\ref{t3},
\item[($iii$)] ``$\mu({\rm VBF}+{\rm VH},\tau\tau)$''
channel in Table~\ref{t5} under the assumption that the ggF and
ttH production modes do not contribute.
\end{itemize}

For the decomposition coefficients 
for ``0/1 jet'' and ``VBF tag'' in Table~\ref{t5},
we take the results 
for the three search 
channels $\mu \tau_h+X$, $e \tau_h+X$,  and $e \mu +X$,
presented in Tables 1, 2, and 3 of Ref.\cite{eff}.

For the CMS ``VBF tagged'' channel  in Table~\ref{t1} and 
the ``0/1 jet'' and ``VBF tag'' channels in Table~\ref{t3}, 
we borrow the numbers found in Ref.~\cite{r17}. 

The Tevatron decomposition coefficients
in Tables \ref{t1} and \ref{t3} are from the ratios of the SM Higgs
production cross sections.
Note that we do not use the Tevatron $\tau\tau$ data
upon the large uncertainty recently reported in Ref.~\cite{:2013vc}.

\section{CP conserving Fits}
In the CP conserving fits, we have fixed
\begin{equation}
C^P_{u,d,\ell}=\Delta P^{g,\gamma}=0\,,
\end{equation}
while varying
\begin{equation}
C^S_{u,d,\ell}, \ \ \
C_v, \ \ \
\Delta S^{g,\gamma}, \ \ \
\Delta\Gamma_{\rm tot}.
\end{equation}
More precisely we have implemented the following 5 fits:
\begin{itemize}
\item[{\bf A}.] SM fit.
\item[{\bf B}.] One-parameter fit varying $\Delta\Gamma_{\rm tot}$ with
$C^S_{u,d,\ell}=C_v=1$ and $\Delta S^{g,\gamma}=0$.
\item[{\bf C}.] Two/three-parameter fit varying $\Delta S^g$ and
$\Delta S^\gamma$ without/with $\Delta\Gamma_{\rm tot}$ 
taking $C^S_{u,d,\ell}=C_v=1$. 
\item[{\bf D}.] Four-parameter fit varying $C^S_{u,d,\ell}$ and $C_v$
with $\Delta S^{g,\gamma}=0$ and $\Delta\Gamma_{\rm tot}=0$.
\item[{\bf E}.] Six-parameter fit varying $C^S_{u,d,\ell}$, $C_v$,
and $\Delta S^{g,\gamma}$ with $\Delta\Gamma_{\rm tot}=0$.
\end{itemize}

 As mentioned above we are going to show the fitting
results with the data set {\it before } and {\it after } the Moriond
2013 meetings for each fit.  Due to the change in the data, especially the
diphoton data from the CMS, the fitting results change dramatically.
In each of the fits, we first describe the fitting results with
respect to the data after the {\it Hadron Collider Physics Symposium
2012} but before the Moriond 2013; then the fitting results with
respect to the data after the Moriond 2013.  The best-fit values for
the parameters of the above fits are all summarized in
Table~\ref{bestfit}.

The figures are shown for the fits with the most
updated data after Moriond. 
Note also that when we show the 2D $\chi^2$ regions for two of the varying 
parameters in the figures, we are marginalizing over the other parameters
if more than two are allowed to vary. 
This also applies to the next section of CP violating fits.

\subsection{SM fit}
\subsubsection{Before Moriond}
As we have mentioned the SM fit gives a $\chi^2/dof =17.5/22 =0.8$,
It gives a $p$-value of $p=0.74$, which means the SM has a chance of $0.74$
to be the true interpretation of the data.
Contributions of chi-square from each data are shown
in the second last column of Tables~\ref{t1}-\ref{t5}.
The $H\to\gamma\gamma$ data from ATLAS, CMS, and Tevatron give the 
largest contribution to the chi-square,
while the least contribution is from $H\to ZZ^*$.
Specifically, in each decay channel
the largest contribution is from:
ATLAS $\mu_{ggH+ttH}$ and Tevatron ($H\to\gamma\gamma$);
CMS Inclusive ($H\to ZZ^*$);
CMS VBF tag ($H\to WW^*$);
ATLAS VH tag ($H\to b\bar b$);
ATLAS $\mu(ggF,VBF+VH)$ ($H\to \tau\tau$).
As we shall see soon that because the chi-square is dominated by the diphoton
data the most efficient way to fit to the data is using 
$\Delta S^\gamma$ and $\Delta S^g$.

\subsubsection{After Moriond}
The SM fit gives a $\chi^2/dof =18.94/22 =0.86$,
which gives a $p$-value of $p=0.65$. This value shows that the SM description
of the data stays more or less the same as before the Moriond update.
The diphoton data still dominate the total $\chi^2$.
The $\chi^2$ of each decay channel, shown in the last column
of Tables~\ref{t1}-\ref{t5}, is about the same as before.

\subsection{Vary only $\Delta \Gamma_{\rm tot}$ while keeping
$C^S_u=C^S_d=C^S_\ell=C_v = 1$ and 
$\Delta S^\gamma = \Delta S^g = 0$}

\subsubsection{Before Moriond}
We found that varying $\Delta \Gamma$ alone does not improve
the chi-square. Numerically the chi-square per dof is 
$17.5/21$ and  the 95\% allowed
range for $\Delta \Gamma_{\rm tot}$ is 
\[
  -0.022^{+1.44}_{-0.85}\; {\rm MeV} \;.
\]
The central value is consistent with zero and thus the 95\% CL upper limit
for $\Delta \Gamma_{\rm tot}$ is about 1.4 MeV.
Note that the total width of the SM Higgs boson with $M_H = 125.5$ GeV is 
about $4.1-4.2$ MeV. Therefore, the 95\% CL upper limit for 
the nonstandard branching ratio of the Higgs boson is about 25\%. 
The nonstandard decays include invisible decays, decays into other lighter
Higgs bosons, or decays into other exotic particles.

\subsubsection{After Moriond}
The situation remains the same. The $\chi^2/dof = 18.89/21$ and the
95\% allowed range for $\Delta \Gamma_{\rm tot}$ is 
\[
  0.10^{+1.11}_{-0.74}\; {\rm MeV} \;.
\]
The central value is consistent with zero and thus the 95\% CL upper limit
for $\Delta \Gamma_{\rm tot}$ is about 1.2 MeV.
Therefore, the 95\% CL upper limit for 
the nonstandard branching ratio of the Higgs boson is about 22\%.

\subsection{Vary $\Delta S^\gamma$ and $\Delta S^g$ while keeping
$C_u^S=C_d^S=C_\ell^S=C_v = 1$}

\subsubsection{Before Moriond}
In this fit, only the  parameters
$\Delta S^\gamma$ and $\Delta S^g$ vary,
which is simply motivated by the fact that the most deviated Higgs data
are the diphoton signal strengths while all the other data are more or less
consistent with the SM values. It turns out that this is the most efficient way
to fit to the data statistically (the most efficient here 
means that the $\chi^2$ is reduced the most with the best $\chi^2/dof$.)
The best fit values are
\begin{equation}
\Delta S^\gamma = -2.73^{+1.11}_{-1.15},\;\;
\Delta S^g = -0.050^{+0.064}_{-0.065},\;\;
\chi^2/dof = 11.27/20=0.56
\end{equation}

There are two solutions of $\Delta S^g$, which can be easily
understood from the expression of $S^g$ in Eq.~(\ref{8}). 
Since the signal strengths depend on the absolute value of $S^g$,
numerically, $ \pm 0.6 \simeq 0.65 + \Delta S^g$,
which gives $\Delta S^g \simeq -0.05$ or 
$-1.25$. They both give $C_g \simeq 0.92$.
Also, we were to extend the range of $\Delta  S^\gamma$ further, there
would have been a solution around $\Delta S^\gamma \simeq 16$, 
according to the expression of $S^\gamma$ in Eq.~(\ref{5}). 
Nevertheless, it is unrealistic to
find a model that can generate such a large $\Delta S^\gamma$. 
The quantities $C_\gamma$ and $C_g$ are very close to physical observables,
and so their best values are 
\[
 C_\gamma \simeq 1.41 ,\;\;\; C_g \simeq 0.92 \;.
\]
The chi-square per dof for this two-parameter fit is $0.56$ compared to
$0.8$ of the SM, which shows some real improvement.

If we further allow $\Delta \Gamma_{\rm tot}$ to vary simultaneously with 
$\Delta S^\gamma$ and $\Delta S^g$, the best fit is
\[
 \Delta S^\gamma = -2.93^{+1.19}_{-1.31} ,\;\;
 \Delta S^g      = 0.0063^{+0.15}_{-0.11} ,\;\;
 \Delta \Gamma_{\rm tot} = 0.79^{+2.0}_{-1.1} \; {\rm MeV},\;\;
\chi^2/dof = 10.83/19 = 0.57  .
\]
It is clear that including $\Delta \Gamma_{\rm tot}$ in the fit does not improve
the chi-square per dof. Since $\Delta \Gamma_{\rm tot}$ is still consistent with
zero in this case, we will fix $\Delta \Gamma_{\rm tot} = 0$ in the later fits.

\subsubsection{After Moriond}
The most obvious difference between the data set before and after the Moriond
2013 can be seen here in this fit. Before the Moriond the $\chi^2$
is dominated by the diphoton data, in which both ATLAS and CMS showed 
$1.5-2\sigma$ excesses, and so the dynamics of the fit will push to
the direction to substantially reduce the $\chi^2$ of the diphoton data.
However, with the new CMS diphoton data ($0.78^{+0.28}_{-0.26}$) 
the whole fit changes. The dynamics of the fit cannot force the
parameters to go into one direction, because the ATLAS data is still 
about $1.5\sigma$ larger than the SM while the CMS one is about $1\sigma$
smaller.

\begin{figure}[th!]
\centering
\includegraphics[width=3.2in]{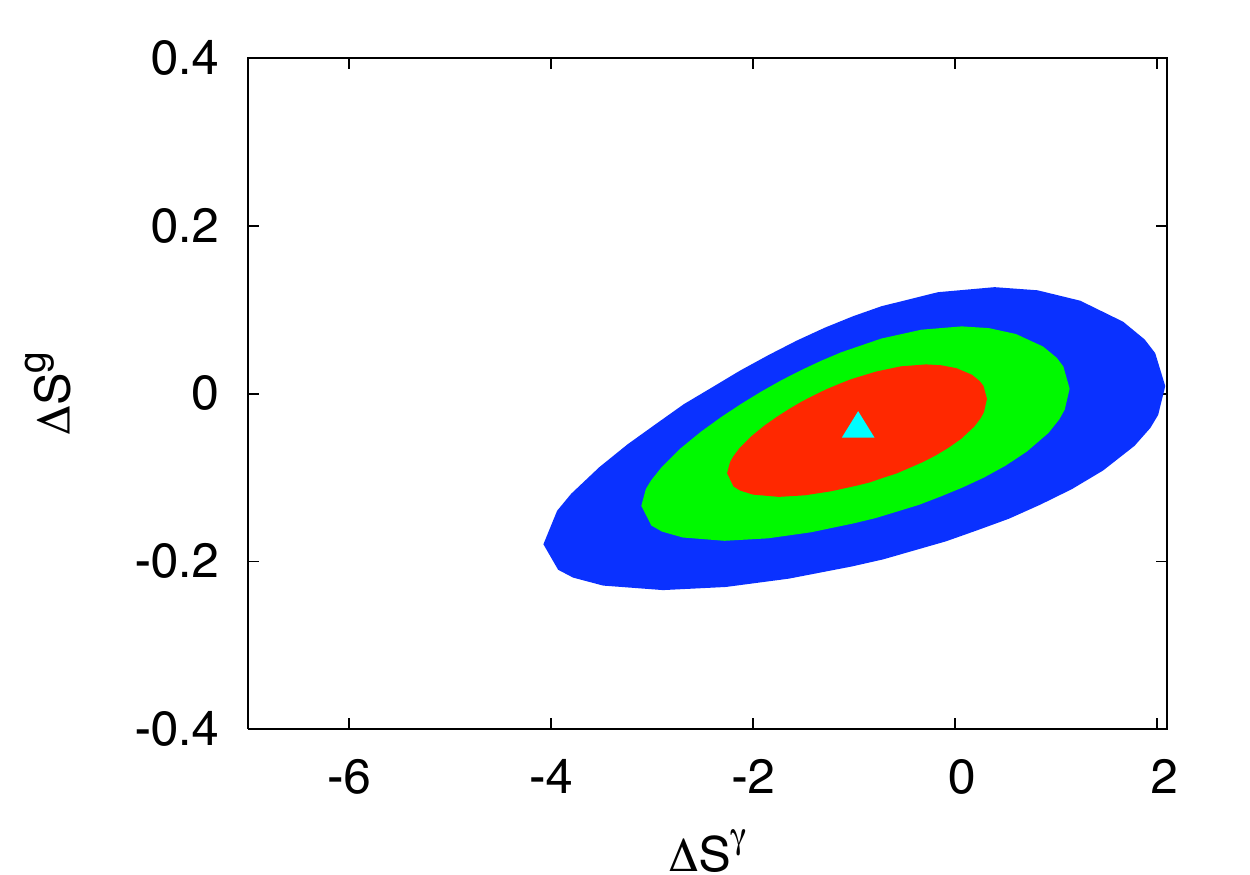}
\includegraphics[width=3.2in]{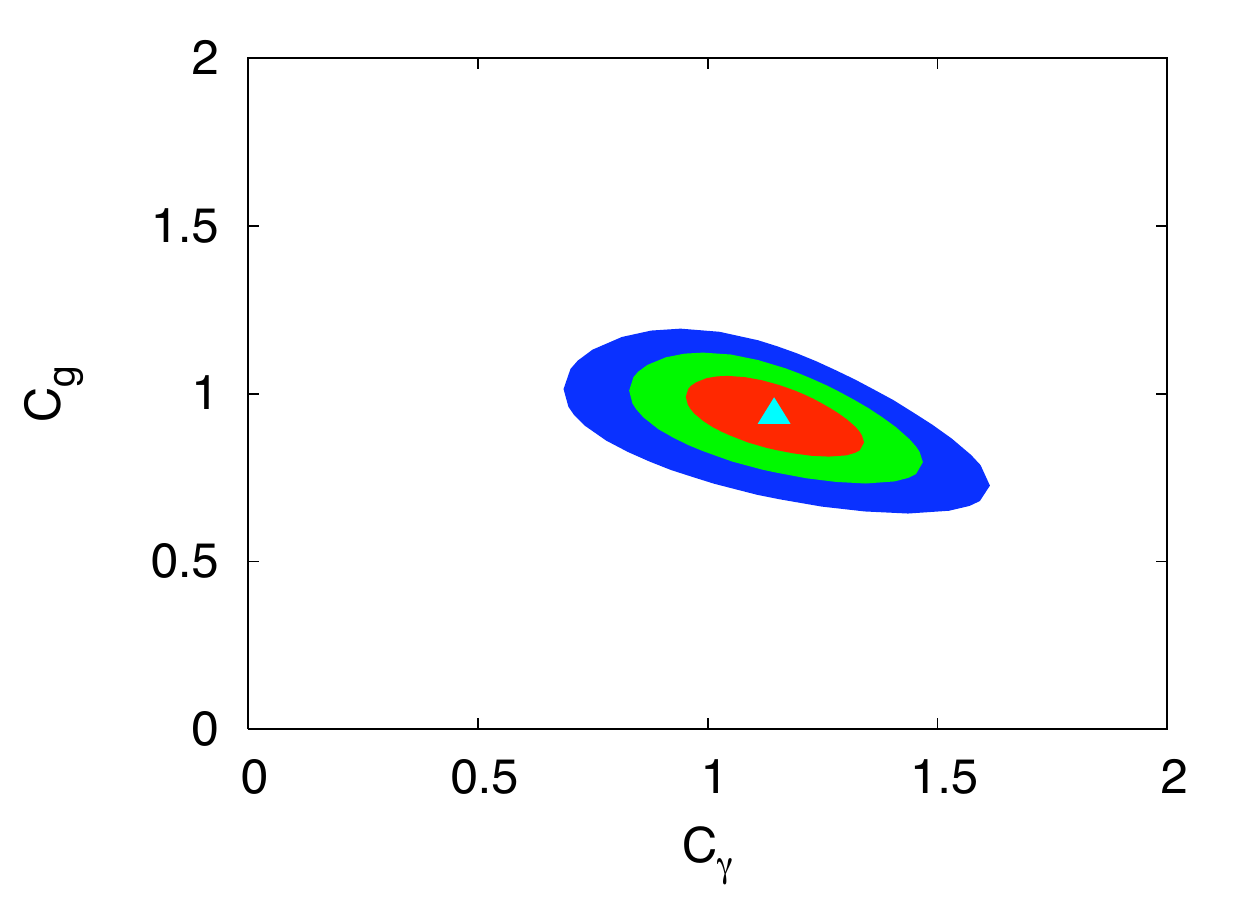}
\caption{\small \label{case1}
The confidence-level regions of the fit by varying $\Delta S^\gamma$ and
$\Delta S^g$ only, 
(a) in the $(\Delta S^\gamma,\Delta S^g)$ plane and (b) in the 
corresponding $(C_\gamma, C_g)$ plane.
The contour regions shown are for 
$\Delta \chi^2 \le 2.3$ (red), $5.99$ (green), and $11.83$ (blue) 
above the minimum, which 
correspond to confidence levels of
$68.3\%$, $95\%$, and $99.7\%$, respectively.
The best-fit point is denoted by the triangle.
}
\end{figure}

The best fit values for $\Delta S^\gamma$ and $\Delta S^g$ are
\begin{equation}
\Delta S^\gamma = -0.96^{+0.84}_{-0.85},\;\;
\Delta S^g = -0.043\pm {0.052},\;\;
\chi^2/dof = 17.55/20=0.88
\end{equation}
We notice that the sizes of the errors are reduced reflecting
more precise measurements of the Higgs data.
The distributions of chi-square in the plane of 
$(\Delta S^\gamma,\Delta S^g)$ and the corresponding $(C_\gamma, C_g)$ plane
are shown in Fig.~\ref{case1}(a) and (b), respectively.
The quantities $C_\gamma$ and $C_g$ are very close to physical observables,
and so their best values are 
\[
 C_\gamma \simeq 1.14 ,\;\;\; C_g \simeq 0.93 \;.
\]
The chi-square per dof for this two-parameter fit is $0.88$ compared to
$0.86$ of the SM. The $p$-values are very similar. 

\begin{figure}[th!]
\centering
\includegraphics[width=3.2in]{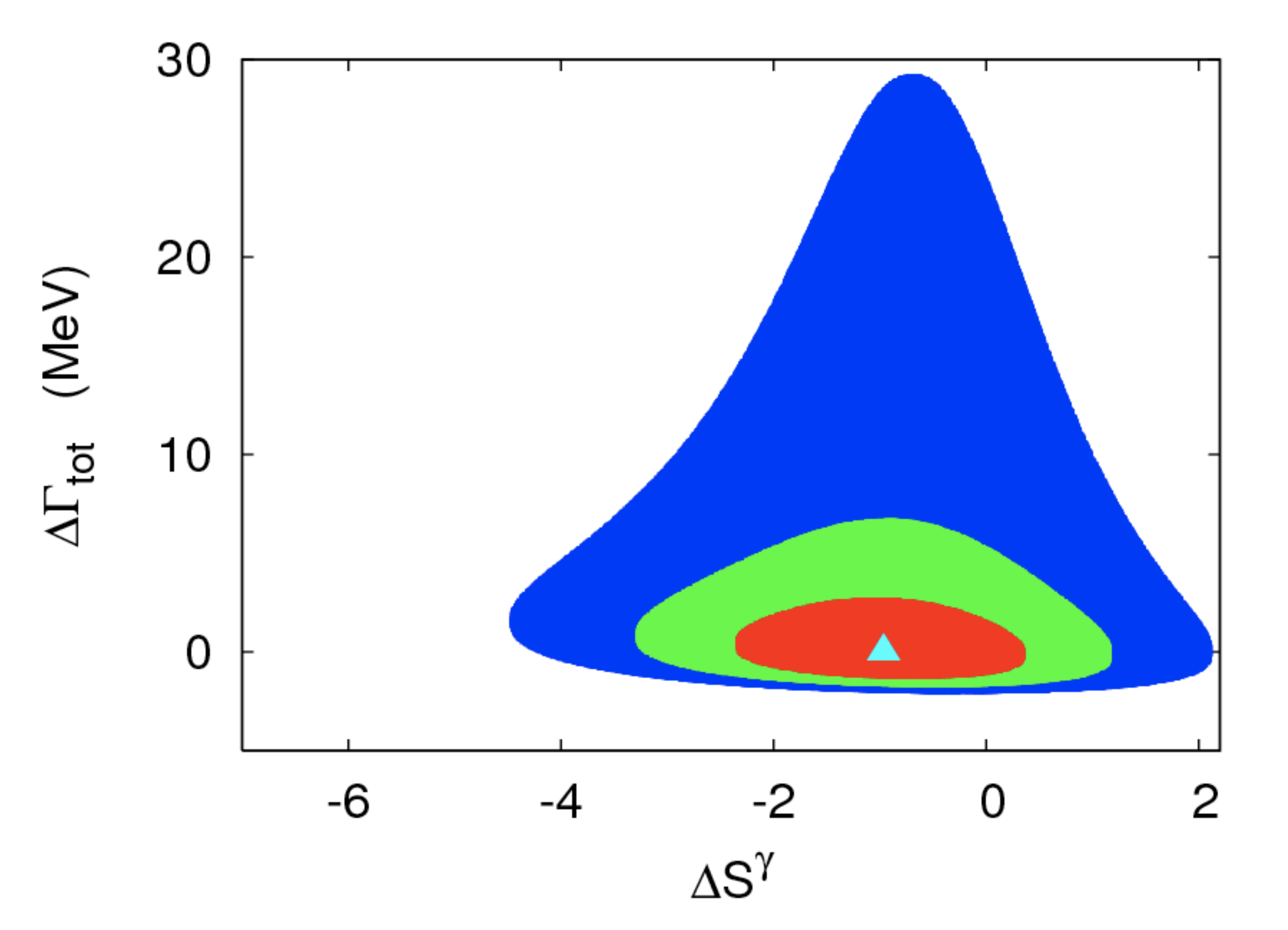}
\includegraphics[width=3.2in]{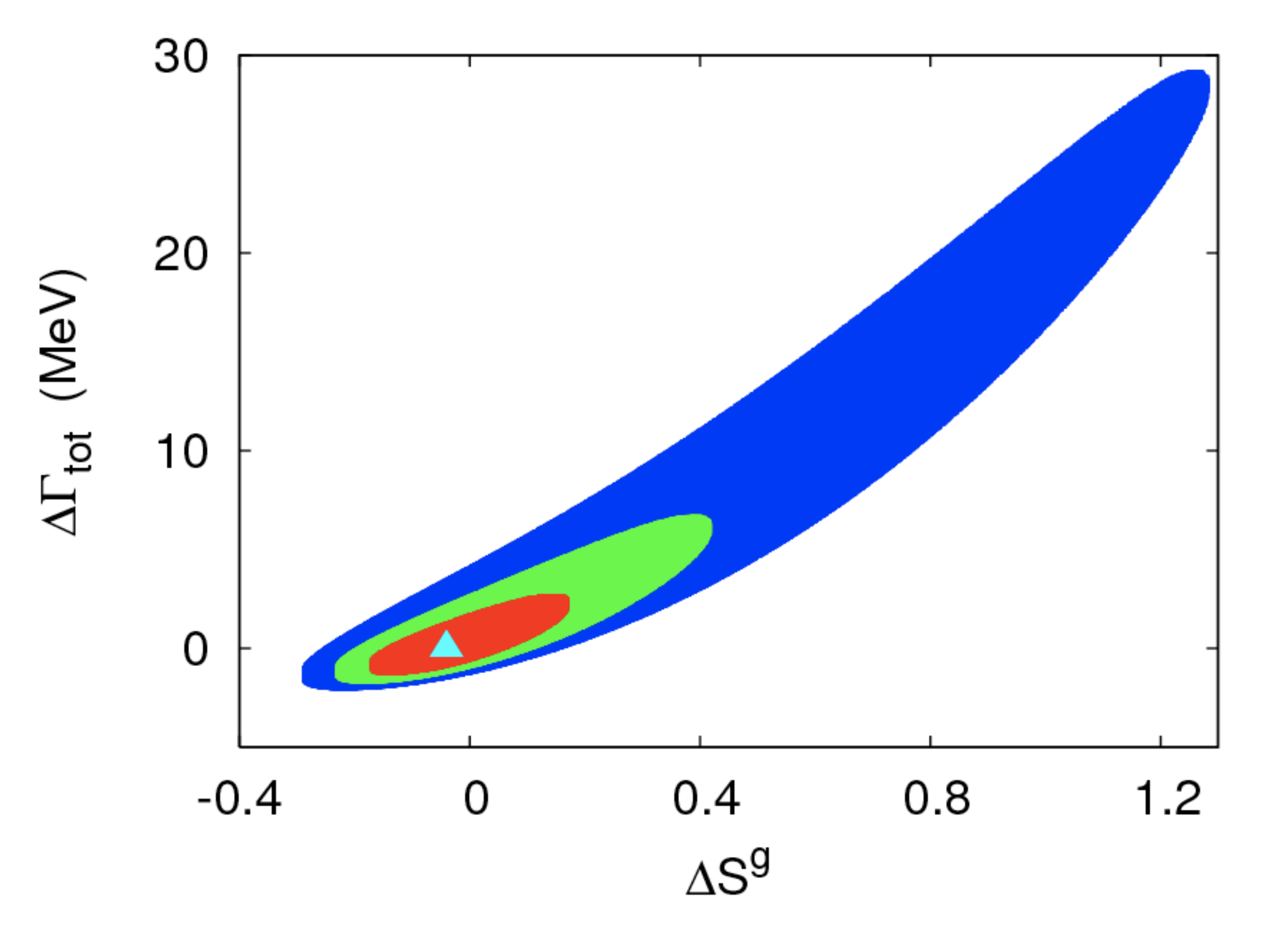}
\includegraphics[width=3.2in]{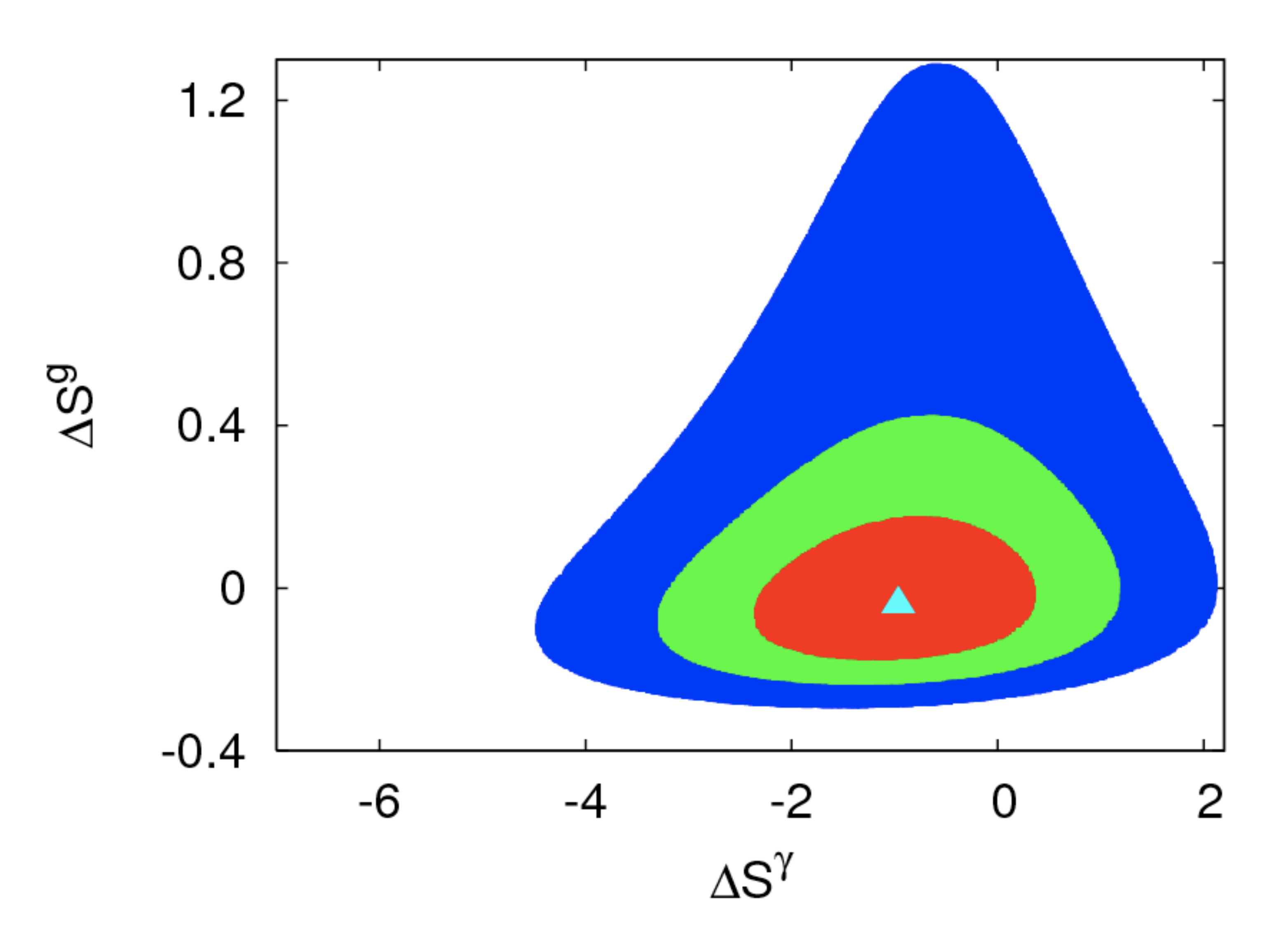}
\includegraphics[width=3.2in]{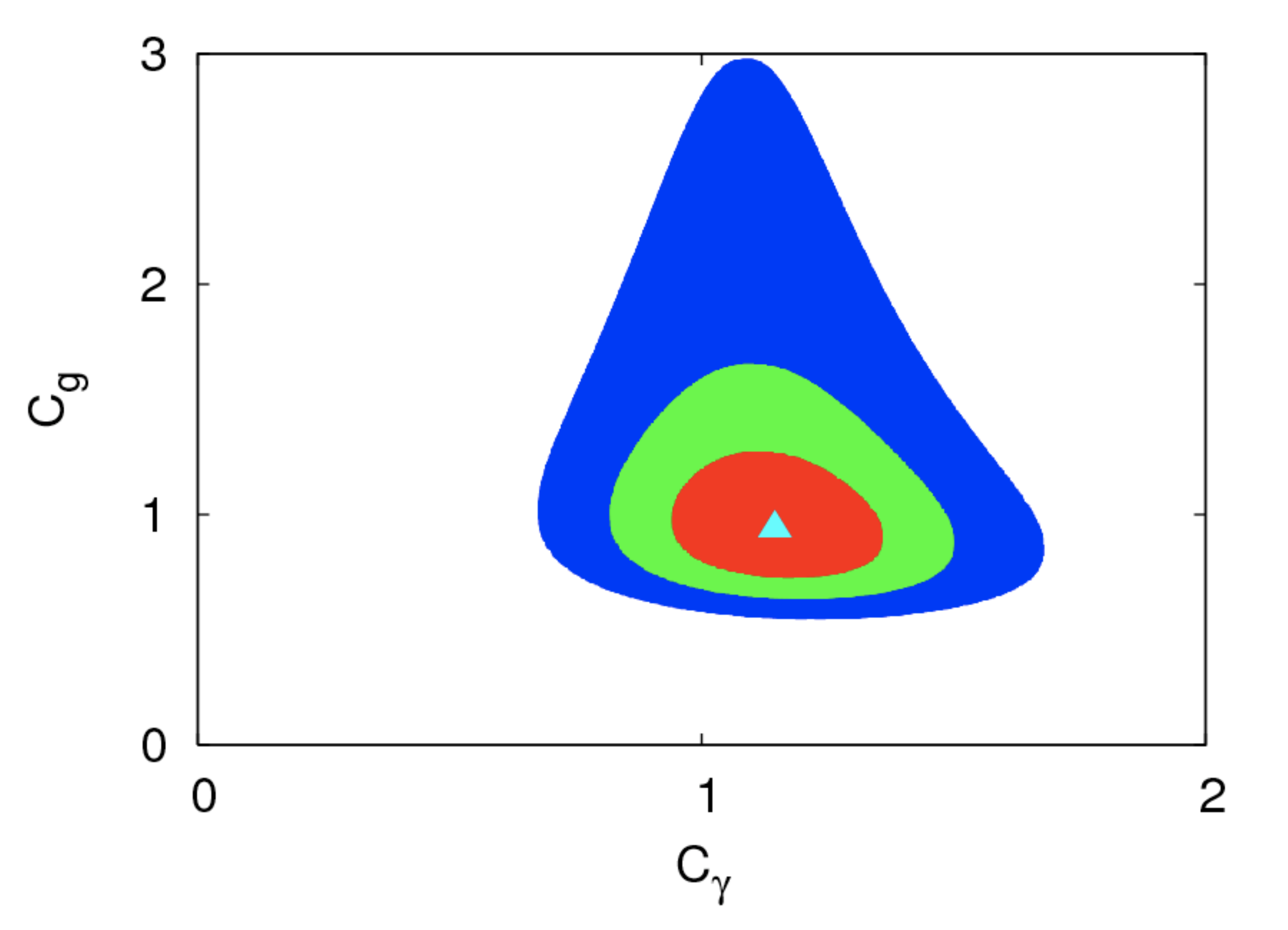}
\caption{\small \label{case1-3}
The confidence-level regions of the fit by varying $\Delta S^\gamma$,
$\Delta S^g$, and $\Delta \Gamma_{\rm tot}$, 
(a) in the $(\Delta S^\gamma,\Delta \Gamma_{\rm tot})$ plane,
(b) in the $(\Delta S^g,\Delta \Gamma_{\rm tot})$ plane,
(c) in the $(\Delta S^\gamma,\Delta S^g)$ plane,
(d) in the corresponding $( C_\gamma , C_g)$ plane.
The description of contour regions is the same as Fig.~\ref{case1}.
}
\end{figure}

Again, we further allow $\Delta \Gamma_{\rm tot}$ to vary simultaneously with 
$\Delta S^\gamma$ and $\Delta S^g$, the best fit is
\[
 \Delta S^\gamma = -0.96^{+0.84}_{-0.87} ,\;\;
 \Delta S^g      = -0.040^{+0.12}_{-0.086} ,\;\;
 \Delta \Gamma_{\rm tot} = 0.027^{+1.33}_{-0.80} \; {\rm MeV},\;\;
\chi^2/dof = 17.55/19 = 0.92  .
\]
The 2-dim contours for the correlations among the 3 parameters are shown in
Fig.~\ref{case1-3}.
Note that anticorrelation between $C_\gamma$ and $C_g$ shown in 
Fig.~\ref{case1}(b) is modified to the shape shown in Fig.~\ref{case1-3}(d).
The elongation along the $C_g$ is allowed with the increase in 
$\Delta \Gamma_{\rm tot}$ such that the production in ggF increases but the 
decays in various channels decrease.
It is clear that including $\Delta \Gamma_{\rm tot}$ in the fit does not improve
the fit.

It is easy to notice that $\Delta S^\gamma$ and $\Delta S^g$ were very
efficient in reducing the $\chi^2$ of the Higgs data {\it before} the
Moriond update, because both the ATLAS and CMS had the diphoton data
on the excess side of the SM value. However, {\it after} the Moriond
update the CMS diphoton data is smaller than the SM value while the
ATLAS is still larger, and therefore the $\chi^2$ cannot be reduced
effectively no matter how $\Delta S^\gamma$ and $\Delta S^g$ are
varied.  

\subsection{Vary 
$C^S_u$, $C^S_d$, $C^S_\ell$, $C_v$  while keeping 
$\Delta S^\gamma = \Delta S^g = 0$}

This choice is motivated in the scenario where there are no
new particles running in the triangular loops
of Higgs boson decaying into $gg$ or $\gamma\gamma$,
but only modifications of Yukawa couplings. It can be realized
in a two-Higgs doublet model with a very heavy charged Higgs boson
and modifications of Yukawa couplings can be expressed in terms 
of the mixing angle $\alpha$ and $\tan\beta$. 

First, we notice that there is an overall symmetry:
\[
C_u^S \leftrightarrow - C_u^S, \; C_d^S \leftrightarrow - C_d^S,\;
C_\ell^S \leftrightarrow - C_\ell^S,\; C_v \leftrightarrow - C_v 
\]
simply obtained by flipping the overall sign in Eqs.~(\ref{eq1}) and 
(\ref{eq2}). 
Furthermore, from Eq.~(\ref{5}) the diphoton production rate depends on 
$|S^\gamma|^2 + |P^\gamma|^2$, and so only the relative signs of gauge and
Yukawa couplings are important.
Therefore, in the following we fix the sign of $C_v$ to
be positive, while the other 3 parameters can be either negative or
positive.

Since the contributions of the bottom and the charged-lepton sectors
to the diphotons are very small,
we expect an approximate symmetry: $C_d^S \leftrightarrow - C_d^S$ and 
$C_\ell^S \leftrightarrow - C_\ell^S$. Even if we change the best-fit point
by $C_\ell^S \to - C_\ell^S$ the total $\chi^2$ only changes by $O(+0.01)$.
This simply means that changing the sign of bottom- and charged-lepton Yukawa
couplings would not affect significantly the loop contributions of 
$H\gamma\gamma$ and $Hgg$.
We show in Fig.~\ref{case2-1} the 2-dim contours for the correlations
of any 2 of the 4 parameters ($C_u^S,C_d^S,C_\ell^S,C_v$). In the figure,
we can see an approximate symmetry: $C_d^S \leftrightarrow - C_d^S$ and 
$C_\ell^S \leftrightarrow - C_\ell^S$. Note that this figure is for
the data after the Moriond update.

On the other hand, the sign of $C_u^S$ is important,
as shown by the two islands in the panel (a).
It is well known that
the SM $W$ boson and top quark loop contributions to $H\gamma\gamma$ come
in opposite sign. Therefore, by flipping the sign of the top quark
contribution ($C_u^S \to - C_u^S$) it can enhance the $H\gamma\gamma$ vertex.

\subsubsection{Before Moriond}

Since before the Moriond update both the CMS and ATLAS diphoton data are
in excess, the dynamics of the fit indeed prefers $C_u^S <0$ 
for positive $C_v$, shown in the fifth column of the upper half of 
Table~\ref{bestfit}.  In this way, the diphoton rate is pushed up to fit
well with the data and significantly reduces the $\chi^2$. 
Thus results in $\chi^2/dof = 10.46/18$.

\subsubsection{After Moriond}

Nevertheless, the new CMS diphoton data affect the fit significantly.
The dynamics of the fit cannot force the parameters to go into one
direction to reduce the $\chi^2$, because the ATLAS diphoton data is
on the opposite side of the CMS data.  Thus, the top-Yukawa $C_u^S
\approx 0.8$ (see the fifth column of the lower half of
Table~\ref{bestfit}), which means that the top contribution to the
$H\gamma\gamma$ vertex is only reduced by a small amount. Therefore,
we only obtain an overall $\chi^2/dof = 17.82/18 = 0.99$, which is
worse than the SM fit.

\begin{figure}[th!]
\centering
\includegraphics[width=3.1in]{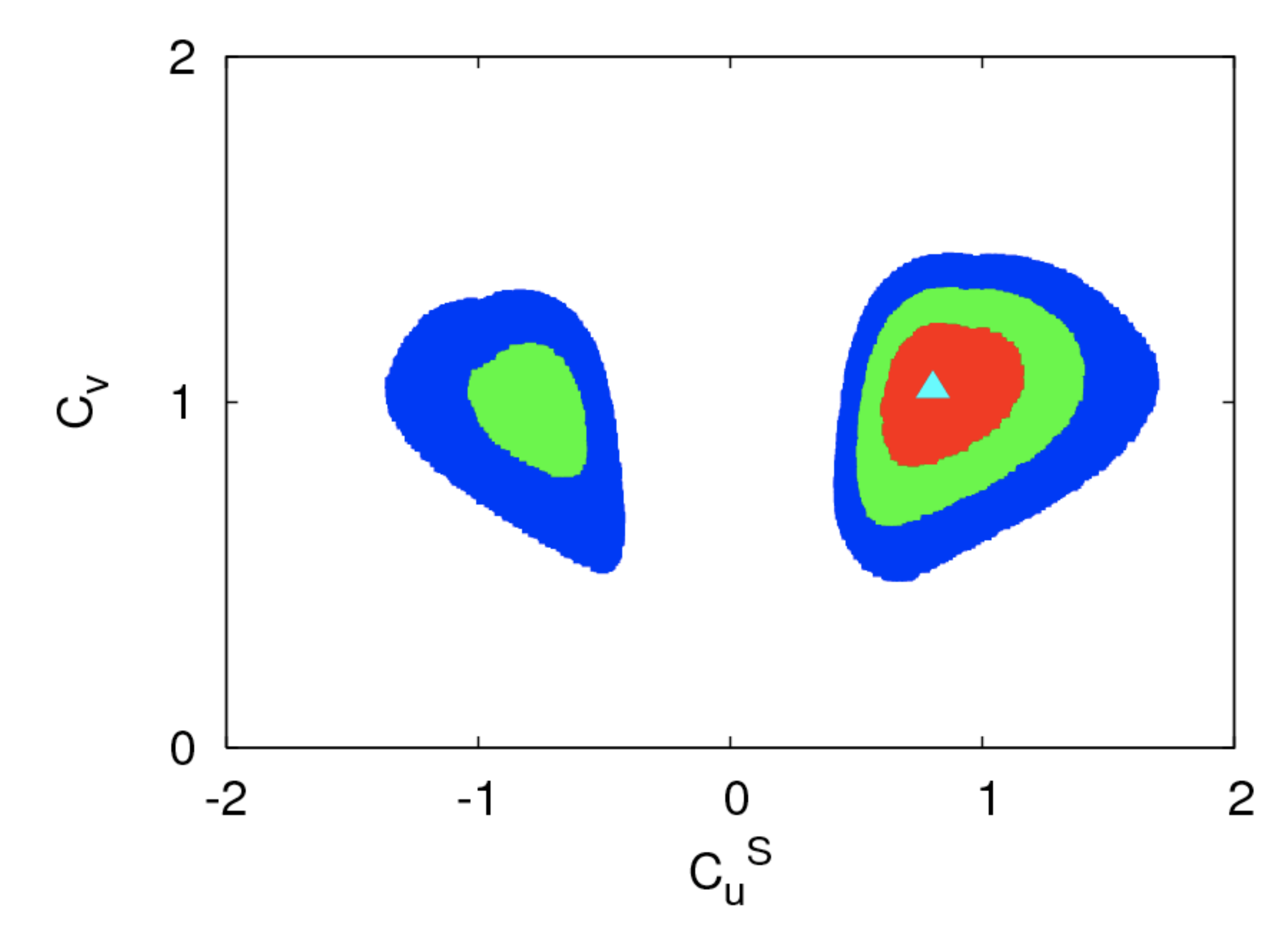}
\includegraphics[width=3.1in]{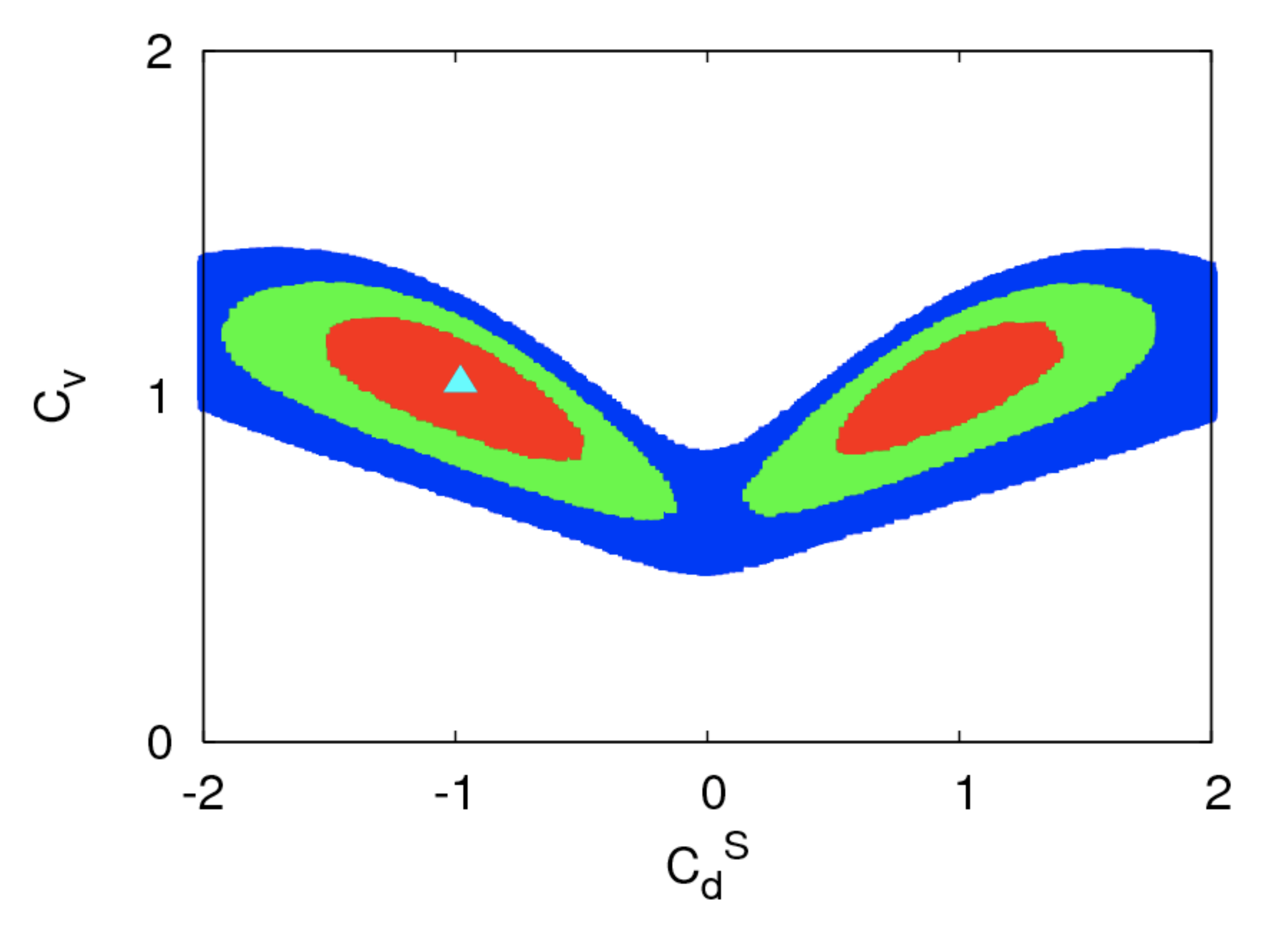}
\includegraphics[width=3.1in]{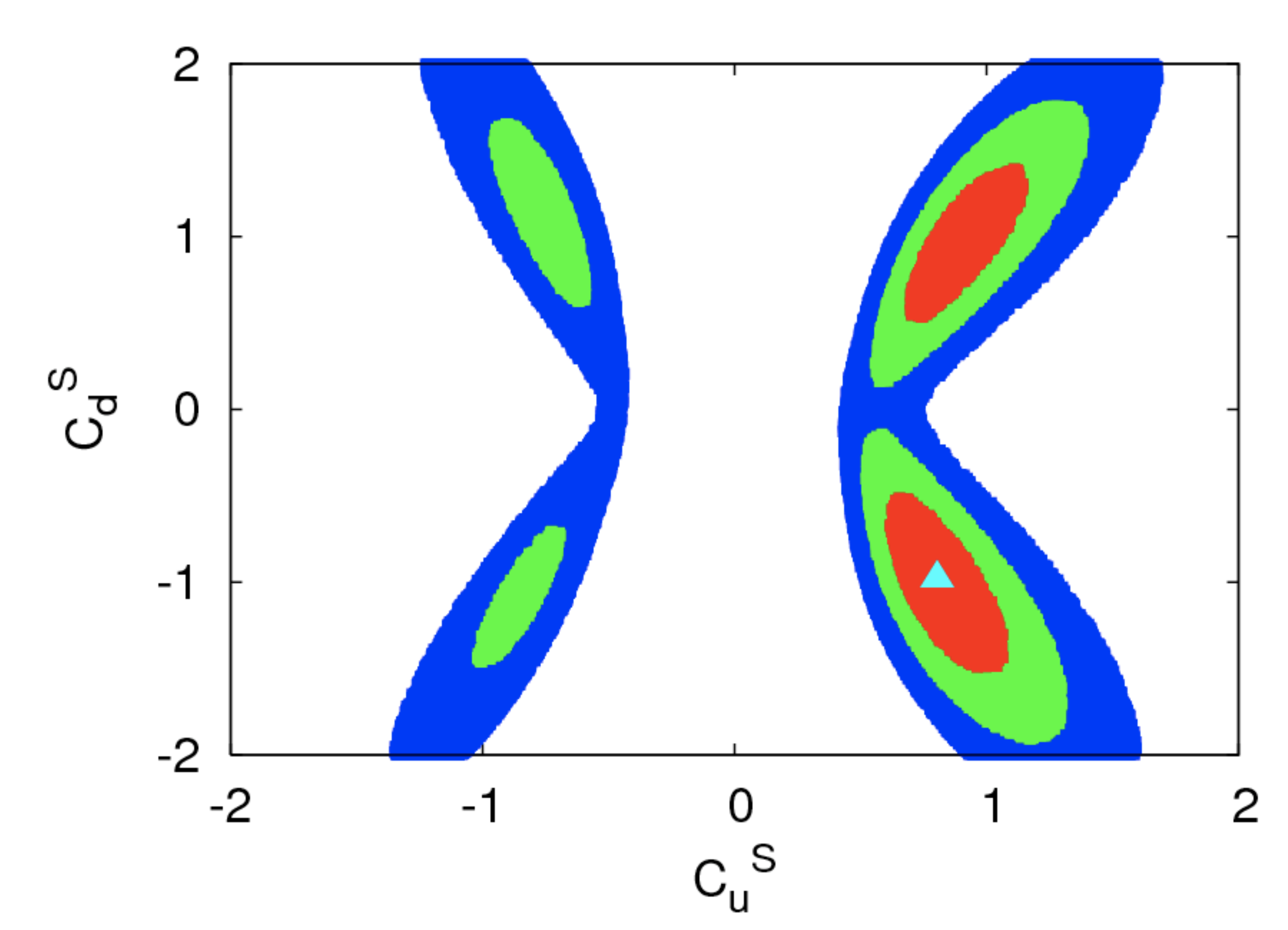}
\includegraphics[width=3.1in]{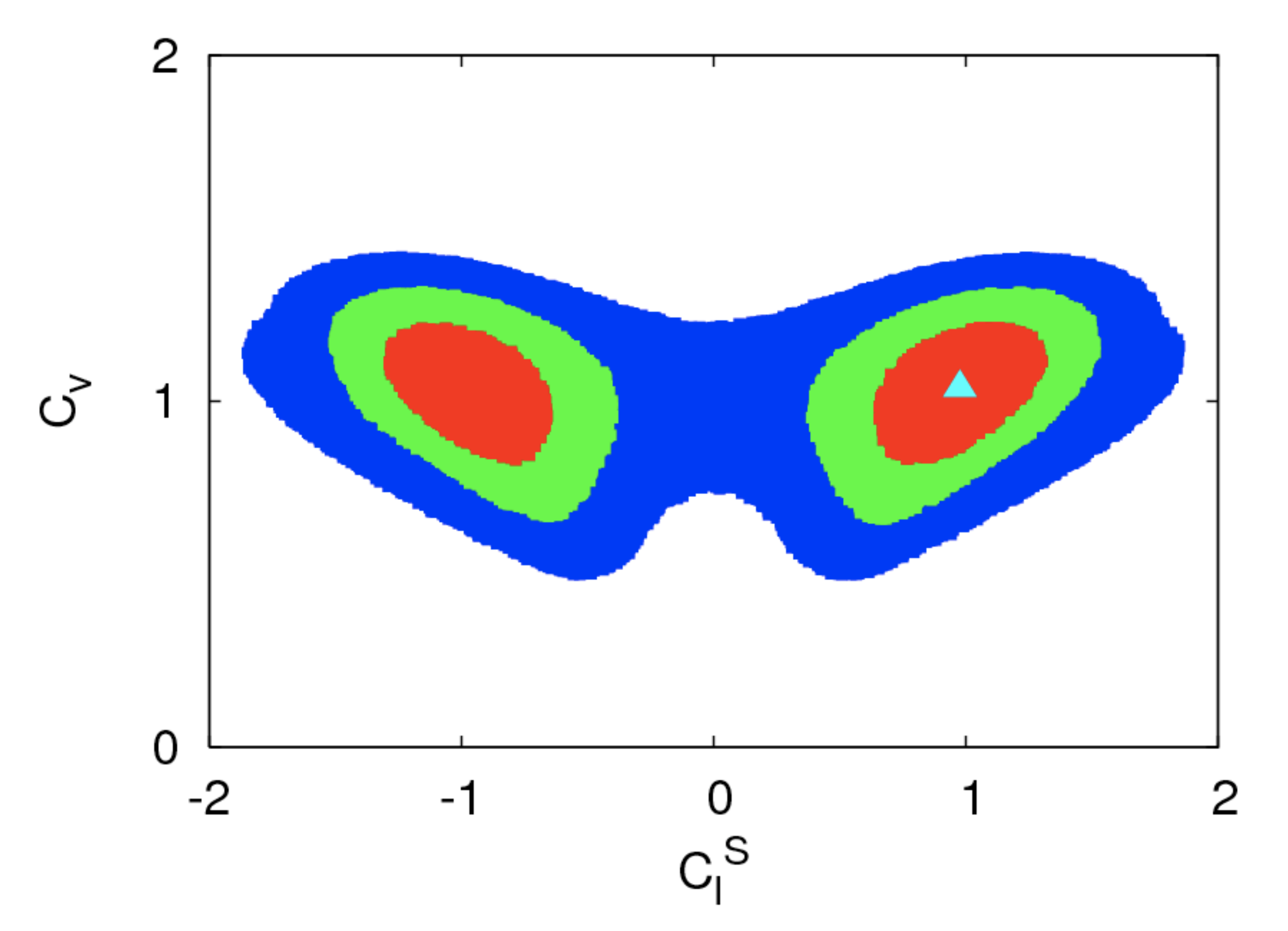}
\includegraphics[width=3.1in]{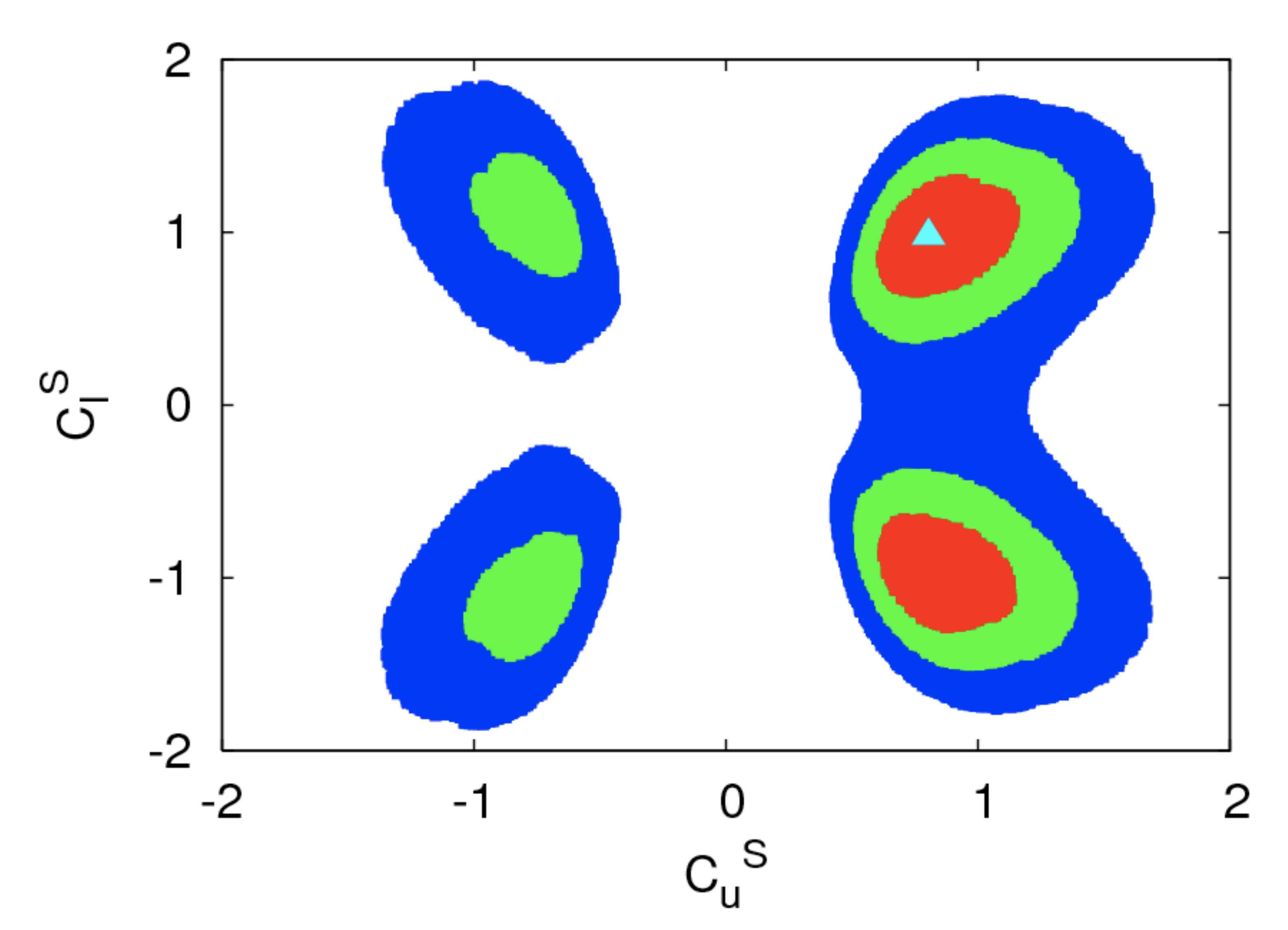}
\includegraphics[width=3.1in]{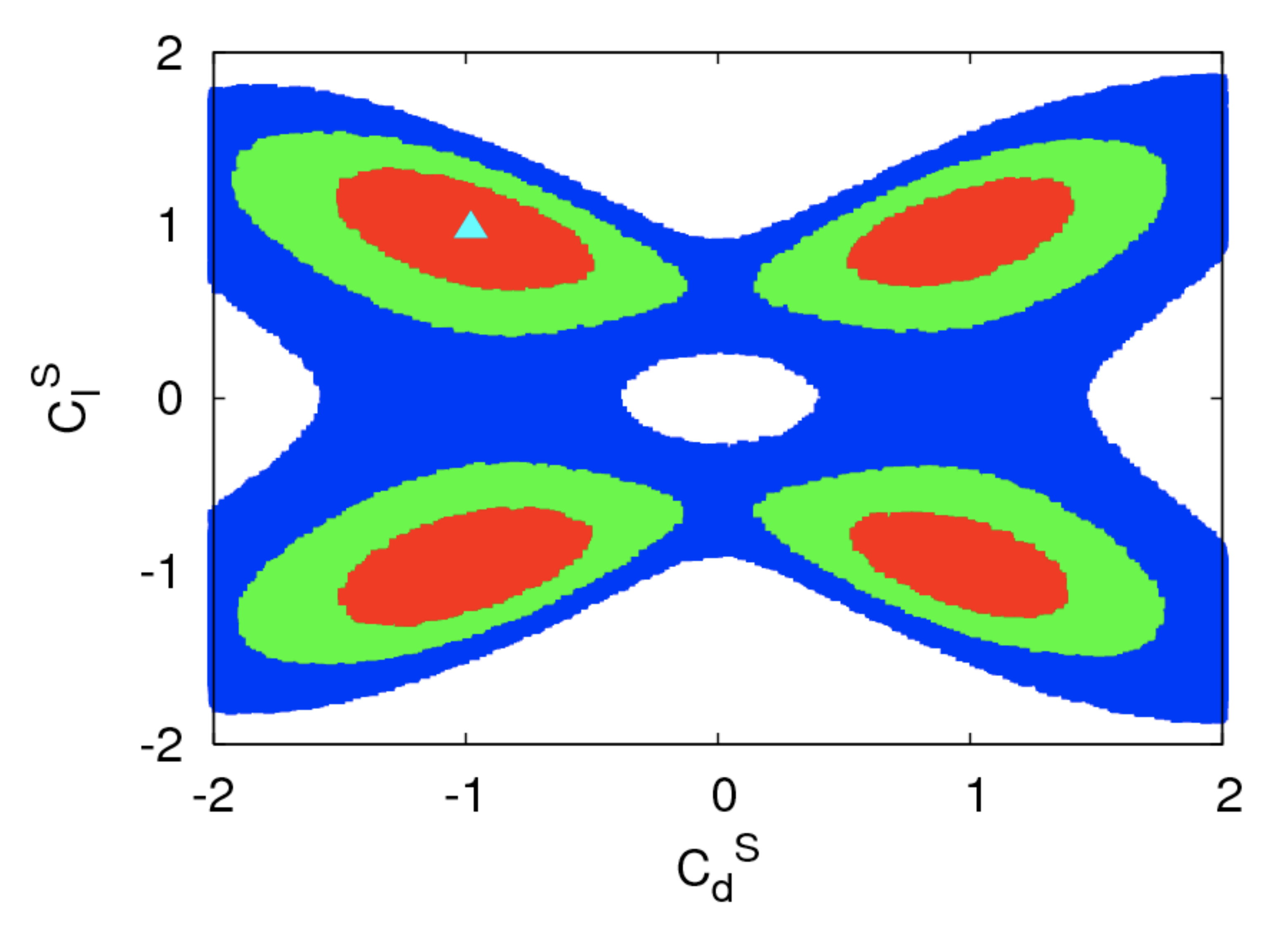}
\caption{ \small \label{case2-1}
The confidence-level regions of the fit by varying $C_u^S$, $C_d^S$,
$C_\ell^S$ and $C_v$ while keeping $\Delta S^\gamma = \Delta S^g =\Delta 
\Gamma_{\rm tot} =0$.
The description of contour regions is the same as Fig.~\ref{case1}.
}
\end{figure}

\begin{figure}[th!]
\centering
\includegraphics[width=3.2in]{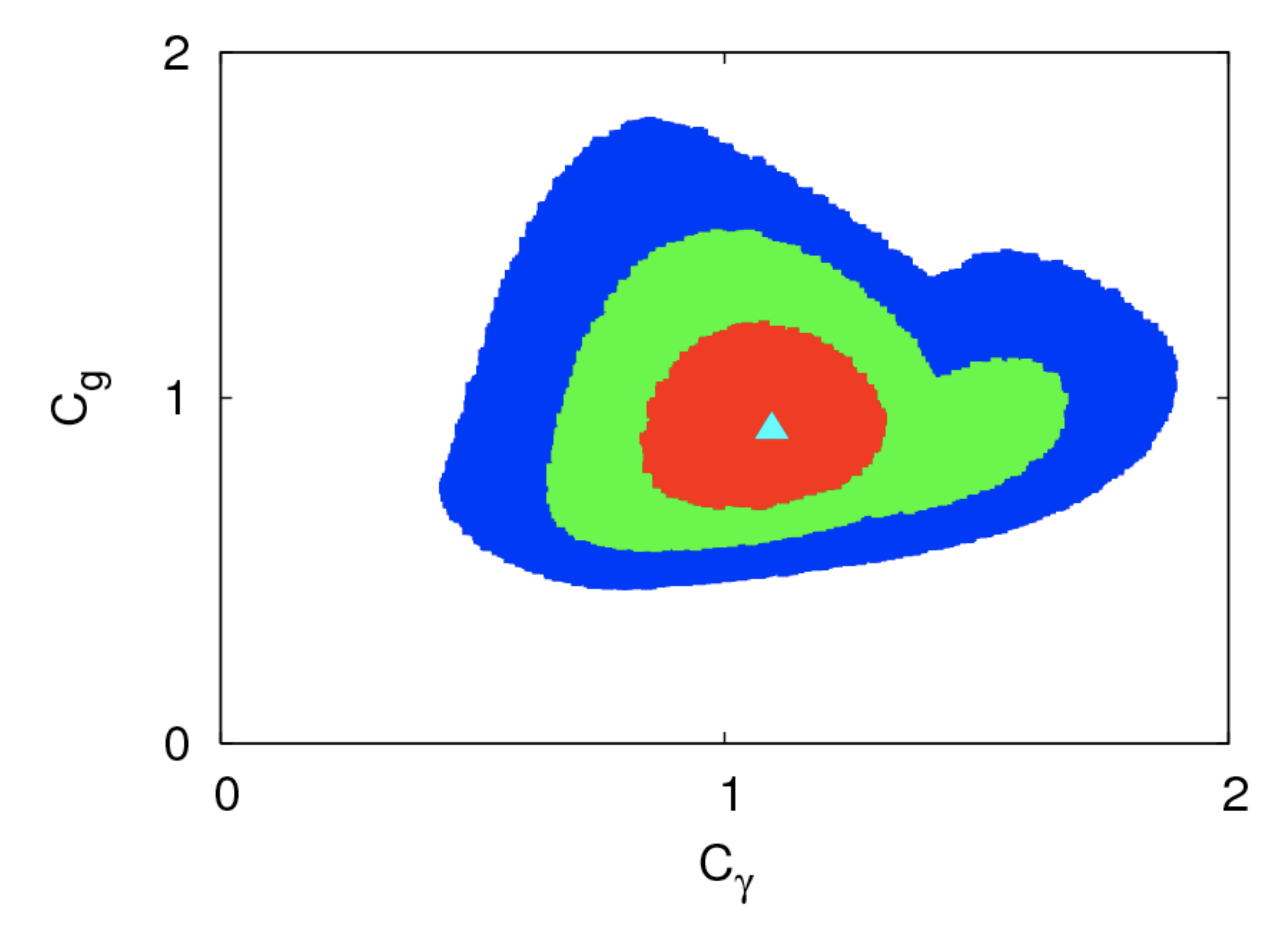}
\includegraphics[width=3.2in]{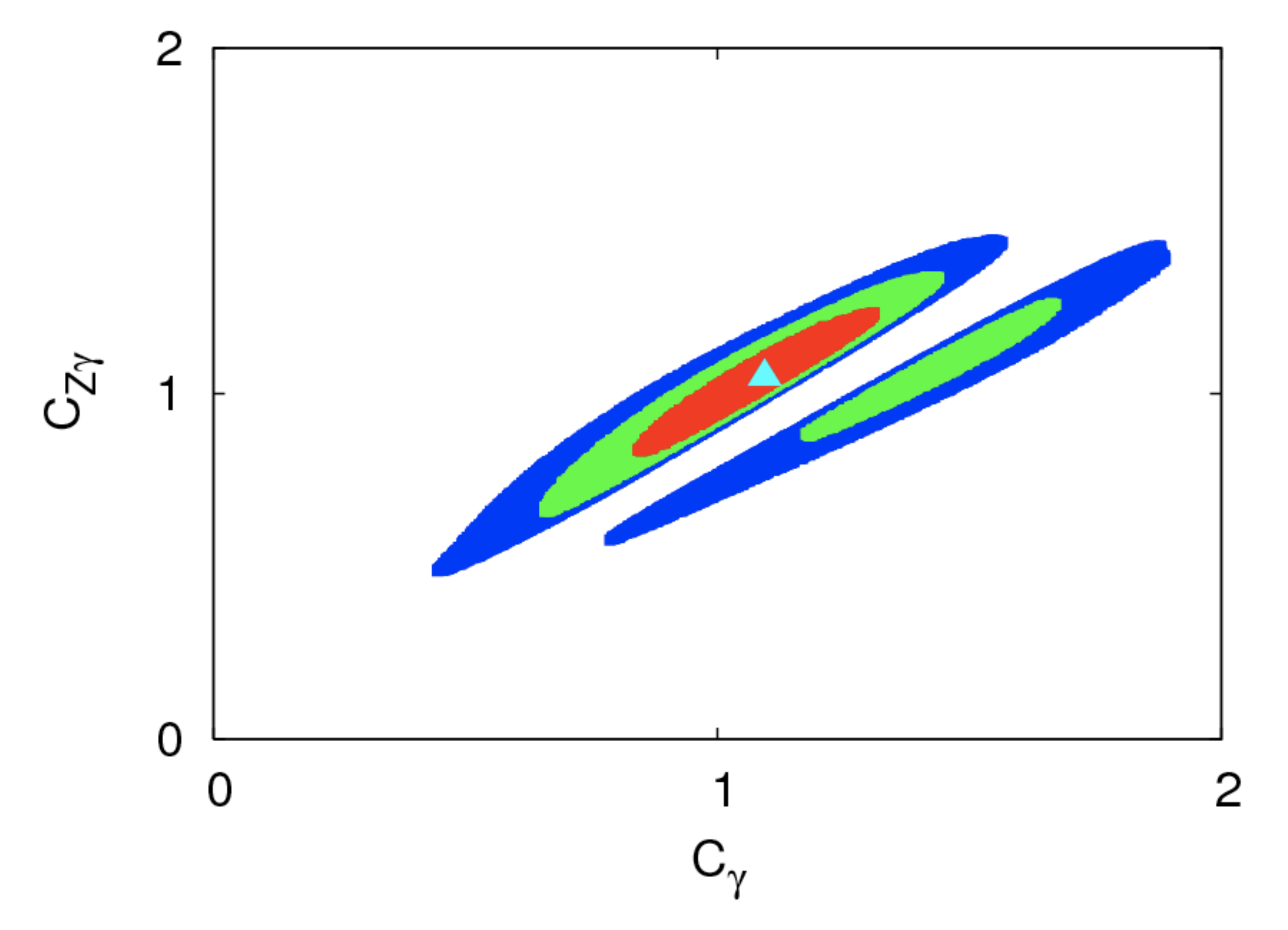}
\caption{ \small \label{case2-2}
(a) Same as Fig.~\ref{case2-1} but in the corresponding plane of 
$(C_\gamma, C_{g})$. (b) Prediction in the corresponding
$(C_\gamma, C_{Z\gamma})$ plane.
The description of contour regions is the same as Fig.~\ref{case1}.
}
\end{figure}

In Fig.~\ref{case2-2}(a), we show the corresponding confidence-level
regions in the $(C_\gamma,C_g)$ plane. 
 The central values are $C_\gamma = 1.09, C_g = 0.91$. 
Note that original anticorrelation between $C_\gamma$ and $C_g$ shown in 
Fig.~\ref{case1}(b) is modified to the shape shown in Fig.~\ref{case2-2}(a).
The enlargement region in $C_g$ can be understood when 
$C_u^S$ increases, $C_g$ will increase but $C_\gamma$ is reduced such that
the diphoton rate stays about the same. Another enlargement region in 
$C_\gamma$ direction can be understood as the left island of 
Fig.~\ref{case2-1}(a),
in which $C_u^S$ is negative, such that $C_\gamma$ is large and $C_g$ is about 
the same.
In part (b), we show the correlation
between $C_\gamma$ and $C_{Z\gamma}$. There are 2 islands
corresponding to those shown in 
Fig.~\ref{case2-1}(a).
Both $C_\gamma$ and $C_{Z\gamma}$ increase or decrease in the same
direction, though the enhancement in $C_{Z\gamma}$ is always smaller
than $C_\gamma$. At the best-fit point, we find $C_{Z\gamma} = 1.05$.

\subsection{Vary 
$C_u^S$, $C_d^S$, $C_\ell^S$, $C_v$, $\Delta S^\gamma$, $ \Delta S^g $}

\begin{figure}[th!]
\centering
\includegraphics[width=3.1in]{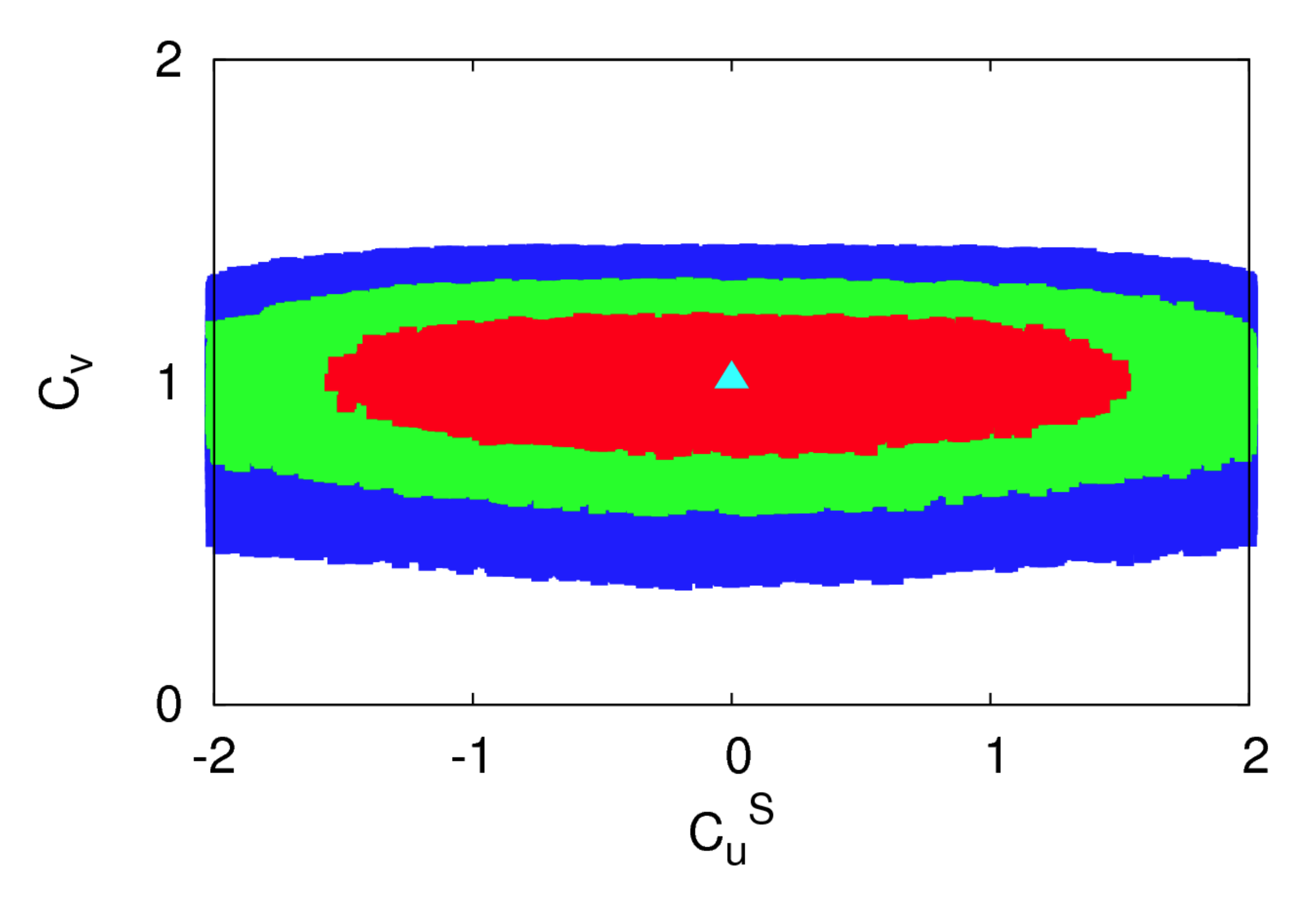}
\includegraphics[width=3.1in]{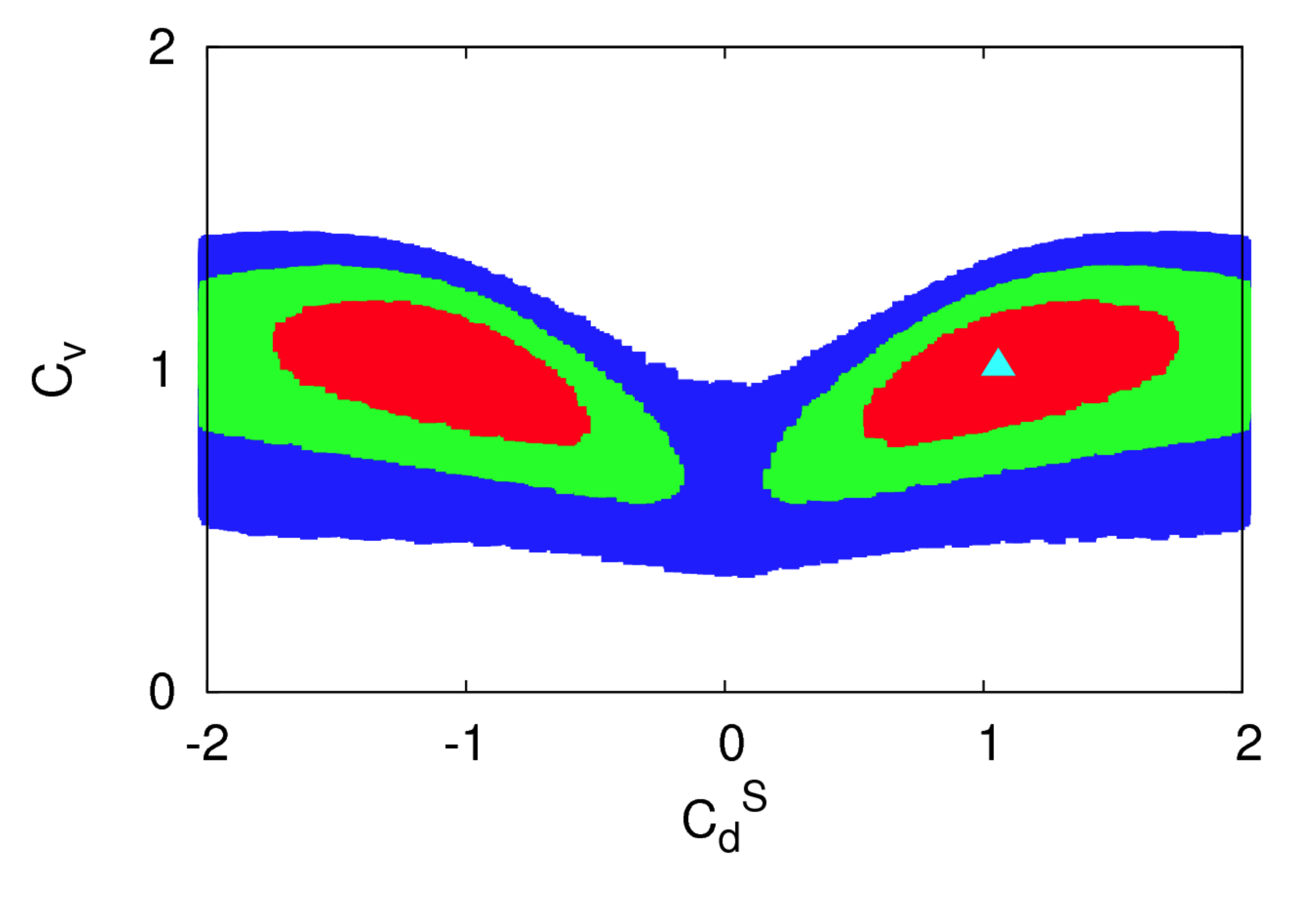}
\includegraphics[width=3.1in]{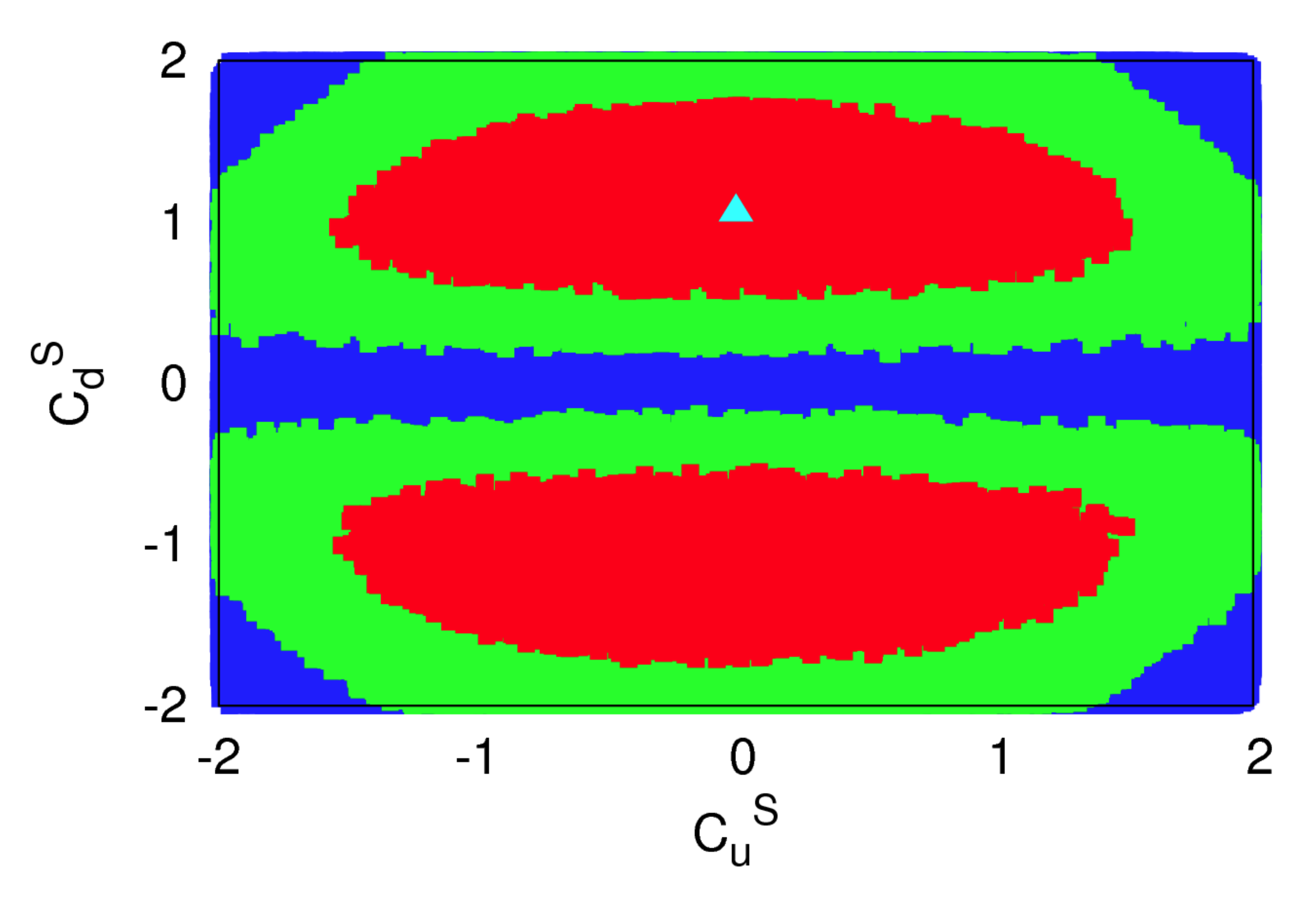}
\includegraphics[width=3.1in]{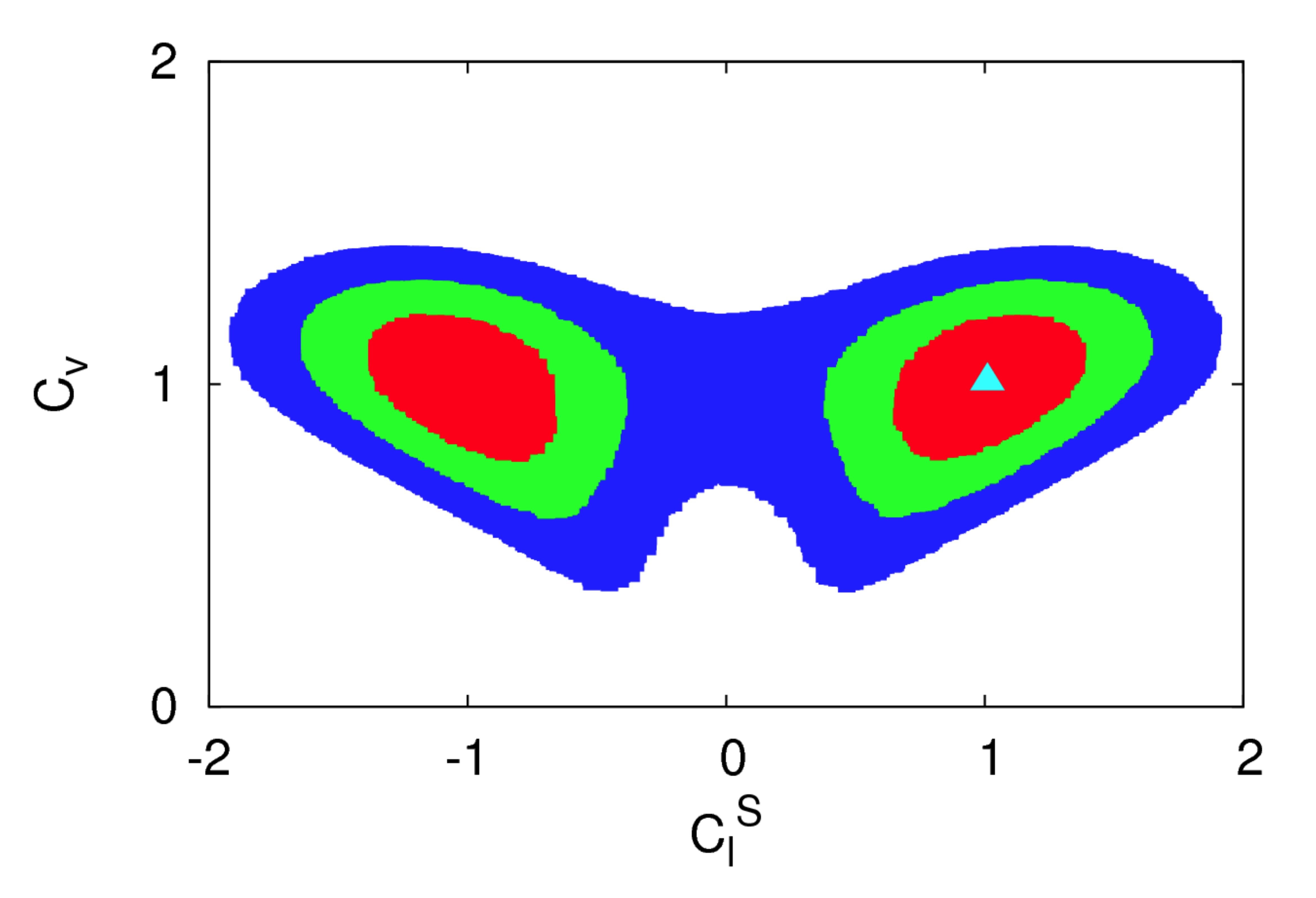}
\includegraphics[width=3.1in]{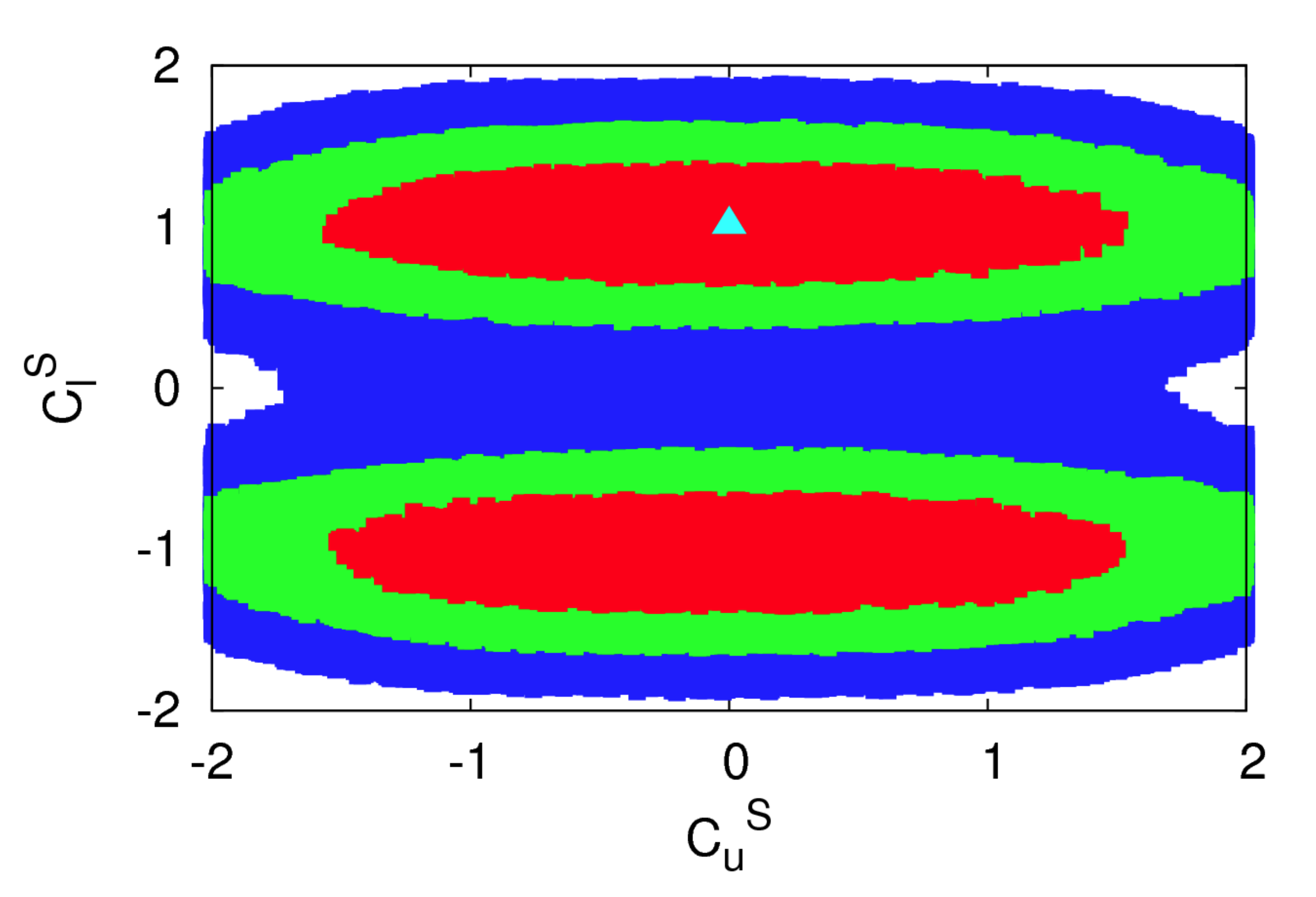}
\includegraphics[width=3.1in]{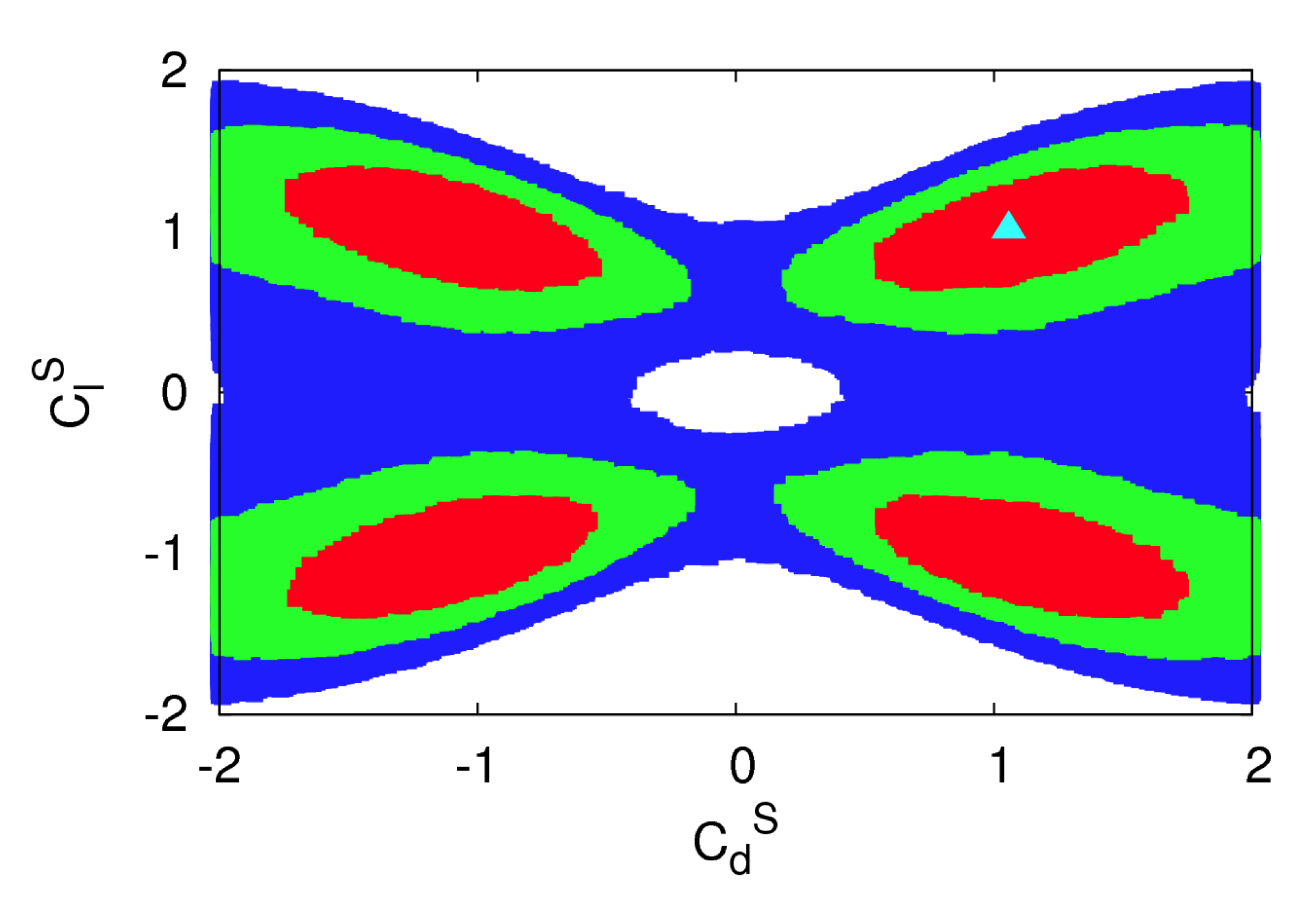}
\caption{\small \label{3-1}
The confidence-level regions of the fit by varying $C_u^S$, $C_d^S$,
$C_\ell^S$, $C_v$, $\Delta S^\gamma$ and $\Delta S^g$ while keeping 
$\Delta \Gamma_{\rm tot} =0$.
Shown are the correlations among 
($C_u^S$, $C_d^S$, $C_\ell^S$, $C_v$).
The description of contour regions is the same as Fig.~\ref{case1}.}
\end{figure}

\begin{figure}[th!]
\centering
\includegraphics[width=3.1in]{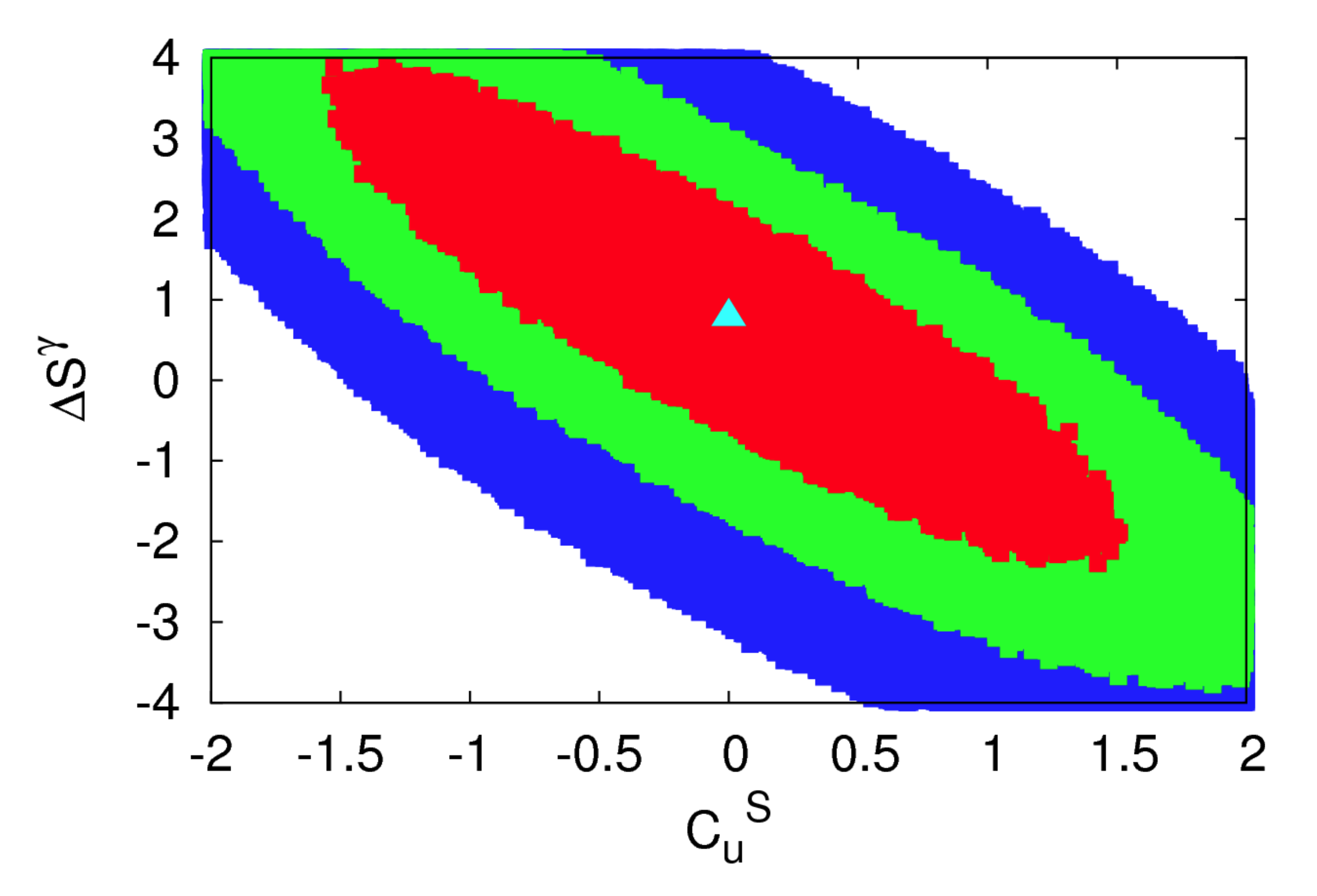}
\includegraphics[width=3.1in]{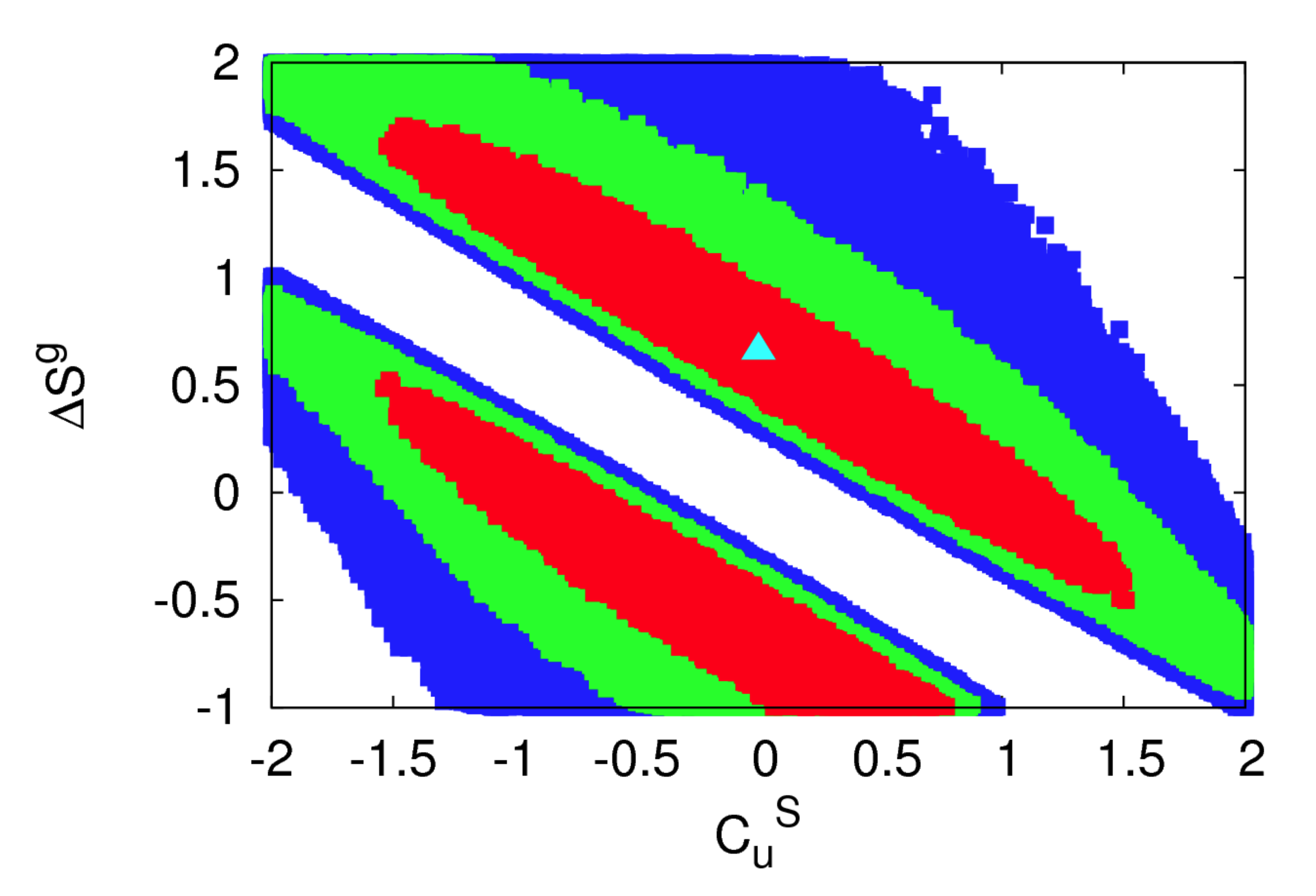}
\includegraphics[width=3.1in]{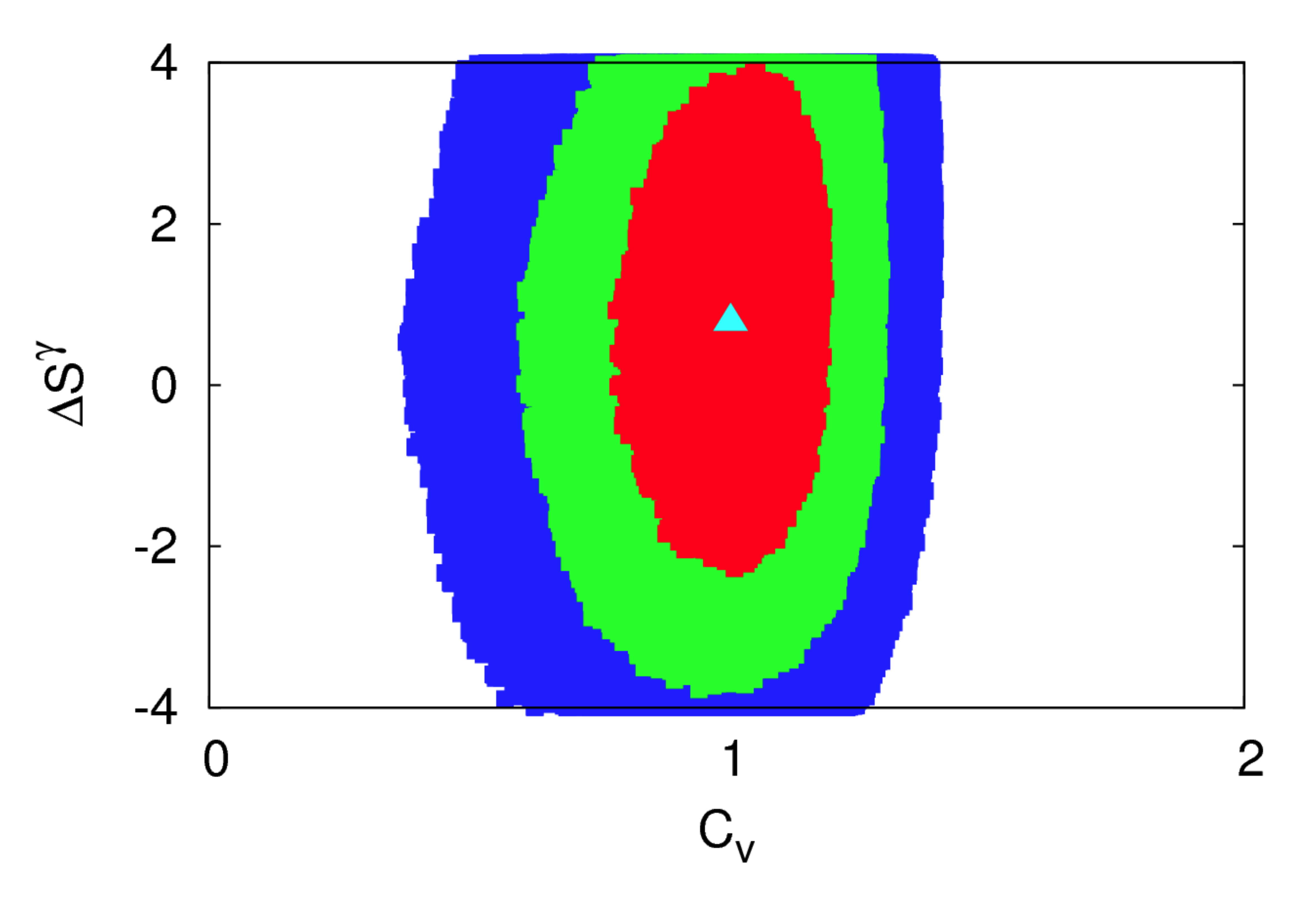}
\includegraphics[width=3.1in]{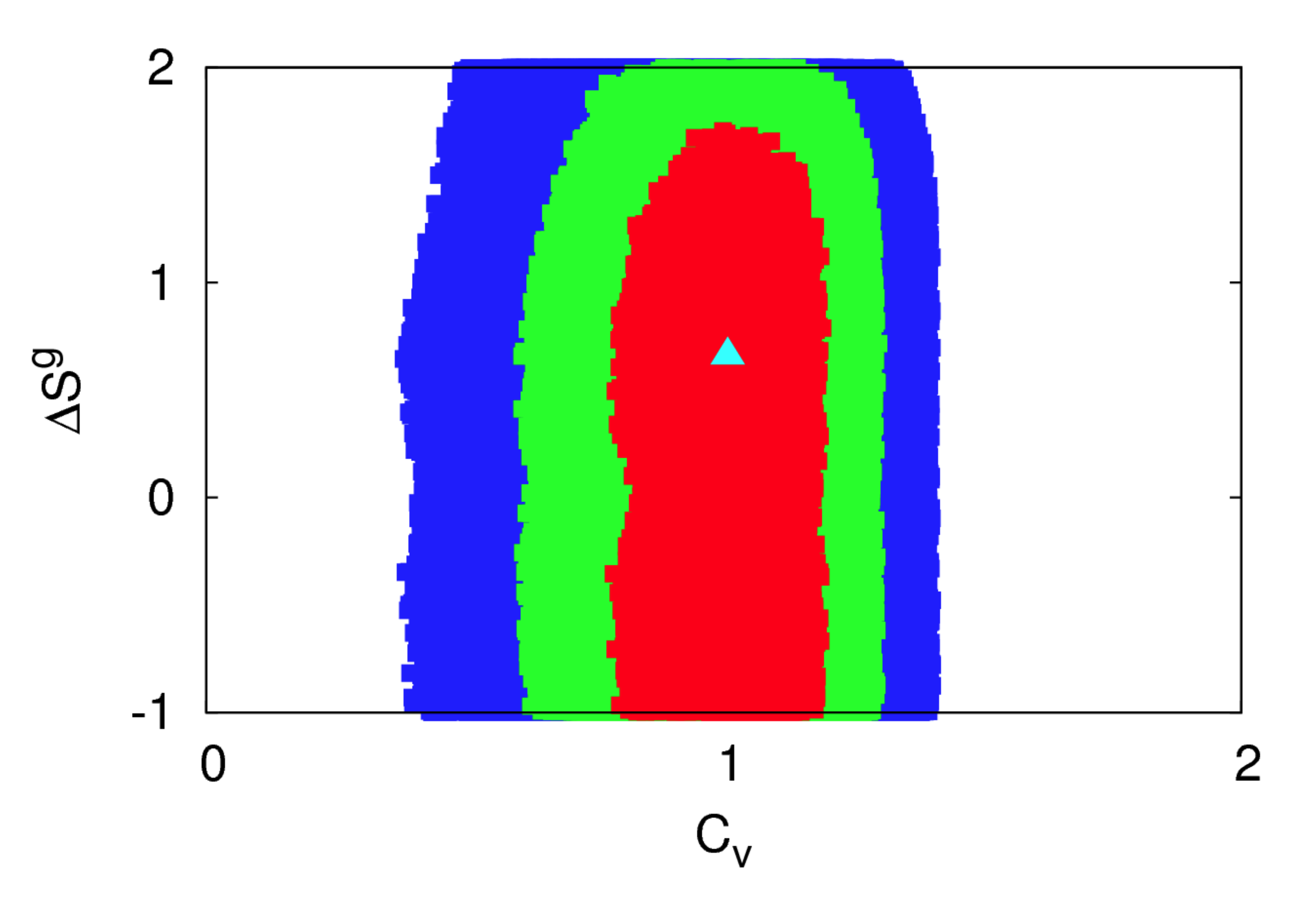}
\caption{\small \label{3-2}
Same as Fig.~\ref{3-1}. 
Shown are the correlations between
($C_u^S$, $C_v$) and ($\Delta S^\gamma$, $\Delta S^g$).
}
\end{figure}

\begin{figure}[th!]
\centering
\includegraphics[width=3.1in]{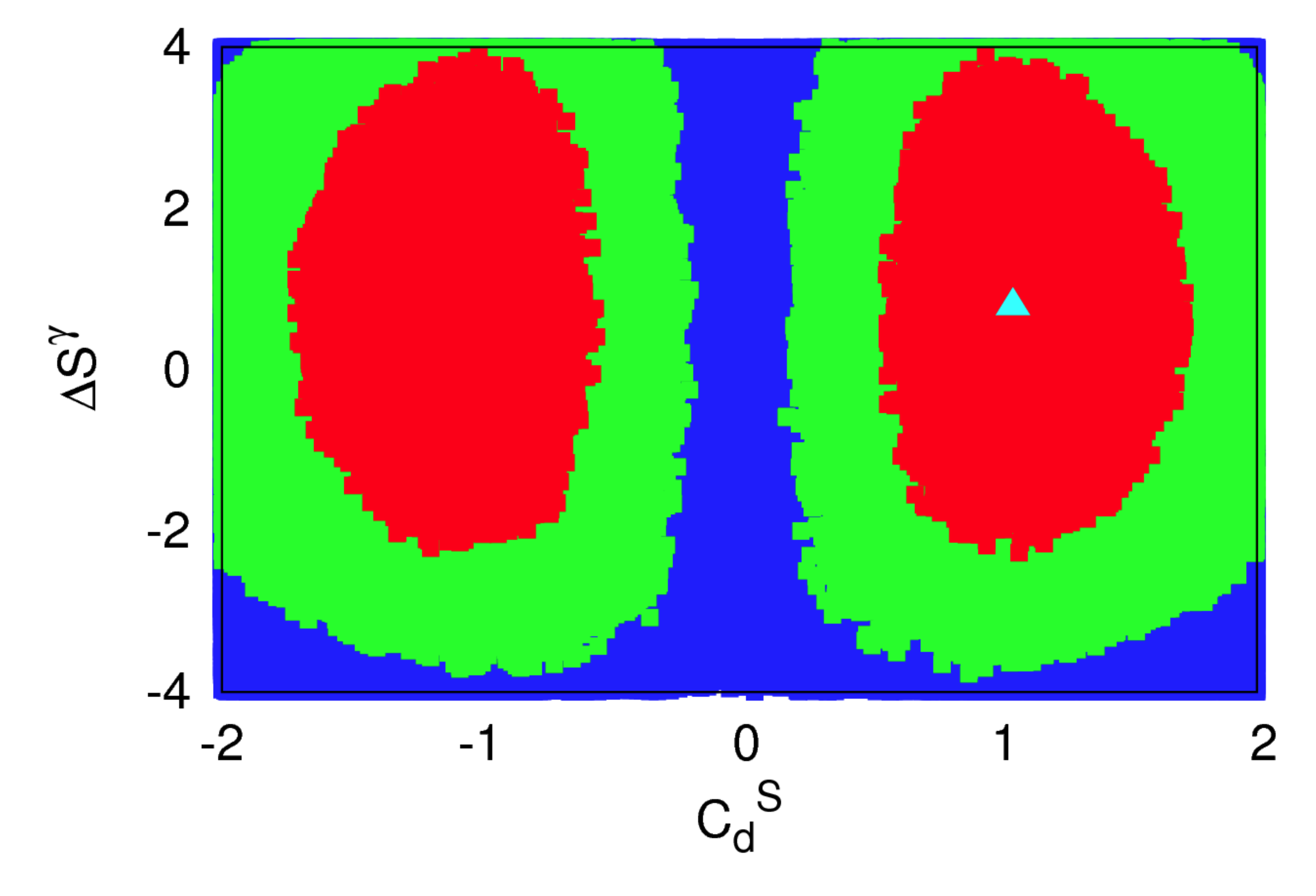}
\includegraphics[width=3.1in]{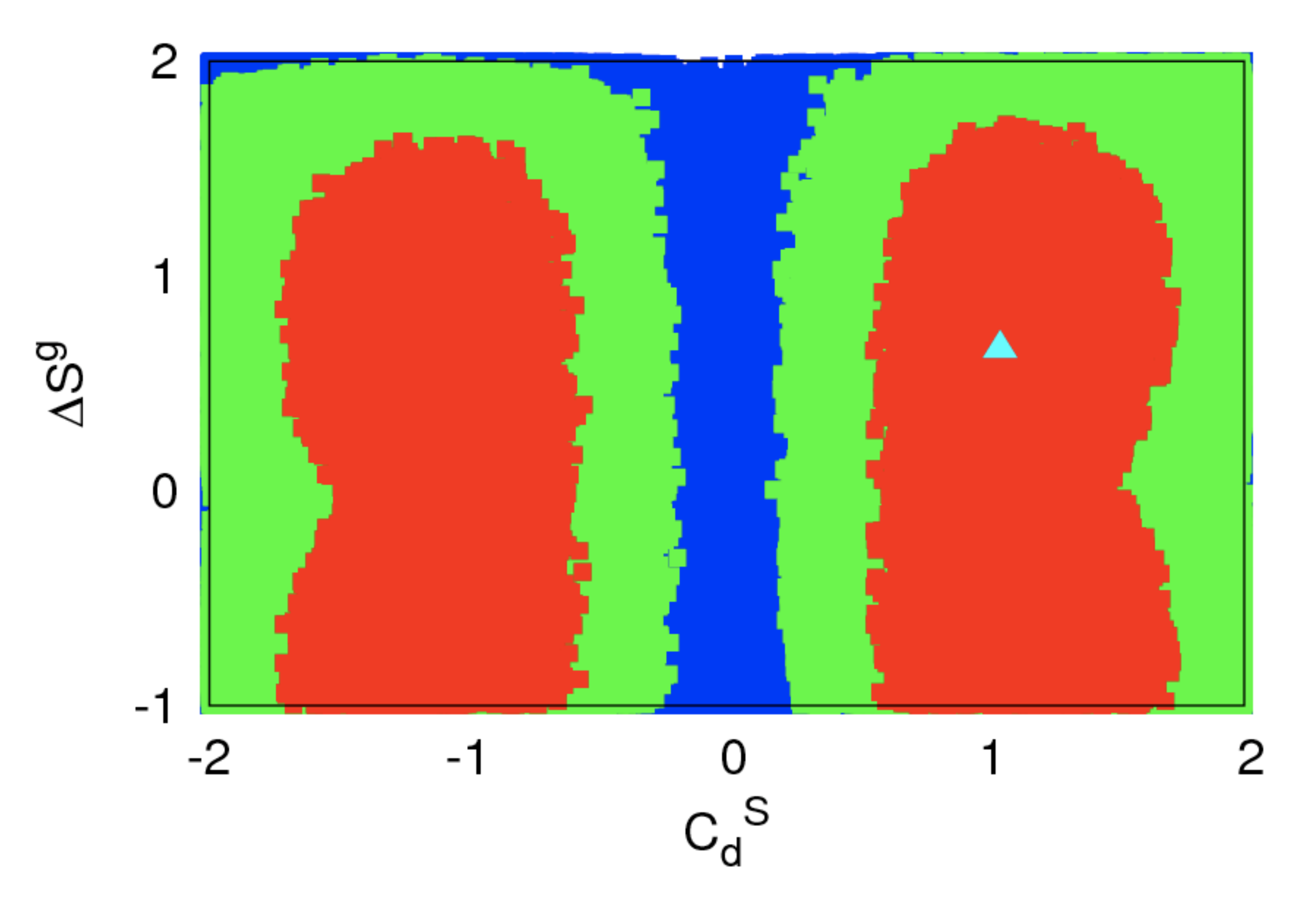}
\includegraphics[width=3.1in]{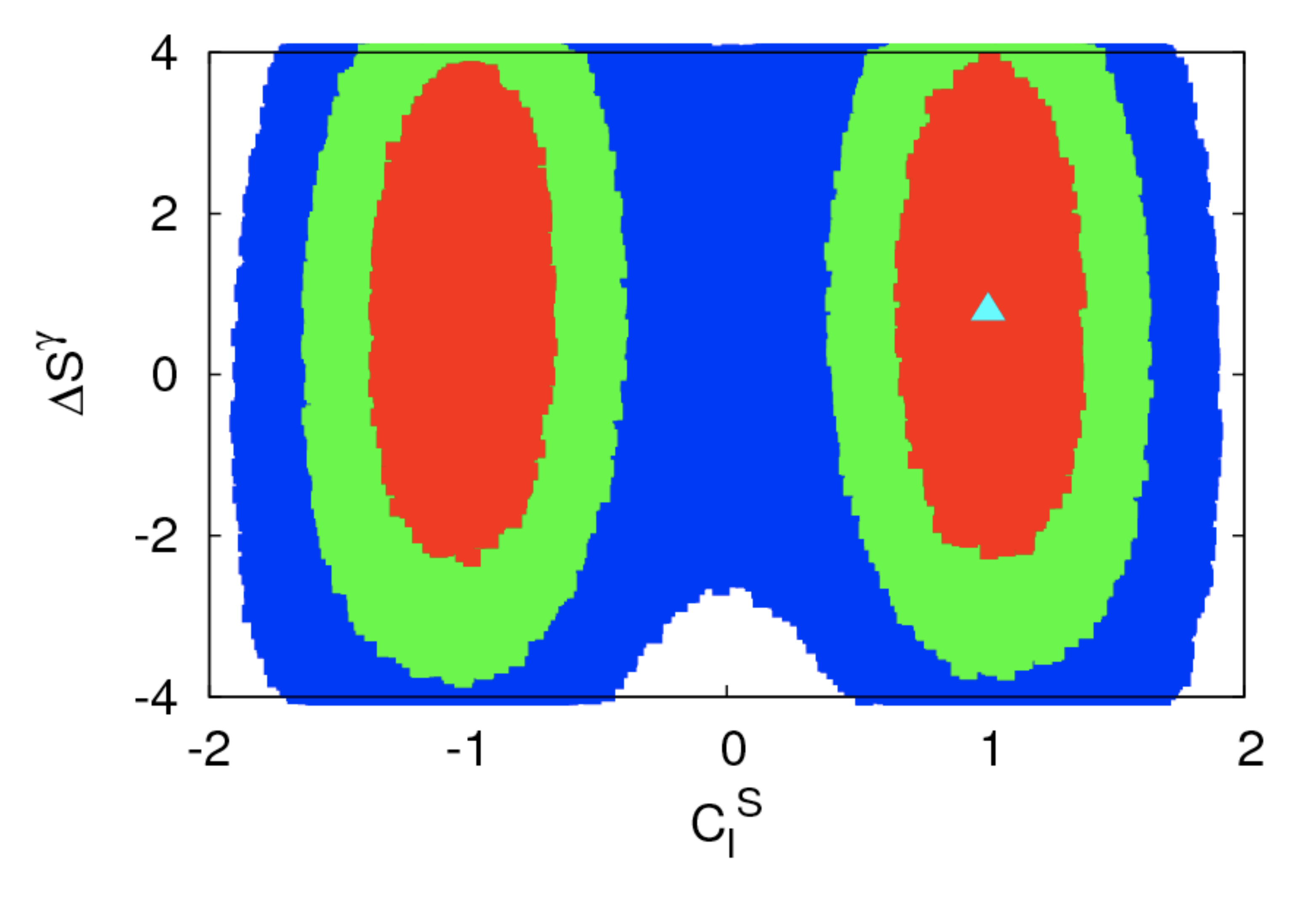}
\includegraphics[width=3.1in]{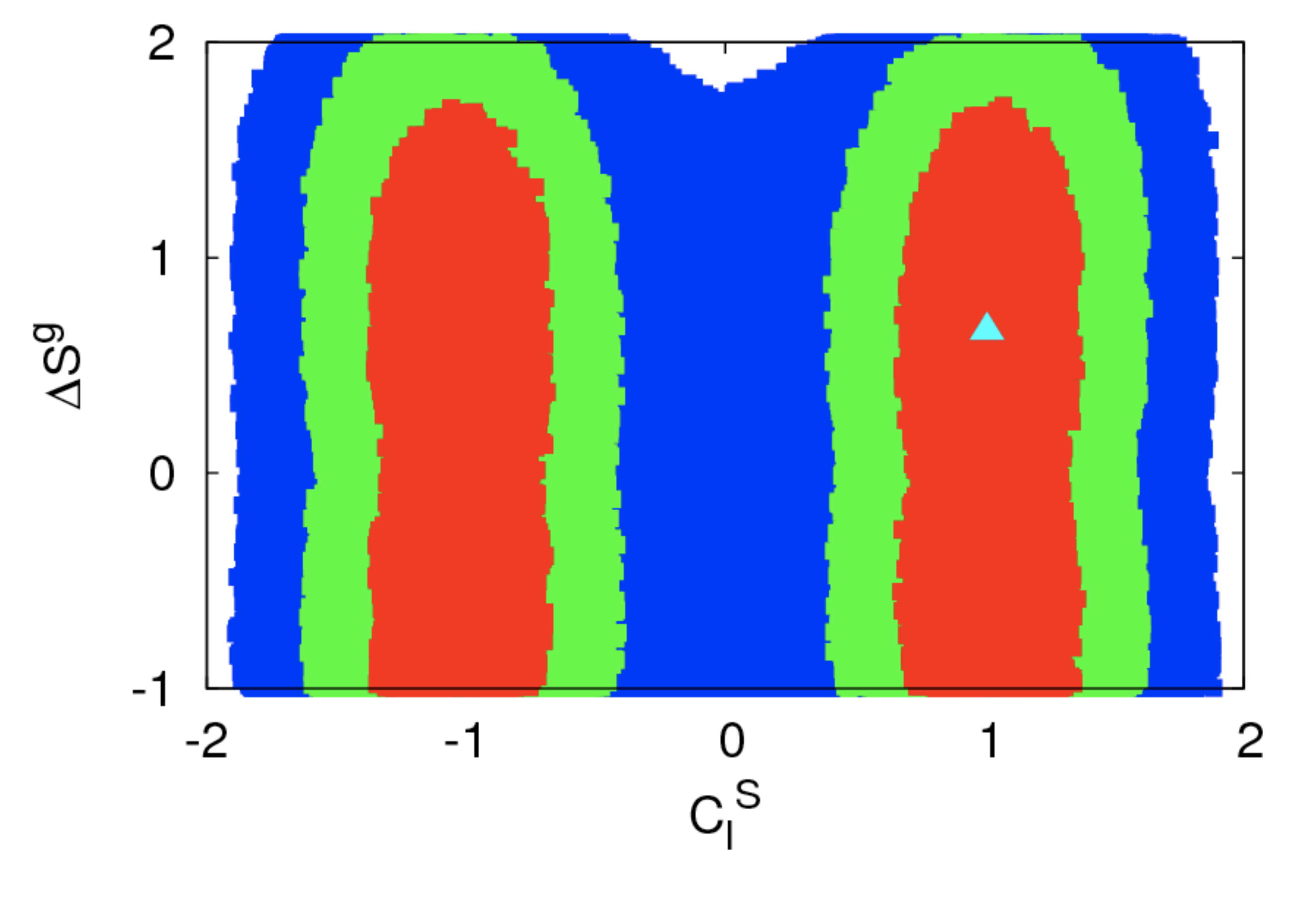}
\caption{\small \label{3-3}
Same as Fig.~\ref{3-1}.
Shown are the correlations between
($C_d^S$, $C_\ell^S$) and ($\Delta S^\gamma$, $\Delta S^g$).
}
\end{figure}

\begin{figure}[th!]
\centering
\includegraphics[width=3.1in]{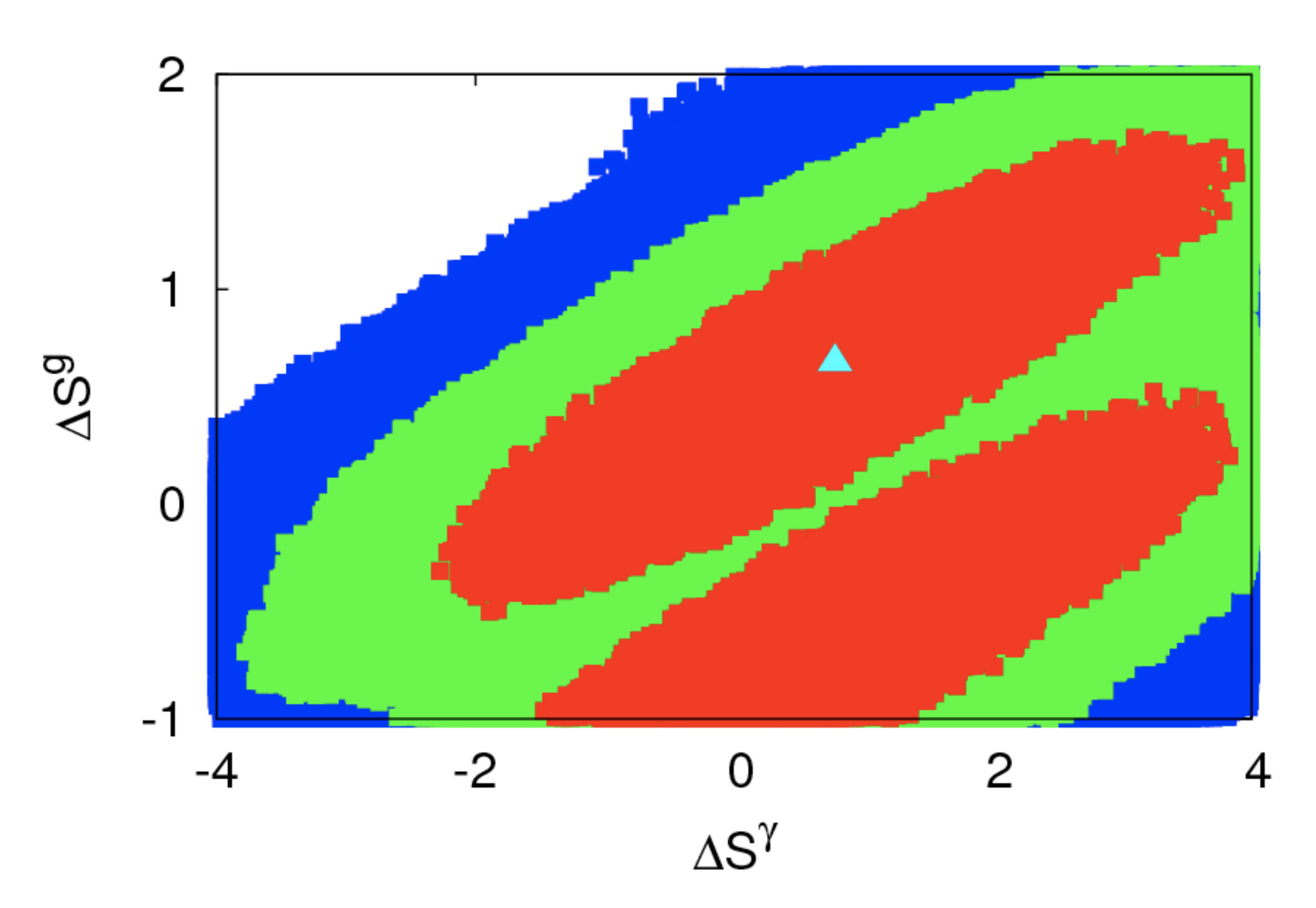}

\includegraphics[width=3.1in]{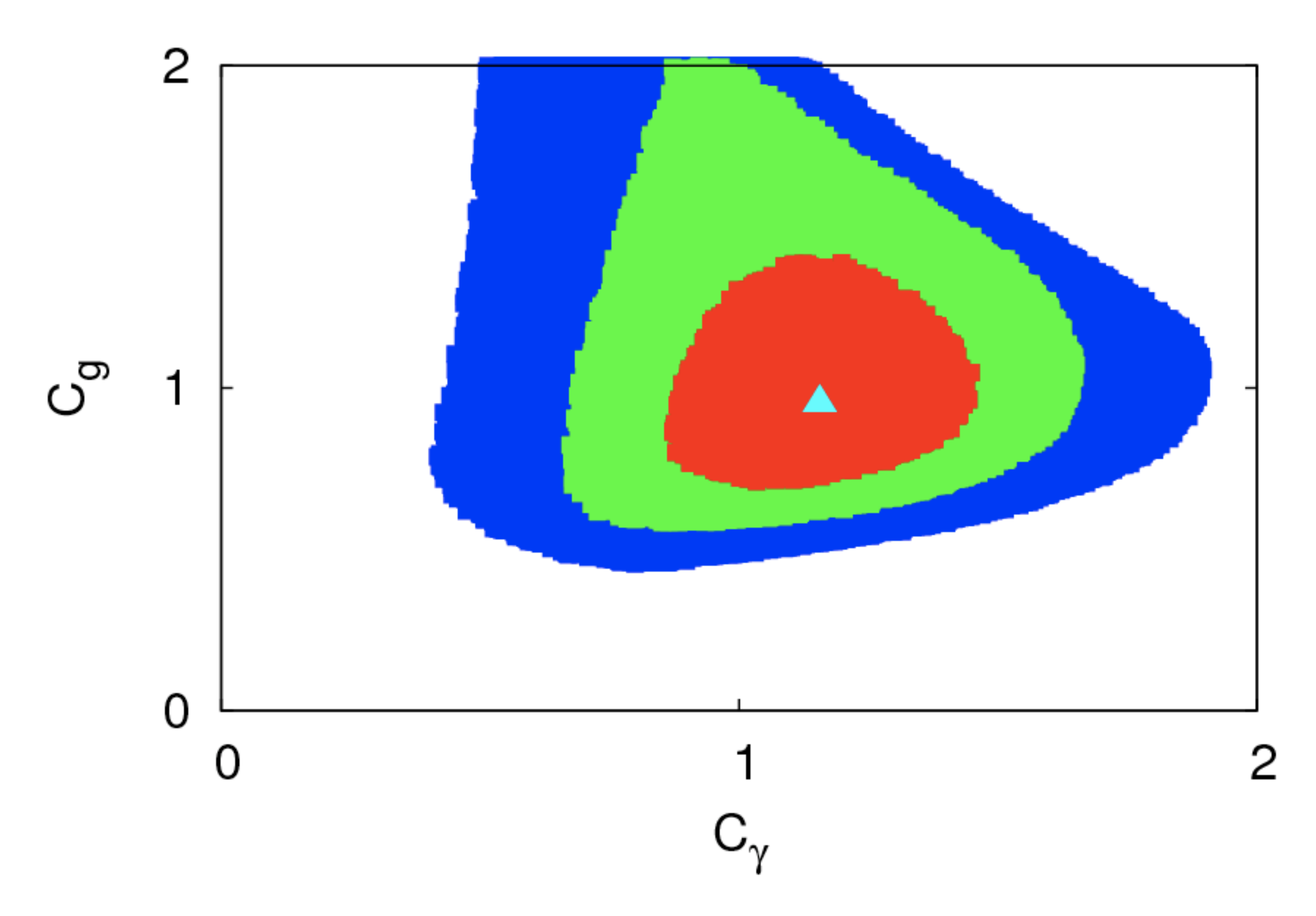}
\includegraphics[width=3.1in]{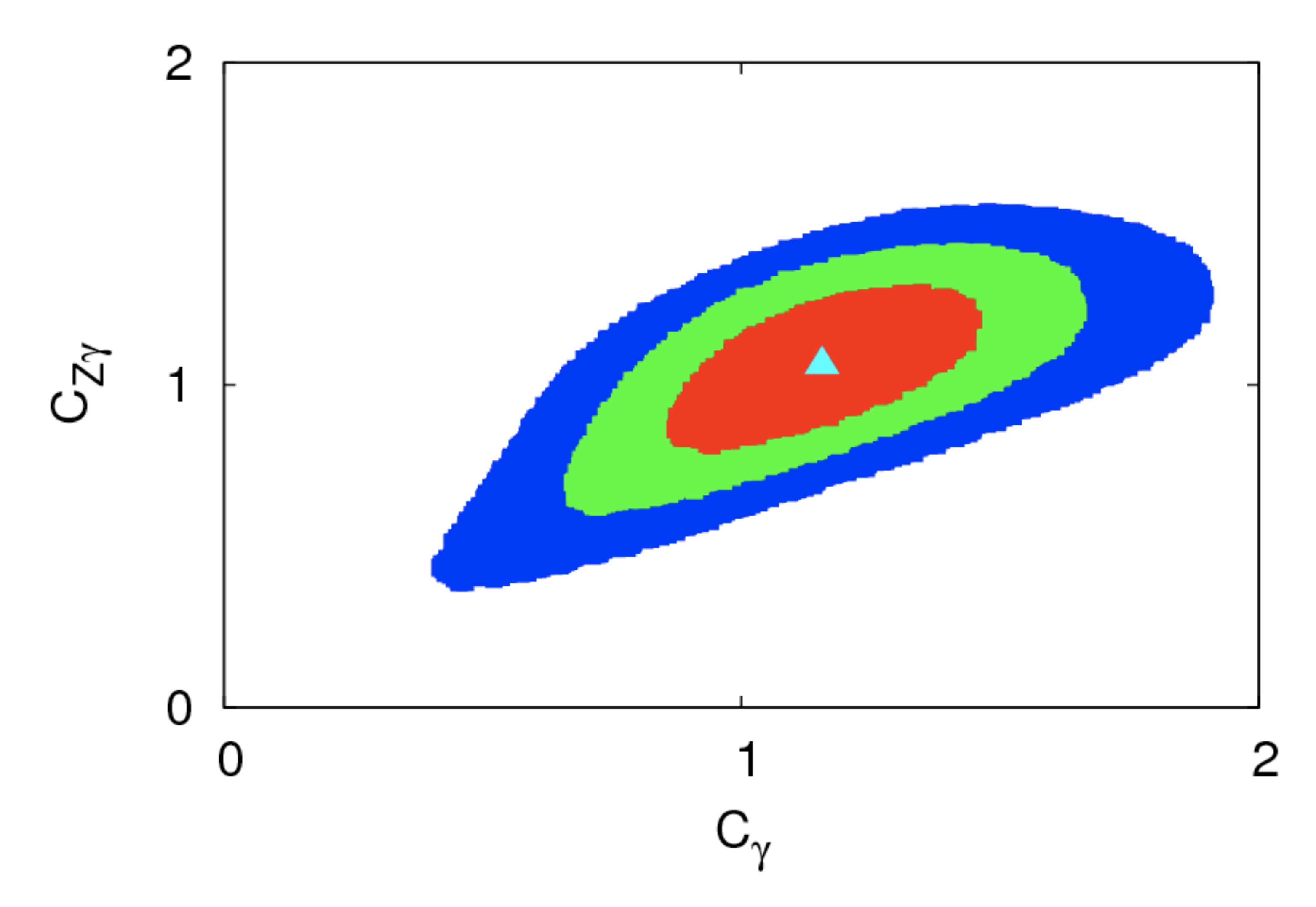}
\caption{\small \label{3-4}
Same as Fig.~\ref{3-1}.
Shown are the correlations between
$\Delta S^\gamma$ and $\Delta S^g$, between $C_\gamma$ and $C_g$,
and between $C_\gamma$ and $C_{Z\gamma}$.
}
\end{figure}

In this fit, we group these 6 parameters into 2 sets: 
$(C^S_u$, $C^S_d$, $C^S_\ell$, $C_v)$ and $(\Delta S^\gamma$, $ \Delta S^g)$.
We first show the correlations among the first set in Fig.~\ref{3-1},
which can be compared directly to the corresponding panels in
Fig.~\ref{case2-1}. 
It is easy to see that all the confidence-level
regions are enlarged due to more dof in 
$(\Delta S^\gamma$, $ \Delta S^g)$. 
The next correlations between ($C^S_u$, $C_v$) and 
($\Delta S^\gamma$,$ \Delta S^g $) are shown in Fig.~\ref{3-2}.
The correlations between
($C^S_d$, $C_\ell^S$) and ($\Delta S^\gamma$,$ \Delta S^g $)
are shown in Fig.~\ref{3-3}.
The corresponding confidence-level regions in 
$(\Delta S^\gamma, \Delta S^g)$ and $(C_\gamma, C_g)$ planes are
shown in Fig.~\ref{3-4}. 
Note that these figures used the data after the Moriond update.

\subsubsection{Before Moriond}
The fit before Moriond is shown in the last column of the upper half
in Table~\ref{bestfit}. 
The most significant changes from the previous fit are
the widening of $C_u^S$ and the shift of the best value of 
$C_u^S$ from approximately $-0.9 \to 0$. We shall explain it shortly.
The $C_v$ and $C_d^S$ remains
approximately the same. Although the $C_\ell^S$ flips the sign, the
absolute value is about the same. The flipping of the sign of 
$C_\ell^S$ is simply a numerical artifact that the difference in 
$\chi^2$ is only $O(10^{-3})$.  In this 6-parameter fit, the signs 
of the best values of
$C^S_u$, $C^S_d$, $C^S_\ell$, $C_v$ are all positive, in accord with the SM.

The shift of $C_u^S$ from $-0.9 \to 0$ can be understood from the numerical
expression of $S^\gamma$ in Eq.~(\ref{5}). In the 4-parameter fit where
$\Delta S^\gamma=0$, the top-Yukawa coupling $C_u^S$ is made negative in 
order to increase numerically the $S^\gamma$; whereas in the 6-parameter
fit the $C_u^S$ goes to zero and $\Delta S^\gamma$ goes to a negative value
to enhance $S^\gamma$. This explains the shift of $C_u^S$ and anti-correlation
between $C_u^S$ and $\Delta S^\gamma$.  

The resulting $\chi^2/dof = 9.89/16$, which is pretty good.  The dynamics
of the fit raises the diphoton rate to fit the data well, so that
the $\chi^2$ is reduced substantially.

\subsubsection{After Moriond}
As shown in the last column of the lower half in Table~\ref{bestfit}, 
the $C_u^S$ changes from approximately $0.8 \to 0$,
while the $C_d^S$, $C_\ell^S$, and $C_v$ remains about $1$. 
The Yukawa couplings are almost in good accord with the SM, except for
the top-Yukawa.  Instead, $\Delta S^\gamma$ and $\Delta S^g$ become
nonzero to accommodate the data. Nevertheless, the reduction of 
$\chi^2$ is very small. The resulting $\chi^2/dof = 16.89/16$,
which is worse than all other fits. 

The next correlations between ($C^S_u$, $C_v$) and 
($\Delta S^\gamma$,$ \Delta S^g $) are shown in Fig.~\ref{3-2}.
Both the $\Delta S^\gamma$ and $ \Delta S^g $ are anti-correlated with
$C_u^S$. 
On the other hand, 
$\Delta S^\gamma$ and $ \Delta S^g$ do not correlate with $C_v$.
Similarly, the correlations between
($C^S_d$, $C_\ell^S$) and ($\Delta S^\gamma$,$ \Delta S^g $)
are shown in Fig.~\ref{3-3}, which are also negligibly correlated.

The corresponding confidence-level regions in 
$(\Delta S^\gamma, \Delta S^g)$ and $(C_\gamma, C_g)$ planes are
shown in Fig.~\ref{3-4}. The prediction for $C_{Z\gamma}$ is also shown.
Again, the $C_{Z\gamma}$ increases or decreases in the same direction
as $C_\gamma$, but is always smaller than $C_\gamma$.

\subsection{Concluding remarks}

The best-fit values for various CP-conserving fits 
using the Higgs data before the Moriond 2013 
are shown in the upper half of Table~\ref{bestfit} while 
using the data after the Moriond are shown in the lower half.
We also show the $p$-value of each fit. It is clear that 
most of the fits have better $p$-values than the SM one before
the Moriond; while all the fits are worse than the SM one
after the Moriond.

Before the Moriond update, 
the diphoton signal strength $pp\to H \to \gamma\gamma$ dominates
the chi-square.
Both the ATLAS and CMS diphoton data are on the {\it same} side of excess of 
the SM value.
The signal strength of
$pp\to H \to \gamma\gamma$ depends largely on $S^\gamma$ and $S^g$,
which in turns depend mostly on $C_u^S$ and $\Delta S^\gamma$. Indeed,
we have shown in the 4-parameter analysis and in the 6-parameter analysis,
the $\chi^2$ is mostly sensitive to $C_u^S$ and $\Delta S^\gamma$. 
In the 4-parameter analysis where $\Delta S^\gamma=0$, the best-fit value
of $C_u^S\approx -0.9$ in order to enhance $S^\gamma$; whereas in the 
6-parameter analysis $C_u^S \approx 0$ and $\Delta S^\gamma \approx -1.2$ are 
preferred in order to enhance $S^\gamma$. The value for $\Delta S^g$
also changes according to the change in the value of $C_u^S$ 
as in the expression for $S^g$ in Eq.~(\ref{8}).

The $C_u^S$ and $\Delta S^\gamma$ are the two parameters 
most sensitive to the signal strength of
$pp\to H \to \gamma\gamma$. Keeping
the other parameters as $C_v=C_d^S = C_\ell^S =1, \Delta S^g=0,
\Delta \Gamma_{\rm tot} = 0$, we vary $C_u^S$ and $\Delta S^\gamma$ only
and we find the best-fit values are 
\begin{equation}
C_u^S = 0.92 ^{+0.094}_{-0.095},\;\;
\Delta S^\gamma = -2.62 ^{+1.02}_{-1.04},\;\;
\chi^2/dof = 11.17/20 \;.
\end{equation}
This is the best $\chi^2/dof$ that we found when only two parameters
are allowed to vary, although
the total $\chi^2$ is only 0.1 unit better than the fit with
$\Delta S^\gamma$ and $\Delta S^g$, which is statistically insignificant. 


With the Higgs updates during the Moriond 2013 meetings \cite{moriond}
the uncertainties in most channels are reduced. The decay channels other
than the diphoton also began to play important roles in the global fits.
The most dramatic change is the CMS diphoton data, in which the
central value (the untagged) changes from $1.42$ to $0.78$. Now the CMS
and ATLAS diphoton data are on the {\it opposite} side of the SM value.
The dynamics of the fit cannot do anything to effectively reduce the
$\chi^2$ from the diphoton data.  We found that all the fits give 
a $p$-value worse than the SM one.

\section{CP violating fits}
We devote this section to including the pseudoscalar Yukawa couplings
and the pseudoscalar contributions $\Delta P^\gamma$ and $\Delta P^g$.

\subsection{$\Delta S^\gamma$, $\Delta S^g$,  $\Delta P^\gamma$ and 
$\Delta P^g$ }

\subsubsection{Before Moriond}
We have learned from all CP-conserving fits in the last section that 
the most efficient parameters fitting the data are the deviation
$\Delta S^\gamma$ to the $H\gamma\gamma$ vertex and the up-type Yukawa
coupling $C_u^S$, as well as the corresponding deviation $\Delta S^g$ to the 
$Hgg$ vertex.
In order to understand the effects of pseudoscalar nature of
the Higgs boson, 
we first perform the analysis by varying
the scalar contributions $\Delta S^\gamma$ and $\Delta S^g$, as well as 
the pseudoscalar contributions $\Delta P^\gamma$ and $\Delta P^g$ to the 
$H\gamma\gamma$ and $Hgg$ vertices. 
We keep all other parameters at the SM values,
$C_u^S = C_d^S = C_\ell^S=C_v=1$, $C_u^P = C_d^P = C_\ell^P=0$ and 
$\Delta \Gamma_{\rm tot} =0$.

The best-fit parameters and the corresponding $\chi^2$ for this case are shown
in the second and third columns of the upper half in 
Table~\ref{cbestfit}. The total $\chi^2 =11.26$, 
almost the same as the total $\chi^2=11.27$ of the case varying 
$\Delta S^\gamma, \Delta S^g$ only. Therefore, including the pseudoscalar 
contributions does not improve the fit at all. In fact, the $\chi^2/dof$
is worsened. 

\subsubsection{After Moriond}
\begin{figure}[th!]
\centering
\includegraphics[width=3.1in]{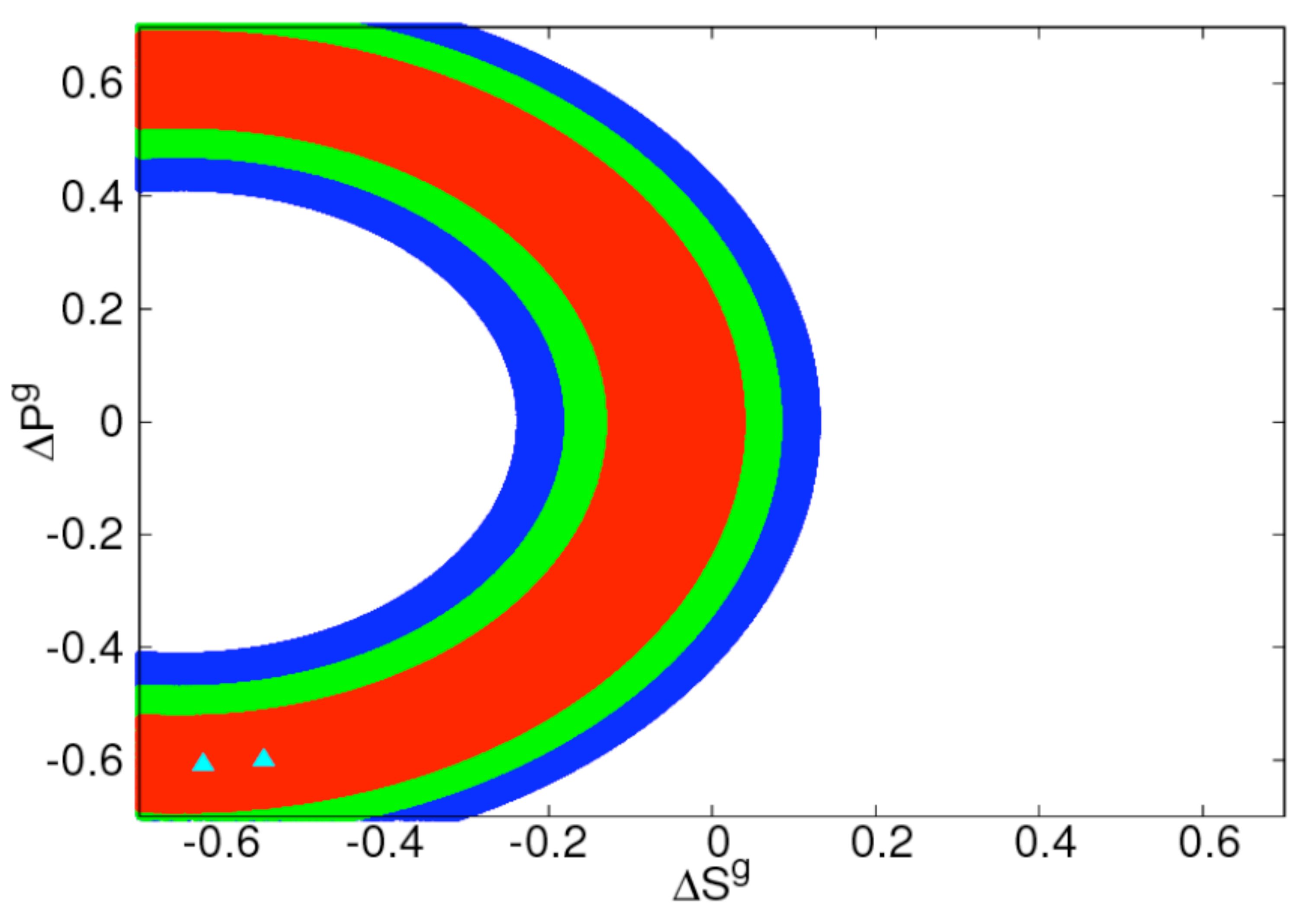} 
\includegraphics[width=3.1in]{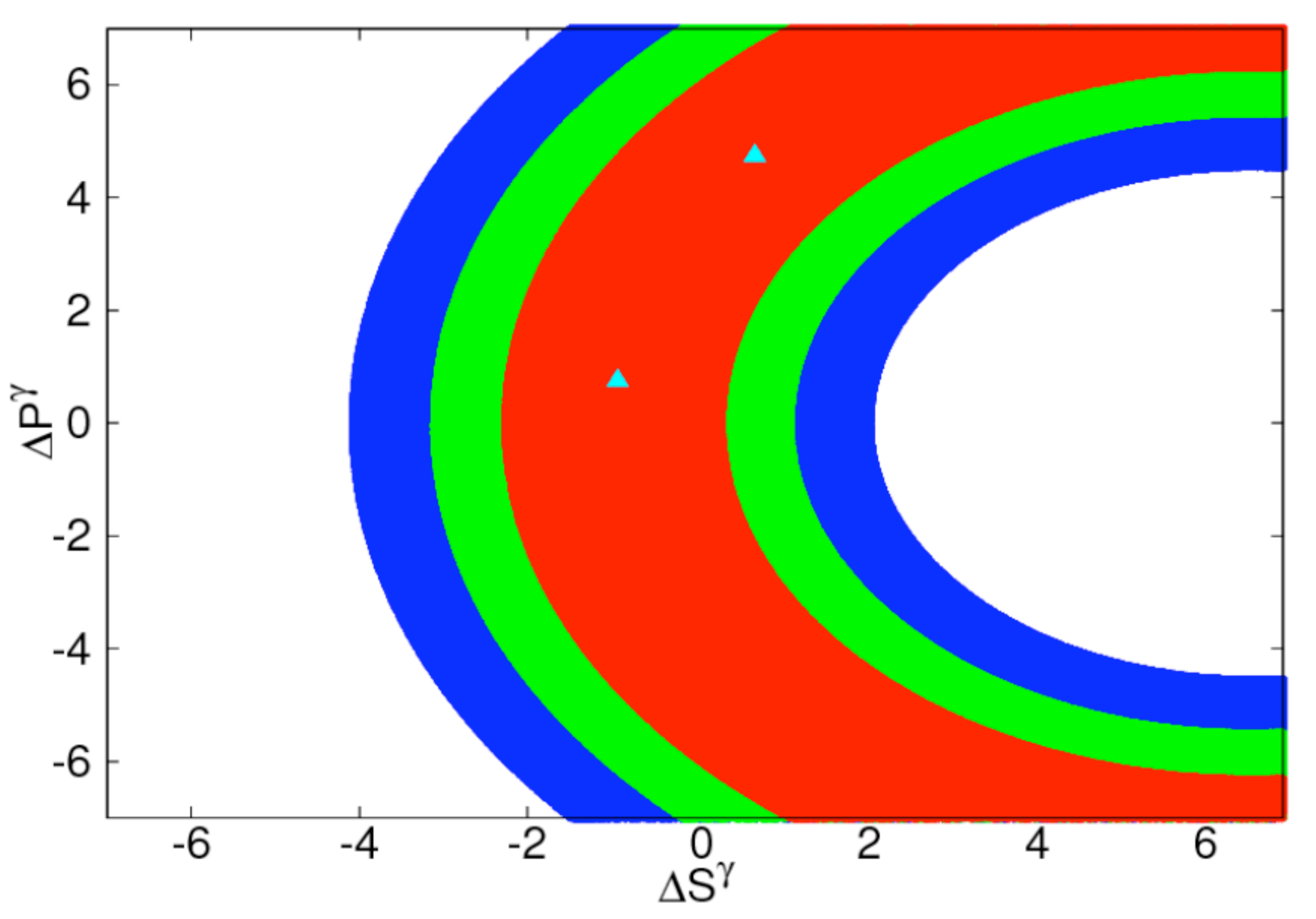}
\includegraphics[width=3.1in]{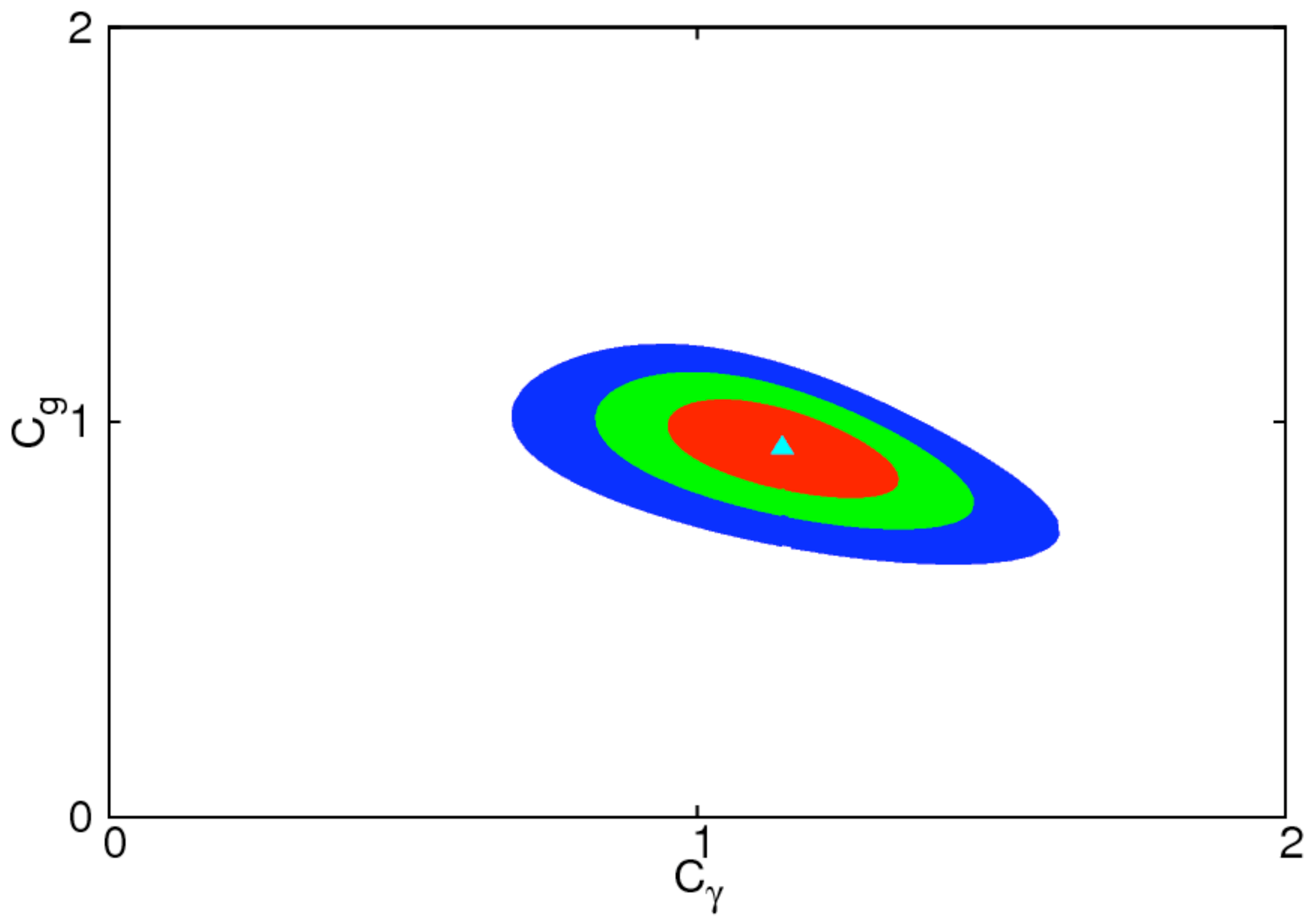}
\caption{\small \label{cpv1}
The confidence-level regions of the fit by varying 
the scalar contributions $\Delta S^\gamma$ and $\Delta S^g$,
and the pseudoscalar contributions $\Delta P^\gamma$ and $\Delta P^g$
while keeping 
$C_u^S = C_d^S = C_\ell^S=1$, $C_u^P = C_d^P = C_\ell^P=0$ and 
$\Delta \Gamma_{\rm tot} =0$.
The description of contour regions is the same as Fig.~\ref{case1}.
}
\end{figure}

The confidence-level regions in the $(\Delta S^g,\Delta P^g)$,
in the $(\Delta S^\gamma, \Delta P^\gamma)$, and in the corresponding
$(C_\gamma, C_g)$ planes are shown in Fig.~\ref{cpv1}. The nearly-physical
values for 
$C_\gamma \approx 1.1$ and $C_g \approx 0.9$, which are
the same as the fit using just $\Delta S^\gamma$ and $\Delta S^g$.  In order
to understand the behavior shown in Fig.~\ref{cpv1}(a) and (b), we can
use the numerical expressions for $S^\gamma,P^\gamma$ in Eq.~(\ref{5})
and $S^g, P^g$ in Eq.~(\ref{8}). 

Numerically, 
\begin{eqnarray}
C_\gamma \approx 1.1 &=& \sqrt{ 
    \frac{ (-6.64 + \Delta S^\gamma)^2 +  (\Delta P^\gamma)^2 }
         { (-6.64)^2 } }\,, \nonumber \\
C_g \approx 0.9 &=& \sqrt{ 
    \frac{ ( 0.65 + \Delta S^g)^2 + (\Delta P^g)^2 }
         { ( 0.65 )^2 } }\,. \nonumber 
\end{eqnarray}
Therefore, we obtain 2 ellipses
\begin{eqnarray}
(7.3)^2 = ( -6.64 + \Delta S^\gamma)^2 + ( \Delta P^\gamma )^2\,, \nonumber \\
(0.59)^2 = ( 0.65 + \Delta S^g )^2 + (\Delta P^g )^2\,. \label{cpveq1}
\end{eqnarray}
that explain the ellipses shown in Fig.~\ref{cpv1}(a) and (b).
It is clear that nonzero values of $\Delta P^\gamma$ and $\Delta P^g$ are
not ruled out, and the data allow for both scalar and pseudoscalar 
values in special combinations for $H\gamma\gamma$ and $Hgg$ vertices.

The best-fit parameters and the corresponding $\chi^2$ for this case are shown
in the second and third columns of the lower half in
Table~\ref{cbestfit}. The total $\chi^2 = 17.55 $, 
the same as the case varying 
$\Delta S^\gamma, \Delta S^g$ only.  Again, including the pseudoscalar 
contributions does not improve the fit at all.

\subsection{ $C_u^S$, $C_u^P$ and $C_v$}

{}From the results of the last section, we can see that among all
the Yukawa couplings
the fitting result is more sensitive to the up-type Yukawa couplings
$C_u^S$. This is easy to understand because the diphoton
data currently dominate the chi-square, and the top-Yukawa and $HWW$ 
couplings are the most important to determine the diphoton signal strength.
Without loss of generality we only consider the up-type scalar and 
pseudoscalar Yukawa couplings and $C_v$ in this subsection. 
The effects of down-type
and charged-lepton pseudoscalar couplings are similar but much milder. 

\subsubsection{Before Moriond}

The best-fit parameters and the corresponding $\chi^2$ for this case are shown
in the fourth and the fifth columns of the upper half
in Table~\ref{cbestfit}. The total $\chi^2 =10.53$, 
almost the same as the total $\chi^2=10.46$ of the case varying 
$C_{u,d,\ell}^S$ and $C_v$. Therefore, including the pseudoscalar 
Yukawa coupling $C_u^P$ does not improve the fit. 
Nevertheless, the $p$-values are about the same, and so the data
cannot rule out the combination of scalar and pseudoscalar Yukawa couplings.

\subsubsection{After Moriond}

\begin{figure}[th!]
\centering
\includegraphics[width=3.1in]{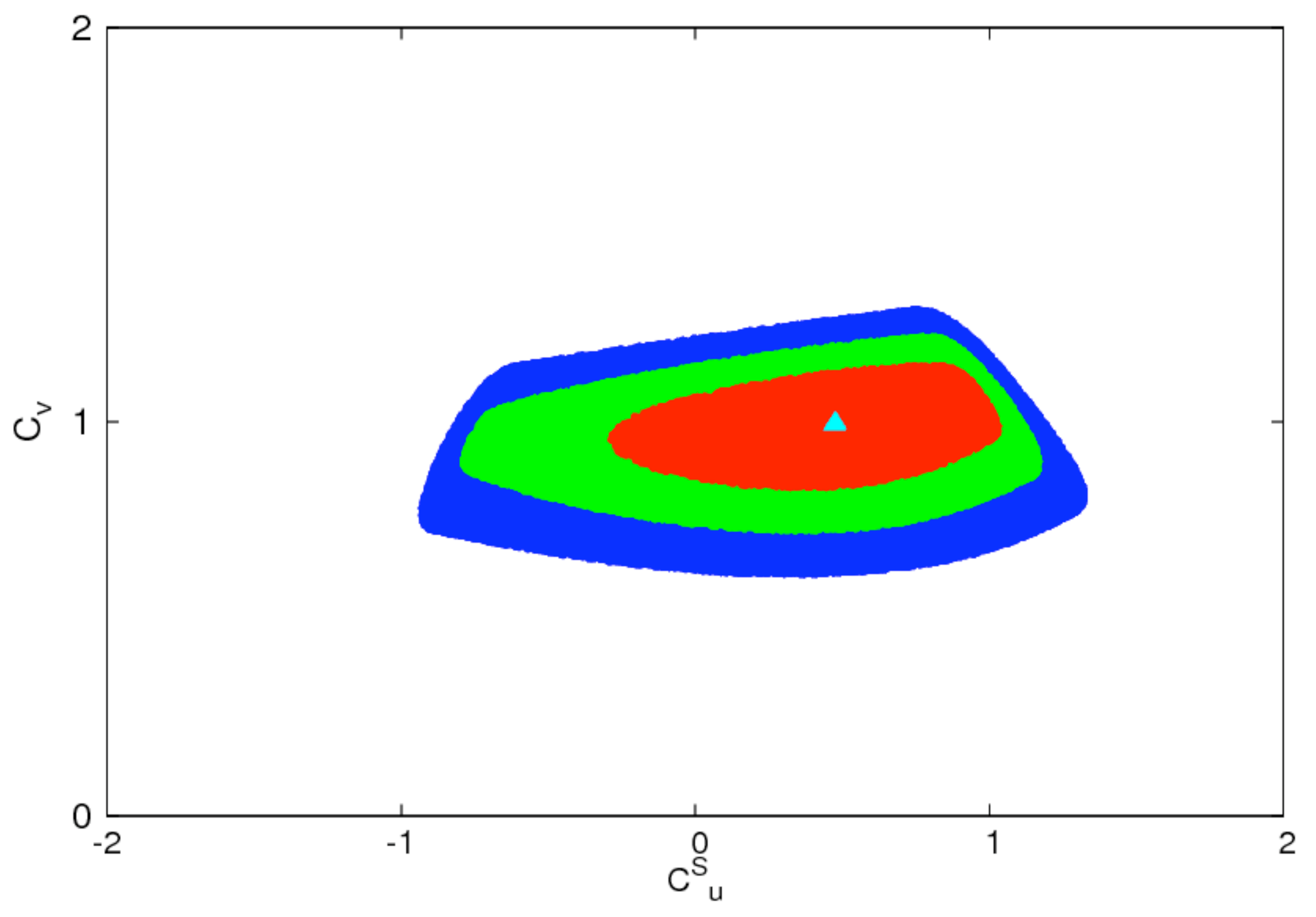}
\includegraphics[width=3.1in]{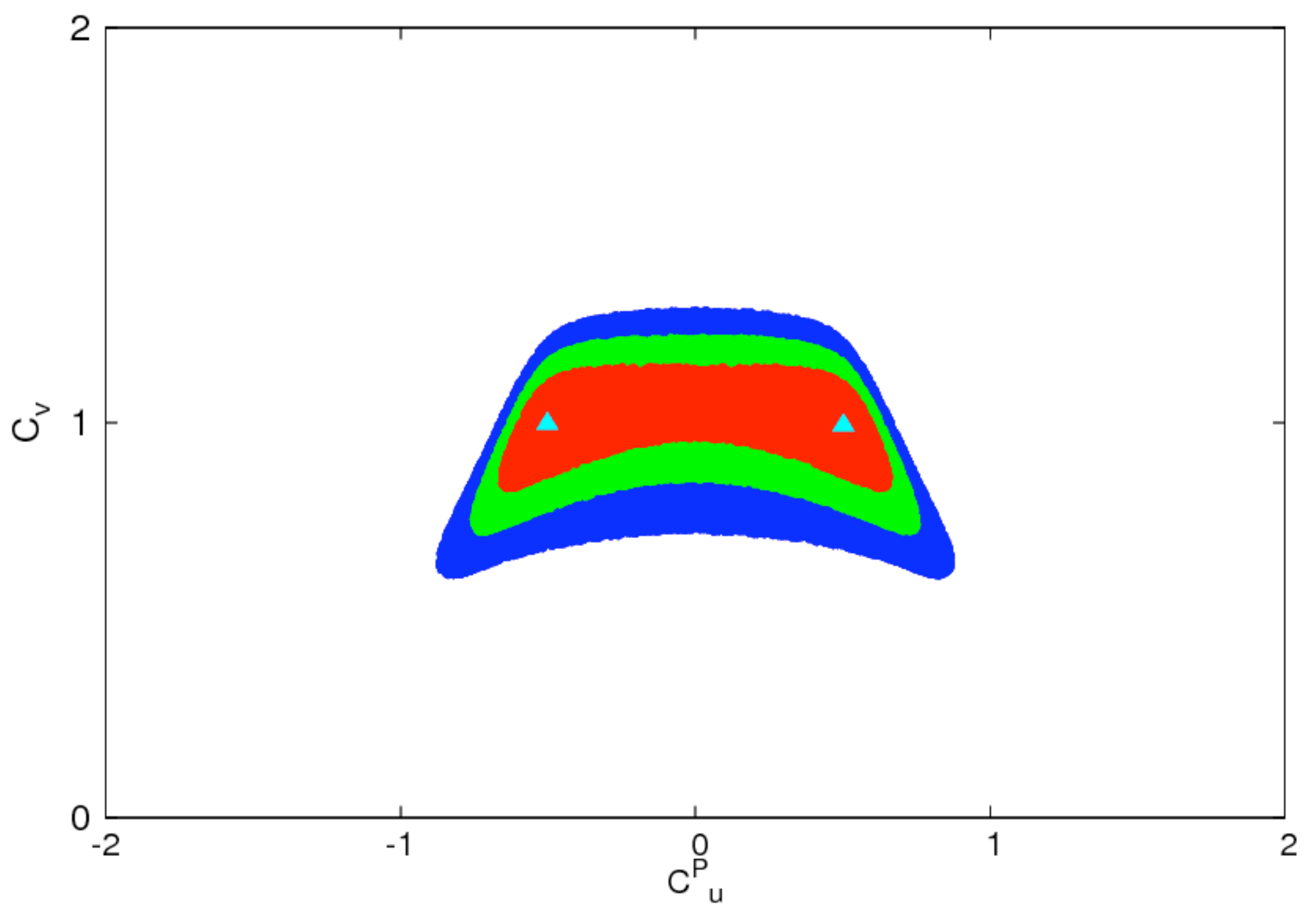}
\includegraphics[width=3.1in]{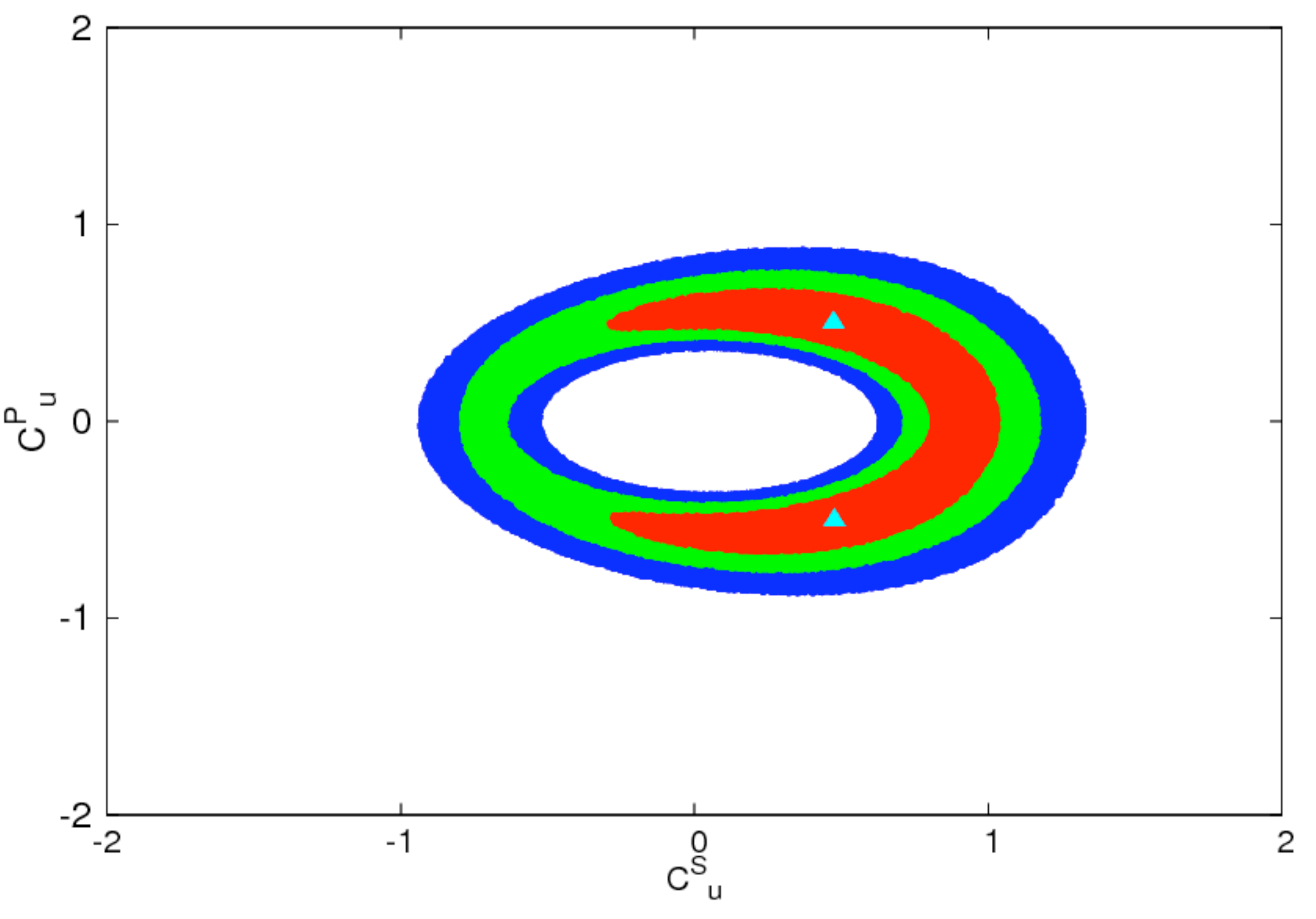}
\includegraphics[width=3.1in]{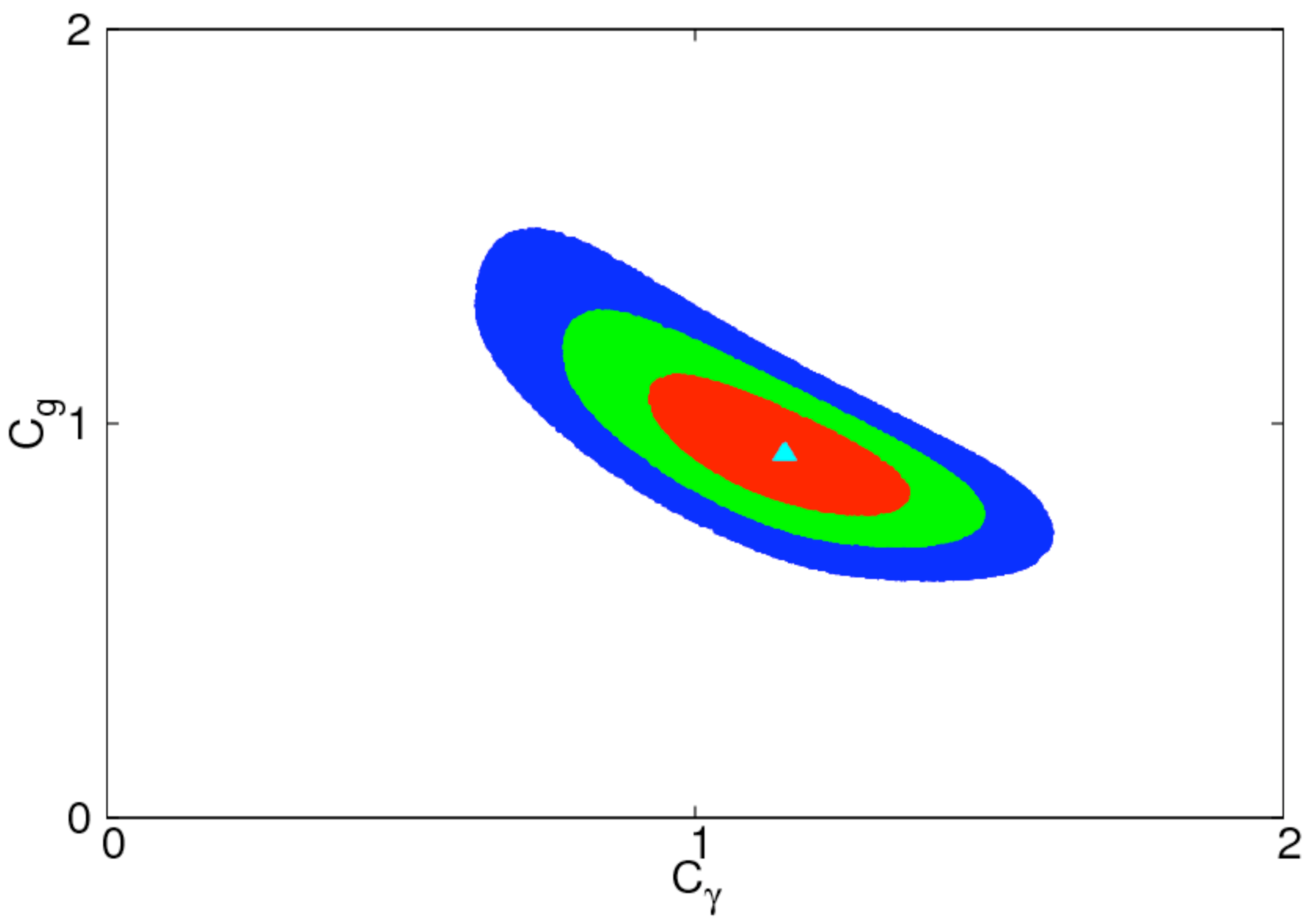}
\includegraphics[width=3.1in]{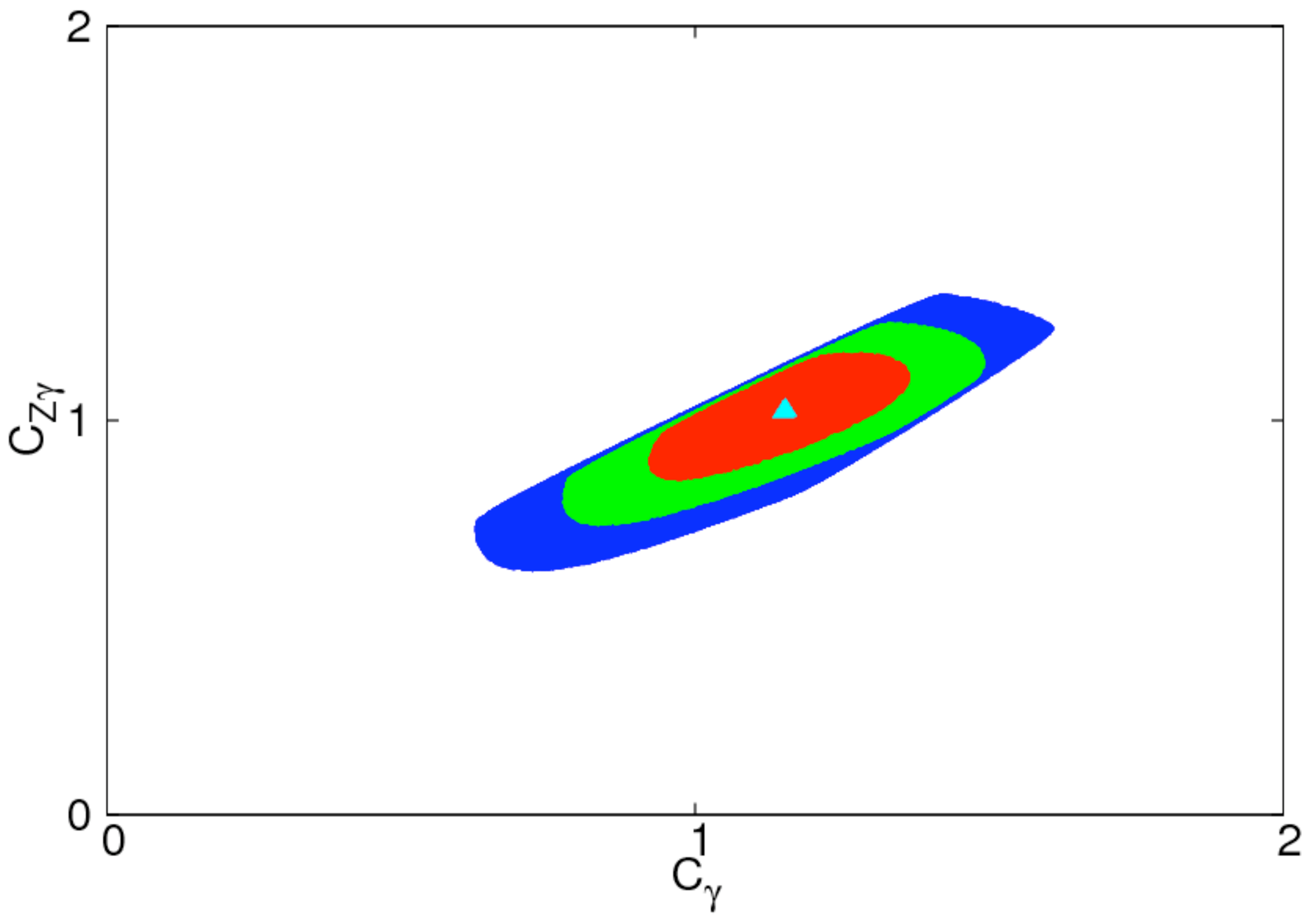}
\includegraphics[width=3.1in]{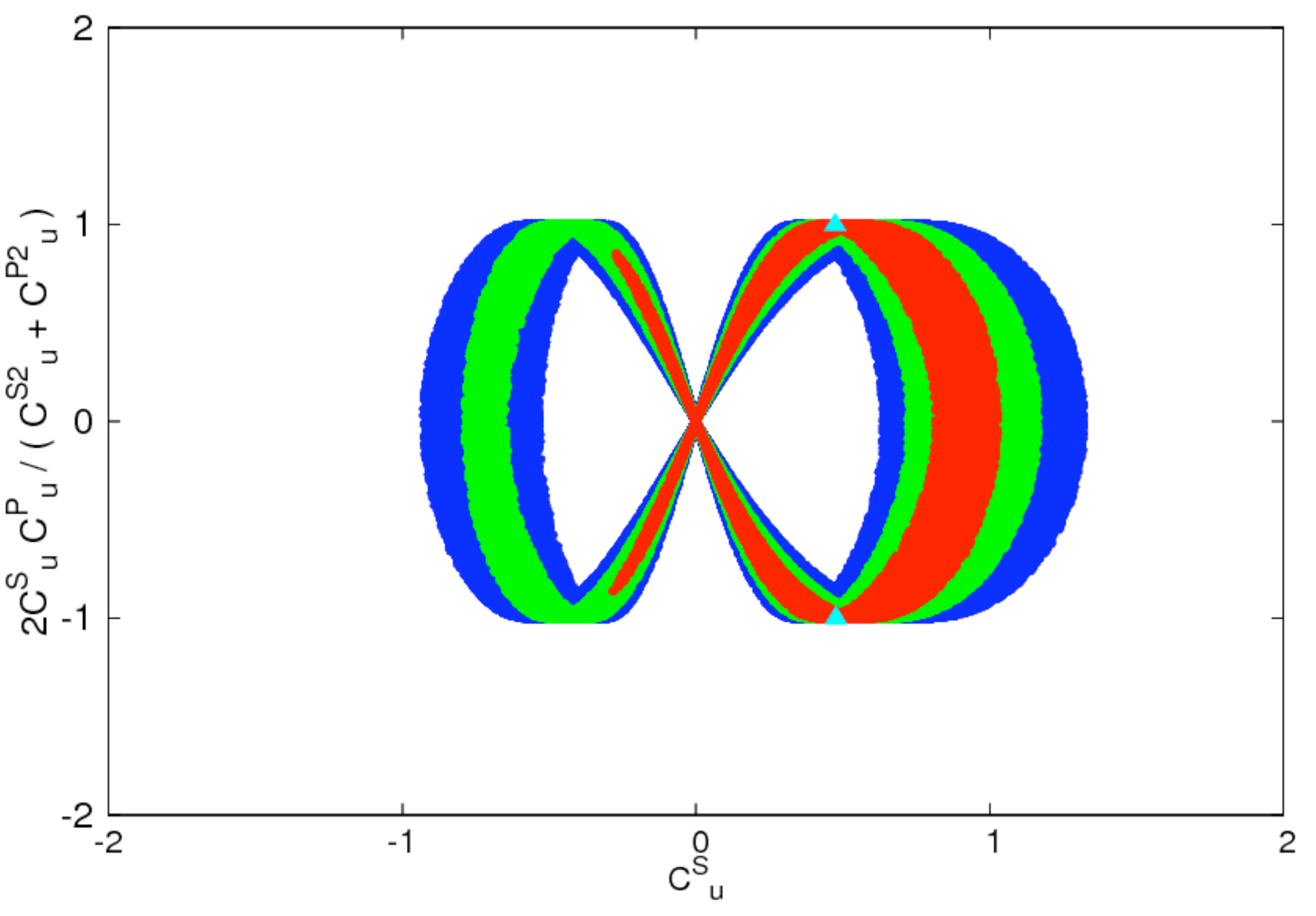}
\caption{\small \label{cpv2}
The confidence-level regions of the fit by varying 
the scalar Yukawa couplings $C_u^S$ and $C_v$,
and the pseudoscalar Yukawa couplings $C_u^P$;
while keeping 
$C_d^S = C_\ell^S=1$, $ C_d^P = C_\ell^P=0$, and 
$\Delta S^\gamma = \Delta S^g 
= \Delta P^\gamma = \Delta P^g = 
\Delta \Gamma_{\rm tot} = 0$.
The description of contour regions is the same as Fig.~\ref{case1}.
}
\end{figure}

The 2-dim confidence-level regions among the parameters $(C_u^S, C_u^P, 
C_v)$ are shown in Fig.~\ref{cpv2}. We can directly compare 
Fig.~\ref{case2-1}(a) and Fig.~\ref{cpv2}(a). 
The 2 islands in Fig.~\ref{case2-1}(a) 
are now linked together in Fig.~\ref{cpv2}(a), 
due to the variation of an additional  parameter $C_u^P$. 
The sickle-shaped region in part (c) indicates that $C_u^S$ and $C_u^P$ satisfy
some equations of ellipses. From the numerical expressions for
$S^\gamma,P^\gamma$ in Eq.~(\ref{5}) and $S^g,P^g$ in Eq.~(\ref{8}), we have 
\begin{eqnarray}
(7.3)^2 = ( - 8.4 + 1.76\, C_u^S )^2 + ( 2.78\, C_u^P)^2 \nonumber \\
(0.59)^2 = ( 0.688\,  C_u^S )^2 + (1.047\, C_u^P )^2 \label{cpveq2}
\end{eqnarray}
which can then explain the shape in part (c). 

The correlation between $C_\gamma$ and $C_{Z\gamma}$
is shown in part (e) and that for the CP-violating observables,
which are proportional to 
$2 C_u^S C_u^P/( {C_u^S}^2 + {C_u^P}^2 )$, is shown in part (f).
The $C_{Z\gamma}$ increases and decreases in the same direction as $C_\gamma$
but always smaller than $C_\gamma$. At the best-fit point, $C_\gamma \approx
1.1$ while $C_{Z\gamma} \approx 1.05$. 
The CP-violating observables arised from the mixing between scalar and 
pseudoscalar contributions are in general proportional to 
$2 C_u^S C_u^P/( {C_u^S}^2 + {C_u^P}^2 )$.

The best-fit parameters and the corresponding $\chi^2$ for this case are shown
in the fourth and the fifth columns of the lower half
in Table~\ref{cbestfit}. The $p$-value is slightly better than
the CP-conserving case of varying $C_{u,d,\ell}^S$ and $C_v$.

\section{Discussion}

In this work, we have established a model-independent framework that
enables one to analyze all the observed Higgs boson signal strengths
and to fit to the Higgs boson couplings to fermions, $W/Z$ bosons,
$\gamma\gamma$, and $gg$. In the future when the $Z\gamma$ data are
available, the current framework can also cover that. Right now we give 
predictions for $Z\gamma$ signal strengths.
We have performed global fits to all the Higgs signal strengths 
recorded by ATLAS, CMS, and at the Tevatron. 
Furthermore, we have also performed fits including the pseudoscalar
up-type Yukawa coupling $C_u^P$ and the pseudoscalar contributions 
$P^\gamma$ and $P^g$ in the $H\gamma\gamma$ and $Hgg$ vertices, respectively.

Furthermore, we have performed fits with respect to all the Higgs
data collected {\it before} and {\it after} the Moriond 2013
meetings.  The main reason why we separately performed them is
because of the dramatic change in the fits due to the shift of the
CMS diphoton data from $1.42$ to $0.78$ of the SM value.  Before the
Moriond update the dynamics of the fit pushes the parameters to
increase the diphoton rate such that the $\chi^2$ is reduced
effectively.  However, with all the data after the Moriond the
dynamics of the fit cannot find the optimal set of parameters so
that the resulting $\chi^2/dof$'s are indeed worse than the SM.

\begin{figure}[th!]
\centering
\includegraphics[width=5in]{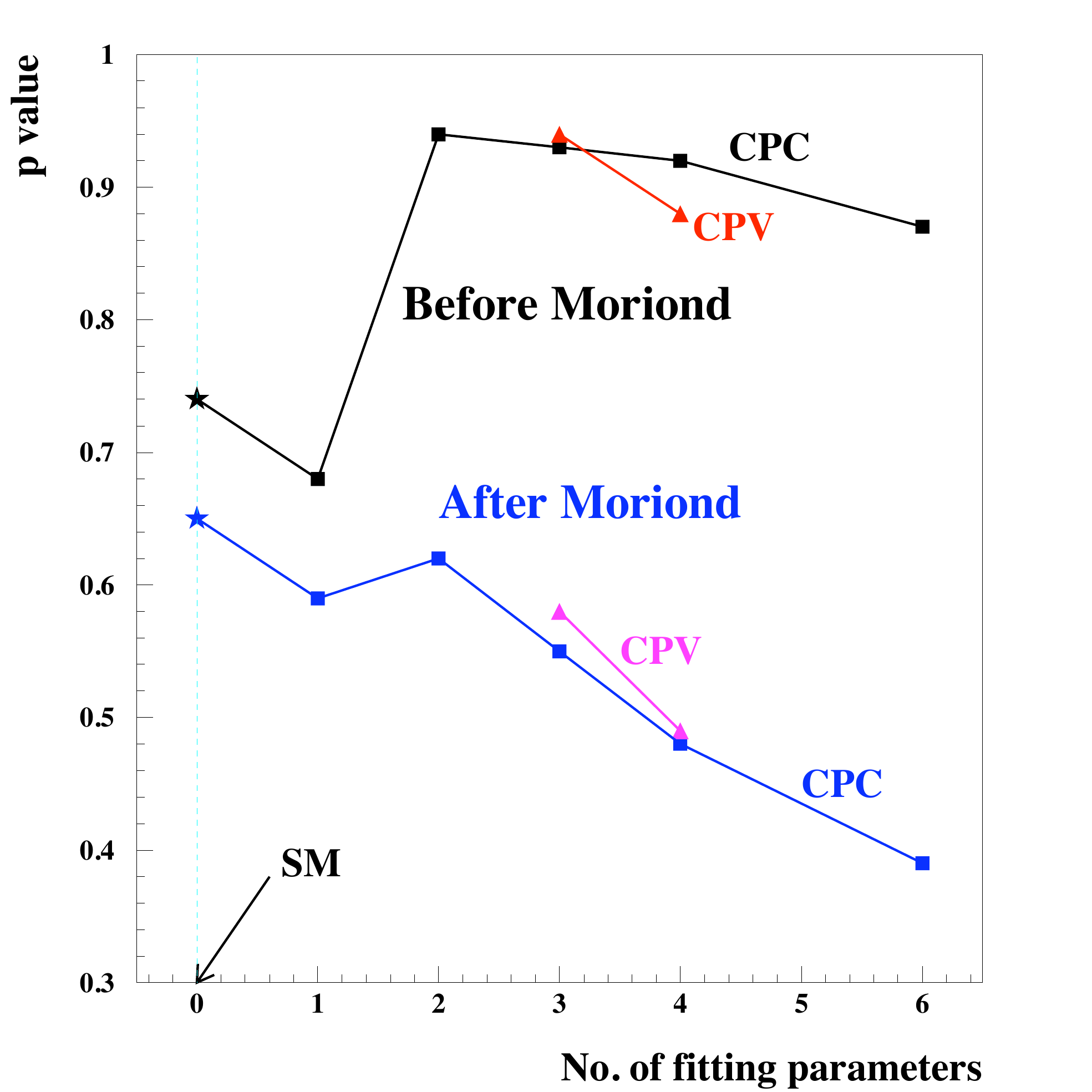}
\caption{\small \label{pvalue}
The $p$-values for various fits considered in this work, including
CP-conserving (CPC) and CP-violating (CPV) ones.
The CPC cases with the No. of fitting parameters equals 
1 denotes the case varying only $\Delta \Gamma_{\rm tot}$; 
2 denotes varying only $\Delta S^\gamma$ and $\Delta S^g$; 
3 denotes varying only $\Delta S^\gamma$, $\Delta S^g$, and 
   $\Delta \Gamma_{\rm tot}$;
4 denotes varying only $C_u^S$, $C_d^S$, $C_\ell^S$, and $C_v$;
6 denotes varying $C_u^S$, $C_d^S$, $C_\ell^S$, $C_v$, 
  $\Delta S^\gamma$, and $\Delta S^g$.
The CPV cases with the No. of fitting parameters equals 
3 denotes varying only $C_u^S$, $C_u^P$, and $C_v$;
4 denotes varying only $\Delta S^\gamma$, $\Delta S^g$, 
       $\Delta P^\gamma$, and $\Delta P^g$.
}
\end{figure}

In summary, the $p$-value of the SM Higgs boson is $0.65$
performed with all the data after the Moriond. Its $p$-value
is higher than any other fits considered in this work, both
the CP conserving ones and the CP-violating ones.
We also plot the $p$-values for all the fits considered in this work
in Fig.~\ref{pvalue}.

Our findings are summarized as follows.
\begin{enumerate}
\item 
Before the Moriond, the SM already enjoyed a good $\chi^2/dof = 0.8$
($p$-value $= 0.74$),
which means that the SM Higgs description of the data is reasonably well. 
Out of the total $\chi^2 = 17.5$ about one half comes from the 
$H\to\gamma\gamma$ data. 
Similarly, after the Moriond the SM has 
$\chi^2/dof = 0.86$ ($p$-value $=0.65$).

\item If only the total Higgs boson width is allowed to vary, we are able to 
constrain the deviation $\Delta \Gamma_{\rm tot}$ to be less than 
$1.2$ MeV
at 95\%CL. Given the SM Higgs boson width is about $4.1-4.2$ MeV for
$M_H = 125-126$ GeV, the nonstandard decay branching ratio of the Higgs
boson is less than 22\% at 95\% CL. 
This is a real improvement from previous estimates of about 40\%
\cite{invisible}.

\item The most efficient set of parameters to fit to the data before
the Moriond 
are the additional particle contributions to the loop functions of 
$H\gamma\gamma$ and $Hgg$ vertices, $\Delta S^\gamma$ and $\Delta S^g$
respectively. This is because both the ATLAS and CMS have diphoton data
above the SM value.
The best $\chi^2/dof$ obtained is about $0.56$. 
This is easy to understand, as the total $\chi^2$ is currently 
dominated by $H\to\gamma\gamma$ signal strength. 
Nevertheless, with the Moriond update the CMS diphoton data is now
below the SM value. No optimal set of parameters can be found to effectively
reduce the total $\chi^2$. 

\item With the data before the Moriond, another efficient set of parameters
are $C_u^S$ and $\Delta S^\gamma$.
We found that they are equally effective as ($\Delta S^\gamma,\Delta S^g$).
Effectively, the modification in $C_u^S$ takes up the place of 
$\Delta S^g$. Again, the reason is the domination of $H\to\gamma\gamma$
in the total $\chi^2$.  
Nevertheless, after the Moriond update no
optimal set of parameters can be found.

\item 
The relative $HVV$ coupling now stands at $C_v = 1.01\,^{+0.13}_{-0.14}$ in the 
6-parameter fit (and a similar value in the 4-parameter fit). This implies
that the observed Higgs boson accounts for most of the EWSB, 
and leaves little rooms for additional Higgs bosons that are
also responsible for EWSB.  Nevertheless, if we take $-2\sigma$ to the
central value, the $C_v$ can be as low as $0.7$.  The vector-boson scattering
could become strong if the UV part of the Higgs sector is very heavy
\cite{partial}.

\item The current data do not rule out 
the pseudoscalar contributions to
the $H\gamma\gamma$ and $Hgg$ vertices nor the pseudoscalar Yukawa couplings.
Nevertheless, including pseudoscalar contributions to $H\gamma\gamma$
and $Hgg$ vertices or pseudoscalar Yukawa couplings do not improve the fits.
The current Higgs observables are not sensitive to CP-violating effects,
and so only combinations of scalar and pseudoscalar contributions are 
constrained, as shown in Eqs.~(\ref{cpveq1}) and (\ref{cpveq2}).
Thus, the current Higgs data do not rule out or favor
pseudoscalar couplings.

\end{enumerate}

Era of Higgs-boson precision studies now begins -- Higgcision. 
In this work, we have already seen dramatic changes in the fits 
with the data collected {\it before} and {\it after} the Moriond 2013. 
We are looking forward to more and more data in the upcoming Summer 
2013 and the following years.

\section*{Acknowledgment}  
This work was supported the National Science
Council of Taiwan under Grants No. 99-2112-M-007-005-MY3 
and the WCU program through the KOSEF 
funded by the MEST (R31-2008-000-10057-0).  
This study was financially supported by Chonnam National University, 2012/2013.
J.S.L thanks National Center for Theoretical Sciences (Hsinchu, Taiwan) for the 
great hospitality
extended to him while this work was being performed.


\begin{table}[thb!]
  \caption{\small \label{t1}
    Data on signal strenghts of $H\rightarrow \gamma \gamma$
    recorded by ATLAS and CMS, and at the Tevatron {\it before} and
    {\it after} Moriond 2013. The luminosity updates at 8 TeV 
    are shown in the parenthesis.
    The percentages of each production mode in each data are given 
    (details are given in the text). 
    The $\chi^2$ of each data with respect to the SM
    is shown in the last two columns for {\it before} and {\it after}
    Moriond. The sub-totals $\chi^2$ of this decay mode
    are shown at the end.
  }
\begin{ruledtabular}
\begin{tabular}{c cc c cccc rr}
Channel & \multicolumn{2}{c}{Signal strength $\mu$} & $M_H$(GeV) & 
\multicolumn{4}{c}{Production mode} & 
   \multicolumn{2}{c}{$\chi^2_{\rm SM}$(each)} \\
        &  {\it Before} & {\it After}  &  & ggF & VBF & VH & ttH & {\it Before}
   & {\it After} \\
\hline
\multicolumn{10}{c}{ATLAS (4.8${\rm fb}^{-1}$ at 7TeV + 13.0 (20.7) 
fb$^{-1}$ at 8TeV):  \cite{atlas_h_aa,atlas_h_aa_2013} } \\
\hline
$\mu_{ggH+ttH}$ & $1.8\pm0.49$ &$1.6\pm 0.4$ & 126.8 & 100\% & - & - & - 
   & 2.67 & 2.25 \\
$\mu_{VBF}$ & $2.0\pm 1.4$ & $1.7\pm 0.9$ & 126.8 & - & 100\% & - & -& 0.53& 
  0.60 \\
$\mu_{VH}$ & $1.9\pm2.6$ & $1.8^{+1.5}_{-1.3}$ & 126.8 & - & - & 100\% & - 
  & 0.12 & 0.38 \\
\hline
\multicolumn{10}{c}{CMS (5.1${\rm fb}^{-1}$ at 7TeV + 5.3 (19.6)${\rm fb}^{-1}$ 
at 8TeV)  \cite{cms_com,cms_aa_2013} } \\
\hline
untagged & $1.42^{+0.55}_{-0.49}$ & $0.78^{+0.28}_{-0.26}$ & 125 
    & 87.5\% & 7.1\% & 4.9\% & 0.5\% & 0.73 & 0.62\\
VBF tagged & $2.25^{+1.34}_{-1.04}$ & $2.25^{+1.34}_{-1.04}$ & 125.8 & 17\% & 83\% & - & - & 1.44 & 1.44 \\
\hline
\multicolumn{10}{c}{Tevatron (10.0${\rm fb}^{-1}$ at 1.96TeV): \cite{tev}}  \\
\hline
Combined & $6.14^{+3.25}_{-3.19}$ & $6.14^{+3.25}_{-3.19}$ & 125 & 78\% & 5\% & 17\% & - & 2.60& 2.60 \\
\hline
&&&&&&& subtot:& 8.09 &  7.89 \\
\end{tabular}
\end{ruledtabular}
\end{table}

\begin{table}[thb!]
\caption{\small \label{t2}
The same as Table~\ref{t1} but for $H\rightarrow Z Z^{(\ast)}$.}
\begin{ruledtabular}
\begin{tabular}{c cc c cccc rr}
Channel & \multicolumn{2}{c}{Signal strength $\mu$} & $M_H$(GeV) & 
\multicolumn{4}{c}{Production mode} & \multicolumn{2}{c}{$\chi^2_{\rm SM}$(each)} 
\\
 & {\it Before} & {\it After} & & ggF & VBF & VH & ttH &{\it Before} & 
  {\it After} \\
\hline
\multicolumn{10}{c}
{ATLAS (4.8${\rm fb}^{-1}$ at 7TeV + 13 (20.7)${\rm fb}^{-1}$
  at 8TeV)  \cite{atlas_com,atlas_com_2013}}\\
\hline
Inclusive & $1.0\pm0.4$ &$1.5\pm 0.4$ & 125.5 
   & 87.5\% & 7.1\% & 4.9\% & 0.5\%  & 0.0& 1.56\\
\hline
\multicolumn{10}{c}
{CMS (5.1${\rm fb}^{-1}$ at 7TeV + 12.2 (19.6) ${\rm fb}^{-1}$ 
at 8TeV) \cite{cms_h_zz,cms_zz_2013} }\\
\hline
Inclusive & $0.80^{+0.35}_{-0.28}$ & $0.91^{+0.30}_{-0.24}$ & 125.8 
        & 87.5\% & 7.1\% & 4.9\% & 0.5\% &0.33 & 0.09 \\
\hline
&&&&&&& subtot: & 0.33 &  1.65
\end{tabular}
\end{ruledtabular}
\end{table}

\begin{table}[thb!]
\caption{\small \label{t3}
The same as Table~\ref{t1} but for $H\rightarrow W W^{(\ast)}$.}
\begin{ruledtabular}
\begin{tabular}{c cc c cccc rr}
Channel & \multicolumn{2}{c}{Signal strength $\mu$} & $M_H$(GeV) & 
\multicolumn{4}{c}{Production mode} & \multicolumn{2}{c}{$\chi^2_{\rm SM}$(each)}
 \\
  & {\it Before} & {\it After} & & ggF & VBF & VH & ttH & {\it Before} &
  {\it After} \\
\hline
\multicolumn{10}{c}
{ATLAS (4.8${\rm fb}^{-1}$ at 7TeV + 13 (20.7) ${\rm fb}^{-1}$ 
at 8TeV)  \cite{atlas_com,atlas_com_2013} }\\
\hline
Inclusive & $1.5\pm0.6$ & $1.0\pm 0.3$ & 125.5 
 & 87.5\% & 7.1\% & 4.9\% & 0.5\% & 0.69 & 0.00 \\
\hline
\multicolumn{10}{c}
{CMS (up to 4.9 ${\rm fb}^{-1}$ at 7TeV + 12.1 (19.5) 
${\rm fb}^{-1}$ at 8TeV) \cite{cms_com,cms_ww_2013} }\\
\hline
0/1 jet & $0.77^{+0.27}_{-0.25}$ & $0.76 \pm 0.21$ & 125 & 97\% & 3\% & - & - 
&0.73  & 1.31 \\
VBF tag & $-0.05^{+0.74}_{-0.55}$ & $-0.05^{+0.74}_{-0.55}$ & 125.8 & 17\% & 83\% & - & - &2.01 &2.01\\
VH tag & $-0.31^{+2.22}_{-1.94}$& $-0.31^{+2.22}_{-1.94}$ & 125.8 & - & - & 100\% & - & 0.35 & 0.35\\
\hline
\multicolumn{10}{c}
{Tevatron (10.0${\rm fb}^{-1}$ at 1.96TeV):  \cite{tev} }\\
\hline
Combined & $0.85^{+0.88}_{-0.81}$  &$0.85^{+0.88}_{-0.81}$ & 125 & 78\% & 5\% & 17\% & - & 0.03 & 0.03\\
\hline
&&&&&&& subtot:& 3.81 & 3.70 
\end{tabular}
\end{ruledtabular}
\end{table}
\begin{table}[thb!]
\caption{\small \label{t4}
The same as Table~\ref{t1} but for $H\rightarrow b\bar{b}$. There are no 
updates for this channel.}
\begin{ruledtabular}
\begin{tabular}{cccccccr}
Channel & Signal strength $\mu$ & $M_H$(GeV) & 
 \multicolumn{4}{c}{Production mode} & $\chi^2_{\rm SM}$(each) \\
        &      &            & ggF & VBF & VH & ttH \\
\hline
\multicolumn{7}{c}
{ATLAS (4.8${\rm fb}^{-1}$ at 7TeV + 13.0${\rm fb}^{-1}$ at 8TeV)
   \cite{atlas_com_2013}}\\
\hline
VH tag & $-0.4\pm 1.0$ & 125.5 & - & - & 100\% & - & 1.96 \\
\hline
\multicolumn{7}{c}
{CMS (up to 5.0${\rm fb}^{-1}$ at 7TeV + 12.1${\rm fb}^{-1}$ at 8TeV)
 \cite{cms_com} }\\
\hline
VH tag & $1.31^{+0.65}_{-0.60}$ & 125.8 & - & - & 100\% & - & 0.27 \\
ttH tag & $-0.80^{+2.10}_{-1.84}$ & 125.8 & - & - & - & 100\% & 0.73\\
\hline
\multicolumn{7}{c}
{Tevatron (10.0${\rm fb}^{-1}$ at 1.96TeV): \cite{tev_h_bb} }\\
\hline
VH tag & $1.56^{+0.72}_{-0.73}$ & 125 & - & - & 100\% & - & 0.59 \\
\hline
&&&&&&& subtot: 3.55 
\end{tabular}
\end{ruledtabular}
\end{table}
\begin{table}[thb!]
\caption{\small \label{t5}
The same as Table~\ref{t1} but for $H\rightarrow \tau \tau$. 
The correlation for the $\tau\tau$ data of ATLAS is $\rho=-0.50$ and $-0.49$
before and after Moriond, respectively.
The percentages of the production modes differ very tiny 
before and after Moriond. }
\begin{ruledtabular}
\begin{tabular}{c cc c cccc rr}
Channel & \multicolumn{2}{c}{Signal strength $\mu$} & $M_H$(GeV) & 
\multicolumn{4}{c}{Production mode} & \multicolumn{2}{c}{$\chi^2_{\rm SM}$(each)}
 \\
  & {\it Before} & {\it After} && ggF & VBF & VH & ttH & {\it Before} & 
  {\it After} \\
\hline
\multicolumn{10}{c}
{ATLAS (4.6${\rm fb}^{-1}$ at 7TeV + 13.0${\rm fb}^{-1}$ at 8TeV)
\cite{atlas_h_tau,atlas_com_2013}}\\
\hline
$\mu(ggF)$ & $2.38\pm 1.57$ & $2.30\pm 1.60$ & 125.5 
 & 100\% & - & - & - & 1.60 & 1.41\\
$\mu(VBF+VH)$ & $-0.25\pm 1.02 $ & $-0.22\pm 1.06$ & 125.5 & - 
  & 59.4\% & 40.6\% & - &  \\
\hline
\multicolumn{10}{c}
{CMS (up to 4.9${\rm fb}^{-1}$ at 7TeV + 12.1 (19.4) 
${\rm fb}^{-1}$ at 8TeV) \cite{cms_com,cms_tau_2013} }\\
\hline
0/1 jet & $0.85^{+0.68}_{-0.66}$ & $0.76^{+0.50}_{-0.52}$ & 125 & 
77.8\% & 13.8\% & 7.6\% & 0.8\% & 0.05 & 0.23 \\
VBF tag & $0.82^{+0.82}_{-0.75}$ & $1.40^{+0.59}_{-0.57}$ & 125 & 20.9\% 
 & 79.1\% & - & - & 0.05 & 0.49 \\
VH tag & $0.86^{+1.92}_{-1.68}$ & $0.77^{+1.49}_{-1.42}$ & 125 & - & - & 100\% &
  - & 0.005 & 0.02 \\
\hline
&&&&&&&subtot: & 1.70 &  2.15
\end{tabular}
\end{ruledtabular}
\end{table}

\begin{table}[th!]
\caption{\small \label{bestfit}
The best fitted values and the $1\sigma$ errors for the parameters in
various CP conserving fits and the corresponding chi-square per degree
of freedom and the $p$-value before and after the Moriond 2013. The $p$-values
for the SM fit are $0.74$ and $0.65$ for the data before and after the
Moriond, respectively.
}
\begin{ruledtabular}
\begin{tabular}{c|ccccc}
            & Vary $\Delta \Gamma_{\rm tot}$ 
            & Vary $\Delta S^\gamma$, 
            & Vary $\Delta S^\gamma$, 
            & Vary $C_u^S$, $C_d^S$, 
    & Vary $C_u^S$, $C_d^S$, $C_\ell^S$, $C_v$\\
Parameters  & & $\Delta S^g$ & $\Delta S^g$, $\Delta \Gamma_{\rm tot}$
   & $C_\ell^S$, $C_v$ &  $\Delta S^\gamma$, $\Delta S^g$ \\
\hline 
 \multicolumn{6}{c}{Before Moriond} \\
\hline
  $C_u^S$     & 1 & 1 & 1 & $-0.88^{+0.16}_{-0.21}$ & $0.00\pm{1.13}$ \\
  $C_d^S$     & 1 & 1 & 1 & $1.12^{+0.45}_{-0.38}$ & $1.19^{+0.57}_{-0.41}$ \\
  $C_\ell^S$ & 1 & 1 & 1 & $-0.97^{+0.30}_{-0.29}$ & $0.98\pm{0.30}$ \\
  $C_v$       & 1 & 1 & 1 & $0.97^{+0.13}_{-0.15}$ & $0.96^{+0.13}_{-0.15}$ \\
$\Delta S^\gamma$&0 & $-2.73^{+1.11}_{-1.15}$ & $-2.93^{+1.19}_{-1.31}$ & 0 &
                                      $-1.23^{+2.44}_{-2.49}$ \\
$\Delta S^g$   & 0 & $-0.050^{+0.064}_{-0.065}$ & $0.0063^{+0.15}_{-0.11}$ & 0&
                                      $0.73^{+0.81}_{-0.80}$ \\
$\Delta \Gamma_{\rm tot}$ (MeV) & $-0.022^{+0.63}_{-0.48}$& 0 &$0.79^{+2.01}_{-1.11}$ &0&
 0 \\
\hline
 $\chi^2/dof$ &$17.48/21$ & $11.27/20$ & $10.83/19$ & $10.46/18$ & $9.89/16$ \\
$p$-value& $0.68$ & $0.94$ & $0.93$ & $0.92$ & $0.87$ \\
\hline
\hline
\multicolumn{6}{c}{After Moriond} \\
\hline
  $C_u^S$     & 1 & 1 & 1 & $0.80^{+0.16}_{-0.13}$ & $0.00\pm{1.18}$ \\
  $C_d^S$     & 1 & 1 & 1 & $-0.98^{+0.31}_{-0.34}$ & $1.06^{+0.41}_{-0.35}$ \\
  $C_\ell^S$ & 1 & 1 & 1 & $0.98^{+0.21}_{-0.21}$ & $1.01\pm{0.23}$ \\
  $C_v$       & 1 & 1 & 1 & $1.04^{+0.12}_{-0.14}$ & $1.01^{+0.13}_{-0.14}$ \\
$\Delta S^\gamma$&0 & $-0.96^{+0.84}_{-0.85}$ & $-0.96^{+0.84}_{-0.87}$ & 0 &
                                      $ 0.78^{+2.34}_{-2.28}$ \\
$\Delta S^g$   & 0 & $-0.043 \pm{0.052}$ & $-0.040^{+0.12}_{-0.086}$ & 0&
                                      $0.66^{+0.42}_{-0.83}$ \\
$\Delta \Gamma_{\rm tot}$ (MeV) & $ 0.10^{+0.51}_{-0.41}$& 0 &
$0.027^{+1.33}_{-0.80}$ &0&  0 \\
\hline
 $\chi^2/dof$ &$18.89/21$ & $17.55/20$ & $17.55/19$ & $17.82/18$ & $16.89/16$ \\
$p$-value & $0.59$ & $0.62$ & $0.55$ & $0.48$ & $0.39$ 
\end{tabular}
\end{ruledtabular}
\end{table}

\begin{table}[th!]
\caption{\small \label{cbestfit}
The best fitted values and the $1\sigma$ errors for the parameters in
the CP-violating fits and the corresponding chi-square
before and after Moriond 2013.
}
\begin{ruledtabular}
\begin{tabular}{c|cc|cc }
Parameters & \multicolumn{2}{c|}
   {Vary $\Delta S^\gamma$, $\Delta S^g$,$\Delta P^\gamma$, $\Delta P^g$}   
    & \multicolumn{2}{c}{Vary $C_u^S$,$C_u^P$, $C_v$}     \\
\hline 
\multicolumn{5}{c}{Before Moriond} \\
\hline
  $C_u^S$   & 1  & 1 & $-0.54^{+0.65}_{-0.39}$ & $-0.54^{+0.65}_{-0.39}$ \\
  $C_d^S$   & 1 & 1 & 1 & 1\\
  $C_\ell^S$ & 1 & 1 & 1 & 1  \\
  $C_v$      & 1 & 1 & $0.93^{+0.10}_{-0.12}$  & $0.93^{+0.10}_{-0.12}$  \\
  $\Delta S^\gamma$&  $0.28^{+16.9}_{-4.16}$ & $-0.28^{+17.4}_{-3.60}$ & 0 & 0 \\
  $\Delta S^g$    & $-0.62^{+0.63}_{-0.70}$ & $-0.62^{+0.63}_{-0.70}$ & 0 & 0 \\
$\Delta \Gamma_{\rm tot}$ (MeV) & 0 & 0 & 0 &0 \\
\hline
$C_u^P$     & 0 & 0 &  $-0.46^{+1.14}_{-0.22}$ & $0.46^{+1.14}_{-0.22}$ \\
$\Delta P^\gamma$&  $-6.88^{+17.4}_{-3.64}$ & $-6.31^{+16.8}_{-4.21}$ & 0 &0 \\
$\Delta P^g$   & $0.60^{+0.065}_{-1.26}$ & $0.60^{+0.065}_{-1.26}$ & 0  &0 \\
\hline
 $\chi^2/dof$ &$11.26/18$ & $11.26/18$ & $10.53/19$ & $10.53/19 $  \\
$p$-value & $0.88$ & $0.88$ & $0.94$ & $0.94$ \\
\hline
\multicolumn{5}{c}{After Moriond}\\
\hline
  $C_u^S$   & 1  & 1 & $0.48^{+0.44}_{-0.48}$ & $0.48^{+0.44}_{-0.48}$ \\
  $C_d^S$   & 1 & 1 & 1 & 1\\
  $C_\ell^S$ & 1 & 1 & 1 & 1  \\
  $C_v$      & 1 & 1 & $0.995^{+0.097}_{-0.104}$  & $0.995^{+0.097}_{-0.104}$  \\
  $\Delta S^\gamma$&  $-0.92^{+16.00}_{-0.89}$ & $0.71^{+14.37}_{-2.51}$ & 0 & 0 \\
  $\Delta S^g$    & $-0.55^{+0.56}_{-0.76}$ & $-0.64^{+0.65}_{-0.67}$ & 0 & 0 \\
$\Delta \Gamma_{\rm tot}$ (MeV) & 0 & 0 & 0 &0 \\
\hline
$C_u^P$     & 0 & 0 &  $0.50^{+0.11}_{-0.40}$ & $-0.50^{+0.44}_{-0.11}$ \\
$\Delta P^\gamma$&  $0.77^{+7.67}_{-9.21}$ & $4.75^{+3.69}_{-13.19}$ & 0 &0 \\
$\Delta P^g$   & $-0.60^{+1.26}_{-0.06}$ & $-0.61^{+1.27}_{-0.052}$ & 0  &0 \\
\hline
 $\chi^2/dof$ &$17.55/18$ & $17.55/18$ & $17.17/19$ & $17.17/19 $  \\
$p$-value & $0.49$ & $0.49$ &  $0.58$ & $0.58$ 
\end{tabular}
\end{ruledtabular}
\end{table}

\newpage

\end{document}